% File jytex.tex, for jyTeX version 2.6M (June 1992)
% Copyright (c) 1991, 1992 by Jonathan P. Yamron
% For full documentation, "get jydoc" from hep-ph@xxx.lanl.gov
%   Problems?  Contact brahm@theory3.caltech.edu.

\catcode`\@=11

%*****************************************************************************

\message{Loading jyTeX fonts...}

%************************************************************
%*
%*             Available fonts
%*
%************************************************************

%************** 5-point fonts *******************************

\font\vptrm=cmr5 \font\vptmit=cmmi5 \font\vptsy=cmsy5 \font\vptbf=cmbx5

\skewchar\vptmit='177 \skewchar\vptsy='60 \fontdimen16 \vptsy=\the\fontdimen17 \vptsy

\def\vpt{\ifmmode\err@badsizechange\else
     \@mathfontinit
     \textfont0=\vptrm  \scriptfont0=\vptrm  \scriptscriptfont0=\vptrm
     \textfont1=\vptmit \scriptfont1=\vptmit \scriptscriptfont1=\vptmit
     \textfont2=\vptsy  \scriptfont2=\vptsy  \scriptscriptfont2=\vptsy
     \textfont3=\xptex  \scriptfont3=\xptex  \scriptscriptfont3=\xptex
     \textfont\bffam=\vptbf
     \scriptfont\bffam=\vptbf
     \scriptscriptfont\bffam=\vptbf
     \@fontstyleinit
     \def\rm{\vptrm\fam=\z@}%
     \def\bf{\vptbf\fam=\bffam}%
     \def\oldstyle{\vptmit\fam=\@ne}%
     \rm\fi}

%************** 6-point fonts *******************************

\font\viptrm=cmr6 \font\viptmit=cmmi6 \font\viptsy=cmsy6 \font\viptbf=cmbx6

\skewchar\viptmit='177 \skewchar\viptsy='60 \fontdimen16 \viptsy=\the\fontdimen17
\viptsy

\def\vipt{\ifmmode\err@badsizechange\else
     \@mathfontinit
     \textfont0=\viptrm  \scriptfont0=\vptrm  \scriptscriptfont0=\vptrm
     \textfont1=\viptmit \scriptfont1=\vptmit \scriptscriptfont1=\vptmit
     \textfont2=\viptsy  \scriptfont2=\vptsy  \scriptscriptfont2=\vptsy
     \textfont3=\xptex   \scriptfont3=\xptex  \scriptscriptfont3=\xptex
     \textfont\bffam=\viptbf
     \scriptfont\bffam=\vptbf
     \scriptscriptfont\bffam=\vptbf
     \@fontstyleinit
     \def\rm{\viptrm\fam=\z@}%
     \def\bf{\viptbf\fam=\bffam}%
     \def\oldstyle{\viptmit\fam=\@ne}%
     \rm\fi}
%************** 7-point fonts *******************************

\font\viiptrm=cmr7 \font\viiptmit=cmmi7 \font\viiptsy=cmsy7 \font\viiptit=cmti7
\font\viiptbf=cmbx7

\skewchar\viiptmit='177 \skewchar\viiptsy='60 \fontdimen16 \viiptsy=\the\fontdimen17
\viiptsy

\def\viipt{\ifmmode\err@badsizechange\else
     \@mathfontinit
     \textfont0=\viiptrm  \scriptfont0=\vptrm  \scriptscriptfont0=\vptrm
     \textfont1=\viiptmit \scriptfont1=\vptmit \scriptscriptfont1=\vptmit
     \textfont2=\viiptsy  \scriptfont2=\vptsy  \scriptscriptfont2=\vptsy
     \textfont3=\xptex    \scriptfont3=\xptex  \scriptscriptfont3=\xptex
     \textfont\itfam=\viiptit
     \scriptfont\itfam=\viiptit
     \scriptscriptfont\itfam=\viiptit
     \textfont\bffam=\viiptbf
     \scriptfont\bffam=\vptbf
     \scriptscriptfont\bffam=\vptbf
     \@fontstyleinit
     \def\rm{\viiptrm\fam=\z@}%
     \def\it{\viiptit\fam=\itfam}%
     \def\bf{\viiptbf\fam=\bffam}%
     \def\oldstyle{\viiptmit\fam=\@ne}%
     \rm\fi}

%************** 8-point fonts *******************************

\font\viiiptrm=cmr8 \font\viiiptmit=cmmi8 \font\viiiptsy=cmsy8 \font\viiiptit=cmti8
%\font\viiiptsl=cmsl8
\font\viiiptbf=cmbx8
%\font\viiipttt=cmtt8
%\font\viiiptss=cmss8

\skewchar\viiiptmit='177 \skewchar\viiiptsy='60 \fontdimen16 \viiiptsy=\the\fontdimen17
\viiiptsy

\def\viiipt{\ifmmode\err@badsizechange\else
     \@mathfontinit
     \textfont0=\viiiptrm  \scriptfont0=\viptrm  \scriptscriptfont0=\vptrm
     \textfont1=\viiiptmit \scriptfont1=\viptmit \scriptscriptfont1=\vptmit
     \textfont2=\viiiptsy  \scriptfont2=\viptsy  \scriptscriptfont2=\vptsy
     \textfont3=\xptex     \scriptfont3=\xptex   \scriptscriptfont3=\xptex
     \textfont\itfam=\viiiptit
     \scriptfont\itfam=\viiptit
     \scriptscriptfont\itfam=\viiptit
     \textfont\bffam=\viiiptbf
     \scriptfont\bffam=\viptbf
     \scriptscriptfont\bffam=\vptbf
     \@fontstyleinit
     \def\rm{\viiiptrm\fam=\z@}%
     \def\it{\viiiptit\fam=\itfam}%
     \def\bf{\viiiptbf\fam=\bffam}%
     \def\oldstyle{\viiiptmit\fam=\@ne}%
     \rm\fi}

%************** Optional 9-point fonts **********************

\def\getixpt{%
     \font\ixptrm=cmr9
     \font\ixptmit=cmmi9
     \font\ixptsy=cmsy9
     \font\ixptit=cmti9
%     \font\ixptsl=cmsl9
     \font\ixptbf=cmbx9
%     \font\ixpttt=cmtt9
%     \font\ixptss=cmss9
     \skewchar\ixptmit='177 \skewchar\ixptsy='60
     \fontdimen16 \ixptsy=\the\fontdimen17 \ixptsy}

\def\ixpt{\ifmmode\err@badsizechange\else
     \@mathfontinit
     \textfont0=\ixptrm  \scriptfont0=\viiptrm  \scriptscriptfont0=\vptrm
     \textfont1=\ixptmit \scriptfont1=\viiptmit \scriptscriptfont1=\vptmit
     \textfont2=\ixptsy  \scriptfont2=\viiptsy  \scriptscriptfont2=\vptsy
     \textfont3=\xptex   \scriptfont3=\xptex    \scriptscriptfont3=\xptex
     \textfont\itfam=\ixptit
     \scriptfont\itfam=\viiptit
     \scriptscriptfont\itfam=\viiptit
     \textfont\bffam=\ixptbf
     \scriptfont\bffam=\viiptbf
     \scriptscriptfont\bffam=\vptbf
     \@fontstyleinit
     \def\rm{\ixptrm\fam=\z@}%
     \def\it{\ixptit\fam=\itfam}%
     \def\bf{\ixptbf\fam=\bffam}%
     \def\oldstyle{\ixptmit\fam=\@ne}%
     \rm\fi}

%************** 10-point fonts ******************************

\font\xptrm=cmr10 \font\xptmit=cmmi10 \font\xptsy=cmsy10 \font\xptex=cmex10
\font\xptit=cmti10 \font\xptsl=cmsl10 \font\xptbf=cmbx10 \font\xpttt=cmtt10
\font\xptss=cmss10 \font\xptsc=cmcsc10 \font\xptbfs=cmb10 \font\xptbmit=cmmib10

\skewchar\xptmit='177 \skewchar\xptbmit='177 \skewchar\xptsy='60 \fontdimen16
\xptsy=\the\fontdimen17 \xptsy

\def\xpt{\ifmmode\err@badsizechange\else
     \@mathfontinit
     \textfont0=\xptrm  \scriptfont0=\viiptrm  \scriptscriptfont0=\vptrm
     \textfont1=\xptmit \scriptfont1=\viiptmit \scriptscriptfont1=\vptmit
     \textfont2=\xptsy  \scriptfont2=\viiptsy  \scriptscriptfont2=\vptsy
     \textfont3=\xptex  \scriptfont3=\xptex    \scriptscriptfont3=\xptex
     \textfont\itfam=\xptit
     \scriptfont\itfam=\viiptit
     \scriptscriptfont\itfam=\viiptit
     \textfont\bffam=\xptbf
     \scriptfont\bffam=\viiptbf
     \scriptscriptfont\bffam=\vptbf
     \textfont\bfsfam=\xptbfs
     \scriptfont\bfsfam=\viiptbf
     \scriptscriptfont\bfsfam=\vptbf
     \textfont\bmitfam=\xptbmit
     \scriptfont\bmitfam=\viiptmit
     \scriptscriptfont\bmitfam=\vptmit
     \@fontstyleinit
     \def\rm{\xptrm\fam=\z@}%
     \def\it{\xptit\fam=\itfam}%
     \def\sl{\xptsl}%
     \def\bf{\xptbf\fam=\bffam}%
     \def\tt{\xpttt}%
     \def\ss{\xptss}%
     \def\sc{\xptsc}%
     \def\bfs{\xptbfs\fam=\bfsfam}%
     \def\bmit{\fam=\bmitfam}%
     \def\oldstyle{\xptmit\fam=\@ne}%
     \rm\fi}

%************** Optional 11-point fonts *********************

\def\getxipt{%
     \font\xiptrm=cmr10  scaled\magstephalf
     \font\xiptmit=cmmi10 scaled\magstephalf
     \font\xiptsy=cmsy10 scaled\magstephalf
     \font\xiptex=cmex10 scaled\magstephalf
     \font\xiptit=cmti10 scaled\magstephalf
     \font\xiptsl=cmsl10 scaled\magstephalf
     \font\xiptbf=cmbx10 scaled\magstephalf
     \font\xipttt=cmtt10 scaled\magstephalf
     \font\xiptss=cmss10 scaled\magstephalf
     \skewchar\xiptmit='177 \skewchar\xiptsy='60
     \fontdimen16 \xiptsy=\the\fontdimen17 \xiptsy}

\def\xipt{\ifmmode\err@badsizechange\else
     \@mathfontinit
     \textfont0=\xiptrm  \scriptfont0=\viiiptrm  \scriptscriptfont0=\viptrm
     \textfont1=\xiptmit \scriptfont1=\viiiptmit \scriptscriptfont1=\viptmit
     \textfont2=\xiptsy  \scriptfont2=\viiiptsy  \scriptscriptfont2=\viptsy
     \textfont3=\xiptex  \scriptfont3=\xptex     \scriptscriptfont3=\xptex
     \textfont\itfam=\xiptit
     \scriptfont\itfam=\viiiptit
     \scriptscriptfont\itfam=\viiptit
     \textfont\bffam=\xiptbf
     \scriptfont\bffam=\viiiptbf
     \scriptscriptfont\bffam=\viptbf
     \@fontstyleinit
     \def\rm{\xiptrm\fam=\z@}%
     \def\it{\xiptit\fam=\itfam}%
     \def\sl{\xiptsl}%
     \def\bf{\xiptbf\fam=\bffam}%
     \def\tt{\xipttt}%
     \def\ss{\xiptss}%
     \def\oldstyle{\xiptmit\fam=\@ne}%
     \rm\fi}

%************** 12-point fonts ******************************

\font\xiiptrm=cmr12 \font\xiiptmit=cmmi12 \font\xiiptsy=cmsy10 scaled\magstep1
\font\xiiptex=cmex10  scaled\magstep1 \font\xiiptit=cmti12 \font\xiiptsl=cmsl12
\font\xiiptbf=cmbx12
%\font\xiipttt=cmtt12
\font\xiiptss=cmss12 \font\xiiptsc=cmcsc10 scaled\magstep1 \font\xiiptbfs=cmb10
scaled\magstep1 \font\xiiptbmit=cmmib10 scaled\magstep1

\skewchar\xiiptmit='177 \skewchar\xiiptbmit='177 \skewchar\xiiptsy='60 \fontdimen16
\xiiptsy=\the\fontdimen17 \xiiptsy

\def\xiipt{\ifmmode\err@badsizechange\else
     \@mathfontinit
     \textfont0=\xiiptrm  \scriptfont0=\viiiptrm  \scriptscriptfont0=\viptrm
     \textfont1=\xiiptmit \scriptfont1=\viiiptmit \scriptscriptfont1=\viptmit
     \textfont2=\xiiptsy  \scriptfont2=\viiiptsy  \scriptscriptfont2=\viptsy
     \textfont3=\xiiptex  \scriptfont3=\xptex     \scriptscriptfont3=\xptex
     \textfont\itfam=\xiiptit
     \scriptfont\itfam=\viiiptit
     \scriptscriptfont\itfam=\viiptit
     \textfont\bffam=\xiiptbf
     \scriptfont\bffam=\viiiptbf
     \scriptscriptfont\bffam=\viptbf
     \textfont\bfsfam=\xiiptbfs
     \scriptfont\bfsfam=\viiiptbf
     \scriptscriptfont\bfsfam=\viptbf
     \textfont\bmitfam=\xiiptbmit
     \scriptfont\bmitfam=\viiiptmit
     \scriptscriptfont\bmitfam=\viptmit
     \@fontstyleinit
     \def\rm{\xiiptrm\fam=\z@}%
     \def\it{\xiiptit\fam=\itfam}%
     \def\sl{\xiiptsl}%
     \def\bf{\xiiptbf\fam=\bffam}%
     \def\tt{\xiipttt}%
     \def\ss{\xiiptss}%
     \def\sc{\xiiptsc}%
     \def\bfs{\xiiptbfs\fam=\bfsfam}%
     \def\bmit{\fam=\bmitfam}%
     \def\oldstyle{\xiiptmit\fam=\@ne}%
     \rm\fi}

%************** Optional 13-point fonts *********************

\def\getxiiipt{%
     \font\xiiiptrm=cmr12  scaled\magstephalf
     \font\xiiiptmit=cmmi12 scaled\magstephalf
     \font\xiiiptsy=cmsy9  scaled\magstep2
     \font\xiiiptit=cmti12 scaled\magstephalf
     \font\xiiiptsl=cmsl12 scaled\magstephalf
     \font\xiiiptbf=cmbx12 scaled\magstephalf
     \font\xiiipttt=cmtt12 scaled\magstephalf
     \font\xiiiptss=cmss12 scaled\magstephalf
     \skewchar\xiiiptmit='177 \skewchar\xiiiptsy='60
     \fontdimen16 \xiiiptsy=\the\fontdimen17 \xiiiptsy}

\def\xiiipt{\ifmmode\err@badsizechange\else
     \@mathfontinit
     \textfont0=\xiiiptrm  \scriptfont0=\xptrm  \scriptscriptfont0=\viiptrm
     \textfont1=\xiiiptmit \scriptfont1=\xptmit \scriptscriptfont1=\viiptmit
     \textfont2=\xiiiptsy  \scriptfont2=\xptsy  \scriptscriptfont2=\viiptsy
     \textfont3=\xivptex   \scriptfont3=\xptex  \scriptscriptfont3=\xptex
     \textfont\itfam=\xiiiptit
     \scriptfont\itfam=\xptit
     \scriptscriptfont\itfam=\viiptit
     \textfont\bffam=\xiiiptbf
     \scriptfont\bffam=\xptbf
     \scriptscriptfont\bffam=\viiptbf
     \@fontstyleinit
     \def\rm{\xiiiptrm\fam=\z@}%
     \def\it{\xiiiptit\fam=\itfam}%
     \def\sl{\xiiiptsl}%
     \def\bf{\xiiiptbf\fam=\bffam}%
     \def\tt{\xiiipttt}%
     \def\ss{\xiiiptss}%
     \def\oldstyle{\xiiiptmit\fam=\@ne}%
     \rm\fi}

%************** 14-point fonts ******************************

\font\xivptrm=cmr12   scaled\magstep1 \font\xivptmit=cmmi12 scaled\magstep1
\font\xivptsy=cmsy10  scaled\magstep2 \font\xivptex=cmex10  scaled\magstep2
\font\xivptit=cmti12 scaled\magstep1 \font\xivptsl=cmsl12  scaled\magstep1
\font\xivptbf=cmbx12 scaled\magstep1
%\font\xivpttt=cmtt12  scaled\magstep1
\font\xivptss=cmss12  scaled\magstep1 \font\xivptsc=cmcsc10 scaled\magstep2
\font\xivptbfs=cmb10  scaled\magstep2 \font\xivptbmit=cmmib10 scaled\magstep2

\skewchar\xivptmit='177 \skewchar\xivptbmit='177 \skewchar\xivptsy='60 \fontdimen16
\xivptsy=\the\fontdimen17 \xivptsy

\def\xivpt{\ifmmode\err@badsizechange\else
     \@mathfontinit
     \textfont0=\xivptrm  \scriptfont0=\xptrm  \scriptscriptfont0=\viiptrm
     \textfont1=\xivptmit \scriptfont1=\xptmit \scriptscriptfont1=\viiptmit
     \textfont2=\xivptsy  \scriptfont2=\xptsy  \scriptscriptfont2=\viiptsy
     \textfont3=\xivptex  \scriptfont3=\xptex  \scriptscriptfont3=\xptex
     \textfont\itfam=\xivptit
     \scriptfont\itfam=\xptit
     \scriptscriptfont\itfam=\viiptit
     \textfont\bffam=\xivptbf
     \scriptfont\bffam=\xptbf
     \scriptscriptfont\bffam=\viiptbf
     \textfont\bfsfam=\xivptbfs
     \scriptfont\bfsfam=\xptbfs
     \scriptscriptfont\bfsfam=\viiptbf
     \textfont\bmitfam=\xivptbmit
     \scriptfont\bmitfam=\xptbmit
     \scriptscriptfont\bmitfam=\viiptmit
     \@fontstyleinit
     \def\rm{\xivptrm\fam=\z@}%
     \def\it{\xivptit\fam=\itfam}%
     \def\sl{\xivptsl}%
     \def\bf{\xivptbf\fam=\bffam}%
     \def\tt{\xivpttt}%
     \def\ss{\xivptss}%
     \def\sc{\xivptsc}%
     \def\bfs{\xivptbfs\fam=\bfsfam}%
     \def\bmit{\fam=\bmitfam}%
     \def\oldstyle{\xivptmit\fam=\@ne}%
     \rm\fi}

%************** 17-point fonts ******************************

\font\xviiptrm=cmr17 \font\xviiptmit=cmmi12 scaled\magstep2 \font\xviiptsy=cmsy10
scaled\magstep3 \font\xviiptex=cmex10 scaled\magstep3 \font\xviiptit=cmti12
scaled\magstep2 \font\xviiptbf=cmbx12 scaled\magstep2 \font\xviiptbfs=cmb10
scaled\magstep3

\skewchar\xviiptmit='177 \skewchar\xviiptsy='60 \fontdimen16 \xviiptsy=\the\fontdimen17
\xviiptsy

\def\xviipt{\ifmmode\err@badsizechange\else
     \@mathfontinit
     \textfont0=\xviiptrm  \scriptfont0=\xiiptrm  \scriptscriptfont0=\viiiptrm
     \textfont1=\xviiptmit \scriptfont1=\xiiptmit \scriptscriptfont1=\viiiptmit
     \textfont2=\xviiptsy  \scriptfont2=\xiiptsy  \scriptscriptfont2=\viiiptsy
     \textfont3=\xviiptex  \scriptfont3=\xiiptex  \scriptscriptfont3=\xptex
     \textfont\itfam=\xviiptit
     \scriptfont\itfam=\xiiptit
     \scriptscriptfont\itfam=\viiiptit
     \textfont\bffam=\xviiptbf
     \scriptfont\bffam=\xiiptbf
     \scriptscriptfont\bffam=\viiiptbf
     \textfont\bfsfam=\xviiptbfs
     \scriptfont\bfsfam=\xiiptbfs
     \scriptscriptfont\bfsfam=\viiiptbf
     \@fontstyleinit
     \def\rm{\xviiptrm\fam=\z@}%
     \def\it{\xviiptit\fam=\itfam}%
     \def\bf{\xviiptbf\fam=\bffam}%
     \def\bfs{\xviiptbfs\fam=\bfsfam}%
     \def\oldstyle{\xviiptmit\fam=\@ne}%
     \rm\fi}

%************** 21-point fonts ******************************

\font\xxiptrm=cmr17  scaled\magstep1
%\font\xxiptmit=cmmi12 scaled\magstep3
%\font\xxiptsy=cmsy10 scaled\magstep4
%\font\xxiptex=cmex10 scaled\magstep4
%\font\xxiptbf=cmbx12 scaled\magstep3

%\skewchar\xxiptmit='177 \skewchar\xxiptsy='60
%\fontdimen16 \xxiptsy=\the\fontdimen17 \xxiptsy

\def\xxipt{\ifmmode\err@badsizechange\else
     \@mathfontinit
%     \textfont0=\xxiptrm  \scriptfont0=\xivptrm  \scriptscriptfont0=\xptrm
%     \textfont1=\xxiptmit \scriptfont1=\xivptmit \scriptscriptfont1=\xptmit
%     \textfont2=\xxiptsy  \scriptfont2=\xivptsy  \scriptscriptfont2=\xptsy
%     \textfont3=\xxiptex  \scriptfont3=\xivptex  \scriptscriptfont3=\xptex
%     \textfont\bffam=\xxiptbf
%     \scriptfont\bffam=\xivptbf
%     \scriptscriptfont\bffam=\xptbf
     \@fontstyleinit
     \def\rm{\xxiptrm\fam=\z@}%
     \rm\fi}

%************** 25-point fonts ******************************

\font\xxvptrm=cmr17  scaled\magstep2
%\font\xxvptmit=cmmi12 scaled\magstep4
%\font\xxvptsy=cmsy10 scaled\magstep5
%\font\xxvptex=cmex10 scaled\magstep5
%\font\xxvptbf=cmbx12 scaled\magstep4

%\skewchar\xxvptmit='177 \skewchar\xxvptsy='60
%\fontdimen16 \xxvptsy=\the\fontdimen17 \xxvptsy

\def\xxvpt{\ifmmode\err@badsizechange\else
     \@mathfontinit
%     \textfont0=\xxvptrm  \scriptfont0=\xviiptrm  \scriptscriptfont0=\xiiptrm
%     \textfont1=\xxvptmit \scriptfont1=\xviiptmit \scriptscriptfont1=\xiiptmit
%     \textfont2=\xxvptsy  \scriptfont2=\xviiptsy  \scriptscriptfont2=\xiiptsy
%     \textfont3=\xxvptex  \scriptfont3=\xviiptex  \scriptscriptfont3=\xiiptex
%     \textfont\bffam=\xxvptbf
%     \scriptfont\bffam=\xviiptbf
%     \scriptscriptfont\bffam=\xiiptbf
     \@fontstyleinit
     \def\rm{\xxvptrm\fam=\z@}%
     \rm\fi}

%************** Other fonts *********************************

%\font\dummy=dummy

%******************************************************************************

\message{Loading jyTeX macros...}

%************************************************************
%*
%*              Simple modifications to plain
%*
%************************************************************
\message{modifications to plain.tex,}

% The "\outer" qualifier is removed from the definitions of \newcount through
% \newif so that they may be used in definitions.  \newif is also changed to
% make \if commands globally defined.

\def\newcount{\alloc@0\count\countdef\insc@unt}
\def\newdimen{\alloc@1\dimen\dimendef\insc@unt}
\def\newskip{\alloc@2\skip\skipdef\insc@unt}
\def\newmuskip{\alloc@3\muskip\muskipdef\@cclvi}
\def\newbox{\alloc@4\box\chardef\insc@unt}
\def\newtoks{\alloc@5\toks\toksdef\@cclvi}
\def\newhelp#1#2{\newtoks#1\global#1\expandafter{\csname#2\endcsname}}
\def\newread{\alloc@6\read\chardef\sixt@@n}
\def\newwrite{\alloc@7\write\chardef\sixt@@n}
\def\newfam{\alloc@8\fam\chardef\sixt@@n}
\def\newinsert#1{\global\advance\insc@unt by\m@ne
     \ch@ck0\insc@unt\count
     \ch@ck1\insc@unt\dimen
     \ch@ck2\insc@unt\skip
     \ch@ck4\insc@unt\box
     \allocationnumber=\insc@unt
     \global\chardef#1=\allocationnumber
     \wlog{\string#1=\string\insert\the\allocationnumber}}
\def\newif#1{\count@\escapechar \escapechar\m@ne
     \expandafter\expandafter\expandafter
          \xdef\@if#1{true}{\let\noexpand#1=\noexpand\iftrue}%
     \expandafter\expandafter\expandafter
          \xdef\@if#1{false}{\let\noexpand#1=\noexpand\iffalse}%
     \global\@if#1{false}\escapechar=\count@}

%************** Some parameter changes **********************

\newlinechar=`\^^J
\overfullrule=0pt

%************** Font-related modifications ******************

% The plain fonts are mapped onto the corresponding jyTeX fonts

% Some control sequences are disabled.

\let\itfam=\undefined

\let\bffam=\undefined

\count18=3

% German sharp s is given a new name (\ss is already taken)

\chardef\sharps="19

% The mathcode assignments of characters in the math italic font are changed to
% allow for switching to boldface.

\mathchardef\alpha="710B \mathchardef\beta="710C \mathchardef\gamma="710D
\mathchardef\delta="710E \mathchardef\epsilon="710F \mathchardef\zeta="7110
\mathchardef\eta="7111 \mathchardef\theta="7112 \mathchardef\iota="7113
\mathchardef\kappa="7114 \mathchardef\lambda="7115 \mathchardef\mu="7116
\mathchardef\nu="7117 \mathchardef\xi="7118 \mathchardef\pi="7119
\mathchardef\rho="711A \mathchardef\sigma="711B \mathchardef\tau="711C
\mathchardef\upsilon="711D \mathchardef\phi="711E \mathchardef\chi="711F
\mathchardef\psi="7120 \mathchardef\omega="7121 \mathchardef\varepsilon="7122
\mathchardef\vartheta="7123 \mathchardef\varpi="7124 \mathchardef\varrho="7125
\mathchardef\varsigma="7126 \mathchardef\varphi="7127 \mathchardef\imath="717B
\mathchardef\jmath="717C \mathchardef\ell="7160 \mathchardef\wp="717D
\mathchardef\partial="7140 \mathchardef\flat="715B \mathchardef\natural="715C
\mathchardef\sharp="715D

%************** Miscellaneous changes ***********************

% The dimension \p@ (1pt) is replaced with \rp@ (relative pt, defined below),
% whose size is determined by the base type size of the document.

\def\angle{{\vbox{\ialign{$\m@th\scriptstyle##$\crcr
     \not\mathrel{\mkern14mu}\crcr
     \noalign{\nointerlineskip}
     \mkern2.5mu\leaders\hrule height.34\rp@\hfill\mkern2.5mu\crcr}}}}
\def\vdots{\vbox{\baselineskip4\rp@ \lineskiplimit\z@
     \kern6\rp@\hbox{.}\hbox{.}\hbox{.}}}
\def\ddots{\mathinner{\mkern1mu\raise7\rp@\vbox{\kern7\rp@\hbox{.}}\mkern2mu
     \raise4\rp@\hbox{.}\mkern2mu\raise\rp@\hbox{.}\mkern1mu}}
\def\overrightarrow#1{\vbox{\ialign{##\crcr
     \rightarrowfill\crcr
     \noalign{\kern-\rp@\nointerlineskip}
     $\hfil\displaystyle{#1}\hfil$\crcr}}}
\def\overleftarrow#1{\vbox{\ialign{##\crcr
     \leftarrowfill\crcr
     \noalign{\kern-\rp@\nointerlineskip}
     $\hfil\displaystyle{#1}\hfil$\crcr}}}
\def\overbrace#1{\mathop{\vbox{\ialign{##\crcr
     \noalign{\kern3\rp@}
     \downbracefill\crcr
     \noalign{\kern3\rp@\nointerlineskip}
     $\hfil\displaystyle{#1}\hfil$\crcr}}}\limits}
\def\underbrace#1{\mathop{\vtop{\ialign{##\crcr
     $\hfil\displaystyle{#1}\hfil$\crcr
     \noalign{\kern3\rp@\nointerlineskip}
     \upbracefill\crcr
     \noalign{\kern3\rp@}}}}\limits}
\def\big#1{{\hbox{$\left#1\vbox to8.5\rp@ {}\right.\n@space$}}}
\def\Big#1{{\hbox{$\left#1\vbox to11.5\rp@ {}\right.\n@space$}}}
\def\bigg#1{{\hbox{$\left#1\vbox to14.5\rp@ {}\right.\n@space$}}}
\def\Bigg#1{{\hbox{$\left#1\vbox to17.5\rp@ {}\right.\n@space$}}}
\def\@vereq#1#2{\lower.5\rp@\vbox{\baselineskip\z@skip\lineskip-.5\rp@
     \ialign{$\m@th#1\hfil##\hfil$\crcr#2\crcr=\crcr}}}
\def\rlh@#1{\vcenter{\hbox{\ooalign{\raise2\rp@
     \hbox{$#1\rightharpoonup$}\crcr
     $#1\leftharpoondown$}}}}
\def\bordermatrix#1{\begingroup\m@th
     \setbox\z@\vbox{%
          \def\cr{\crcr\noalign{\kern2\rp@\global\let\cr\endline}}%
          \ialign{$##$\hfil\kern2\rp@\kern\p@renwd
               &\thinspace\hfil$##$\hfil&&\quad\hfil$##$\hfil\crcr
               \omit\strut\hfil\crcr
               \noalign{\kern-\baselineskip}%
               #1\crcr\omit\strut\cr}}%
     \setbox\tw@\vbox{\unvcopy\z@\global\setbox\@ne\lastbox}%
     \setbox\tw@\hbox{\unhbox\@ne\unskip\global\setbox\@ne\lastbox}%
     \setbox\tw@\hbox{$\kern\wd\@ne\kern-\p@renwd\left(\kern-\wd\@ne
          \global\setbox\@ne\vbox{\box\@ne\kern2\rp@}%
          \vcenter{\kern-\ht\@ne\unvbox\z@\kern-\baselineskip}%
          \,\right)$}%
     \null\;\vbox{\kern\ht\@ne\box\tw@}\endgroup}
\def\endinsert{\egroup
     \if@mid\dimen@\ht\z@
          \advance\dimen@\dp\z@
          \advance\dimen@12\rp@
          \advance\dimen@\pagetotal
          \ifdim\dimen@>\pagegoal\@midfalse\p@gefalse\fi
     \fi
     \if@mid\bigskip\box\z@
          \bigbreak
     \else\insert\topins{\penalty100 \splittopskip\z@skip
               \splitmaxdepth\maxdimen\floatingpenalty\z@
               \ifp@ge\dimen@\dp\z@
                    \vbox to\vsize{\unvbox\z@\kern-\dimen@}%
               \else\box\z@\nobreak\bigskip
               \fi}%
     \fi
     \endgroup}

% \normalbaselines is removed from \cases and \matrix.

\def\cases#1{\left\{\,\vcenter{\m@th
     \ialign{$##\hfil$&\quad##\hfil\crcr#1\crcr}}\right.}
\def\matrix#1{\null\,\vcenter{\m@th
     \ialign{\hfil$##$\hfil&&\quad\hfil$##$\hfil\crcr
          \mathstrut\crcr
          \noalign{\kern-\baselineskip}
          #1\crcr
          \mathstrut\crcr
          \noalign{\kern-\baselineskip}}}\,}

% \raggedbottom modified slightly

\newif\ifraggedbottom

\def\raggedbottom{\ifraggedbottom\else
     \advance\topskip by\z@ plus60pt \raggedbottomtrue\fi}%
\def\normalbottom{\ifraggedbottom
     \advance\topskip by\z@ plus-60pt \raggedbottomfalse\fi}

%************************************************************
%*
%*              Miscellaneous definitions
%*
%************************************************************
\message{hacks,}

%************** Hack registers ******************************

\toksdef\toks@i=1 \toksdef\toks@ii=2

%************** Basic macros ********************************

\def\TeX{T\kern-.1667em \lower.5ex \hbox{E}\kern-.125em X\null}
\def\jyTeX{{\leavevmode
     \raise.587ex \hbox{\it\j}\kern-.1em \lower.048ex \hbox{\it y}\kern-.12em
     \TeX}}

\let\then=\iftrue
\def\ifnoarg#1\then{\def\hack@{#1}\ifx\hack@\empty}
\def\ifundefined#1\then{%
     \expandafter\ifx\csname\expandafter\blank\string#1\endcsname\relax}
\def\useif#1\then{\csname#1\endcsname}
\def\usename#1{\csname#1\endcsname}
\def\useafter#1#2{\expandafter#1\csname#2\endcsname}

% Modify so that I can have \loop's within \loop's?
\long\def\loop#1\repeat{\def\@iterate{#1\expandafter\@iterate\fi}\@iterate
     \let\@iterate=\relax}
%\long\def\loop#1\repeat{\def\@loopbody{#1}\@iterate}
%\def\@iterate{\@loopbody\let\next=\@iterate\else\let\next=\relax\fi\next}

\let\TeXend=\end
\def\begin#1{\begingroup\def\@@blockname{#1}\usename{begin#1}}
\def\end#1{\usename{end#1}\def\hack@{#1}%
     \ifx\@@blockname\hack@
          \endgroup
     \else\err@badgroup\hack@\@@blockname
     \fi}
\def\@@blockname{}

\def\defaultoption[#1]#2{%
     \def\hack@{\ifx\hack@ii[\toks@={#2}\else\toks@={#2[#1]}\fi\the\toks@}%
     \futurelet\hack@ii\hack@}

\def\markup#1{\let\@@marksf=\empty
     \ifhmode\edef\@@marksf{\spacefactor=\the\spacefactor\relax}\/\fi
     ${}^{\hbox{\subscriptfonts#1}}$\@@marksf}

%************** Time registers ******************************

\newtoks\shortyear
\newtoks\militaryhour
\newtoks\standardhour
\newtoks\minute
\newtoks\amorpm

\def\settime{\count@=\time\divide\count@ by60
     \militaryhour=\expandafter{\number\count@}%
     {\multiply\count@ by-60 \advance\count@ by\time
          \xdef\hack@{\ifnum\count@<10 0\fi\number\count@}}%
     \minute=\expandafter{\hack@}%
     \ifnum\count@<12
          \amorpm={am}
     \else\amorpm={pm}
          \ifnum\count@>12 \advance\count@ by-12 \fi
     \fi
     \standardhour=\expandafter{\number\count@}%
     \def\hack@19##1##2{\shortyear={##1##2}}%
          \expandafter\hack@\the\year}

\def\monthword#1{%
     \ifcase#1
          $\bullet$\err@badcountervalue{monthword}%
          \or January\or February\or March\or April\or May\or June%
          \or July\or August\or September\or October\or November\or December%
     \else$\bullet$\err@badcountervalue{monthword}%
     \fi}

\def\monthabbr#1{%
     \ifcase#1
          $\bullet$\err@badcountervalue{monthabbr}%
          \or Jan\or Feb\or Mar\or Apr\or May\or Jun%
          \or Jul\or Aug\or Sep\or Oct\or Nov\or Dec%
     \else$\bullet$\err@badcountervalue{monthabbr}%
     \fi}

\def\militarytime{\the\militaryhour:\the\minute}
\def\standardtime{\the\standardhour:\the\minute}

%************** Number styles *******************************

\def\@setnumstyle#1#2{\expandafter\global\expandafter\expandafter
     \expandafter\let\expandafter\expandafter
     \csname @\expandafter\blank\string#1style\endcsname
     \csname#2\endcsname}
\def\numstyle#1{\usename{@\expandafter\blank\string#1style}#1}
\def\ifblank#1\then{\useafter\ifx{@\expandafter\blank\string#1}\blank}

\def\blank#1{}

\def\Roman#1{\expandafter\uppercase\expandafter{\romannumeral#1}}
\def\alphabetic#1{%
     \ifcase#1
          $\bullet$\err@badcountervalue{alphabetic}%
          \or a\or b\or c\or d\or e\or f\or g\or h\or i\or j\or k\or l\or m%
          \or n\or o\or p\or q\or r\or s\or t\or u\or v\or w\or x\or y\or z%
     \else$\bullet$\err@badcountervalue{alphabetic}%
     \fi}
\def\Alphabetic#1{\expandafter\uppercase\expandafter{\alphabetic{#1}}}
\def\symbols#1{%
     \ifcase#1
          $\bullet$\err@badcountervalue{symbols}%
          \or*\or\dag\or\ddag\or\S\or$\|$%
          \or**\or\dag\dag\or\ddag\ddag\or\S\S\or$\|\|$%
     \else$\bullet$\err@badcountervalue{symbols}%
     \fi}

%************** String macros *******************************

\catcode`\^^?=13 \def^^?{\relax}

\def\trimleading#1\to#2{\edef#2{#1}%
     \expandafter\@trimleading\expandafter#2#2^^?^^?}
\def\@trimleading#1#2#3^^?{\ifx#2^^?\def#1{}\else\def#1{#2#3}\fi}

\def\trimtrailing#1\to#2{\edef#2{#1}%
     \expandafter\@trimtrailing\expandafter#2#2^^? ^^?\relax}
\def\@trimtrailing#1#2 ^^?#3{\ifx#3\relax\toks@={}%
     \else\def#1{#2}\toks@={\trimtrailing#1\to#1}\fi
     \the\toks@}

\def\trim#1\to#2{\trimleading#1\to#2\trimtrailing#2\to#2}

\catcode`\^^?=15

%************** List macros *********************************

\long\def\additemL#1\to#2{\toks@={\^^\{#1}}\toks@ii=\expandafter{#2}%
     \xdef#2{\the\toks@\the\toks@ii}}

\long\def\additemR#1\to#2{\toks@={\^^\{#1}}\toks@ii=\expandafter{#2}%
     \xdef#2{\the\toks@ii\the\toks@}}

\def\getitemL#1\to#2{\expandafter\@getitemL#1\hack@#1#2}
\def\@getitemL\^^\#1#2\hack@#3#4{\def#4{#1}\def#3{#2}}

%************************************************************
%*
%*             Font-related macros
%*
%************************************************************
\message{font macros,}

%************** Font set-up *********************************

\newdimen\rp@
\newcount\@@sizeindex \@@sizeindex=0
\newcount\@@factori
\newcount\@@factorii
\newcount\@@factoriii
\newcount\@@factoriv

\countdef\maxfam=18
\newfam\itfam
\newfam\bffam
\newfam\bfsfam
\newfam\bmitfam

\def\@mathfontinit{\count@=4
     \loop\textfont\count@=\nullfont
          \scriptfont\count@=\nullfont
          \scriptscriptfont\count@=\nullfont
          \ifnum\count@<\maxfam\advance\count@ by\@ne
     \repeat}

\def\@fontstyleinit{%
     \def\it{\err@fontnotavailable\it}%
     \def\bf{\err@fontnotavailable\bf}%
     \def\bfs{\err@bfstobf}%
     \def\bmit{\err@fontnotavailable\bmit}%
     \def\sc{\err@fontnotavailable\sc}%
     \def\sl{\err@sltoit}%
     \def\ss{\err@fontnotavailable\ss}%
     \def\tt{\err@fontnotavailable\tt}}

\def\@parameterinit#1{\rm\rp@=.1em \@getscaling{#1}%
     \let\^^\=\@doscaling\scalingskipslist
     \setbox\strutbox=\hbox{\vrule
          height.708\baselineskip depth.292\baselineskip width\z@}}

\def\@getfactor#1#2#3#4{\@@factori=#1 \@@factorii=#2
     \@@factoriii=#3 \@@factoriv=#4}

\def\@getscaling#1{\count@=#1 \advance\count@ by-\@@sizeindex\@@sizeindex=#1
     \ifnum\count@<0
          \let\@mulordiv=\divide
          \let\@divormul=\multiply
          \multiply\count@ by\m@ne
     \else\let\@mulordiv=\multiply
          \let\@divormul=\divide
     \fi
     \edef\@@scratcha{\ifcase\count@                {1}{1}{1}{1}\or
          {1}{7}{23}{3}\or     {2}{5}{3}{1}\or      {9}{89}{13}{1}\or
          {6}{25}{6}{1}\or     {8}{71}{14}{1}\or    {6}{25}{36}{5}\or
          {1}{7}{53}{4}\or     {12}{125}{108}{5}\or {3}{14}{53}{5}\or
          {6}{41}{17}{1}\or    {13}{31}{13}{2}\or   {9}{107}{71}{2}\or
          {11}{139}{124}{3}\or {1}{6}{43}{2}\or     {10}{107}{42}{1}\or
          {1}{5}{43}{2}\or     {5}{69}{65}{1}\or    {11}{97}{91}{2}\fi}%
     \expandafter\@getfactor\@@scratcha}

\def\@doscaling#1{\@mulordiv#1by\@@factori\@divormul#1by\@@factorii
     \@mulordiv#1by\@@factoriii\@divormul#1by\@@factoriv}

%************* Size-changing commands ***********************

\newskip\headskip
\newskip\footskip

\def\typesize=#1pt{\count@=#1 \advance\count@ by-10
     \ifcase\count@
          \@setsizex\or\err@badtypesize\or
          \@setsizexii\or\err@badtypesize\or
          \@setsizexiv
     \else\err@badtypesize
     \fi}

\def\@setsizex{\getixpt
     \def\subsubscriptfonts{\vpt}%
          \def\subsubscriptsize{\vpt\@parameterinit{-8}}%
     \def\subscriptfonts{\viipt}\def\subscriptsize{\viipt\@parameterinit{-4}}%
     \def\footnotefonts{\viiipt}\def\footnotesize{\viiipt\@parameterinit{-2}}%
     \def\smallfonts{\ixpt}\def\smallsize{\ixpt\@parameterinit{-1}}%
     \def\normalfonts{\xpt}\def\normalsize{\xpt\@parameterinit{0}}%
     \def\bigfonts{\xiipt}\def\bigsize{\xiipt\@parameterinit{2}}%
     \def\Bigfonts{\xivpt}\def\Bigsize{\xivpt\@parameterinit{4}}%
     \def\biggfonts{\xviipt}\def\biggsize{\xviipt\@parameterinit{6}}%
     \def\Biggfonts{\xxipt}\def\Biggsize{\xxipt\@parameterinit{8}}%
     \def\tinyfonts{\vpt}\def\tinysize{\vpt\@parameterinit{-8}}%
     \def\HUGEFONTS{\xxvpt}\def\HUGESIZE{\xxvpt\@parameterinit{10}}%
     \normalsize\fixedskipslist}

\def\@setsizexii{\getxipt
     \def\subsubscriptfonts{\vipt}%
          \def\subsubscriptsize{\vipt\@parameterinit{-6}}%
     \def\subscriptfonts{\viiipt}%
          \def\subscriptsize{\viiipt\@parameterinit{-2}}%
     \def\footnotefonts{\xpt}\def\footnotesize{\xpt\@parameterinit{0}}%
     \def\smallfonts{\xipt}\def\smallsize{\xipt\@parameterinit{1}}%
     \def\normalfonts{\xiipt}\def\normalsize{\xiipt\@parameterinit{2}}%
     \def\bigfonts{\xivpt}\def\bigsize{\xivpt\@parameterinit{4}}%
     \def\Bigfonts{\xviipt}\def\Bigsize{\xviipt\@parameterinit{6}}%
     \def\biggfonts{\xxipt}\def\biggsize{\xxipt\@parameterinit{8}}%
     \def\Biggfonts{\xxvpt}\def\Biggsize{\xxvpt\@parameterinit{10}}%
     \def\tinyfonts{\vpt}\def\tinysize{\vpt\@parameterinit{-8}}%
     \def\HUGEFONTS{\xxvpt}\def\HUGESIZE{\xxvpt\@parameterinit{10}}%
     \normalsize\fixedskipslist}

\def\@setsizexiv{\getxiiipt
     \def\subsubscriptfonts{\viipt}%
          \def\subsubscriptsize{\viipt\@parameterinit{-4}}%
     \def\subscriptfonts{\xpt}\def\subscriptsize{\xpt\@parameterinit{0}}%
     \def\footnotefonts{\xiipt}\def\footnotesize{\xiipt\@parameterinit{2}}%
     \def\smallfonts{\xiiipt}\def\smallsize{\xiiipt\@parameterinit{3}}%
     \def\normalfonts{\xivpt}\def\normalsize{\xivpt\@parameterinit{4}}%
     \def\bigfonts{\xviipt}\def\bigsize{\xviipt\@parameterinit{6}}%
     \def\Bigfonts{\xxipt}\def\Bigsize{\xxipt\@parameterinit{8}}%
     \def\biggfonts{\xxvpt}\def\biggsize{\xxvpt\@parameterinit{10}}%
     \def\Biggfonts{\err@sizetoolarge\Biggfonts\HUGEFONTS}%
          \def\Biggsize{\err@sizetoolarge\Biggsize\HUGESIZE}%
     \def\tinyfonts{\vpt}\def\tinysize{\vpt\@parameterinit{-8}}%
     \def\HUGEFONTS{\xxvpt}\def\HUGESIZE{\xxvpt\@parameterinit{10}}%
     \normalsize\fixedskipslist}

\def\subsubscriptfonts{\vpt} \def\subsubscriptsize{\vpt\@parameterinit{-8}}
\def\subscriptfonts{\viipt}  \def\subscriptsize{\viipt\@parameterinit{-4}}
\def\footnotefonts{\viiipt}  \def\footnotesize{\viiipt\@parameterinit{-2}}
\def\smallfonts{\err@sizenotavailable\smallfonts}
                             \def\smallsize{\ixpt\@parameterinit{-1}}
\def\normalfonts{\xpt}       \def\normalsize{\xpt\@parameterinit{0}}
\def\bigfonts{\xiipt}        \def\bigsize{\xiipt\@parameterinit{2}}
\def\Bigfonts{\xivpt}        \def\Bigsize{\xivpt\@parameterinit{4}}
\def\biggfonts{\xviipt}      \def\biggsize{\xviipt\@parameterinit{6}}
\def\Biggfonts{\xxipt}       \def\Biggsize{\xxipt\@parameterinit{8}}
\def\tinyfonts{\vpt}         \def\tinysize{\vpt\@parameterinit{-8}}
\def\HUGEFONTS{\xxvpt}       \def\HUGESIZE{\xxvpt\@parameterinit{10}}

%************************************************************
%*
%*             Document layout
%*
%************************************************************
\message{document layout,}

%************** Page format *********************************

\newtoks\everyoutput \everyoutput={}
\newdimen\depthofpage
\newcount\pagenum \pagenum=0

\newdimen\oddtopmargin  \newdimen\eventopmargin
\newdimen\oddleftmargin \newdimen\evenleftmargin
\newtoks\oddhead        \newtoks\evenhead
\newtoks\oddfoot        \newtoks\evenfoot

\def\topmargin{\afterassignment\@seteventop\oddtopmargin}
\def\leftmargin{\afterassignment\@setevenleft\oddleftmargin}
\def\head{\afterassignment\@setevenhead\oddhead}
\def\foot{\afterassignment\@setevenfoot\oddfoot}

\def\@seteventop{\eventopmargin=\oddtopmargin}
\def\@setevenleft{\evenleftmargin=\oddleftmargin}
\def\@setevenhead{\evenhead=\oddhead}
\def\@setevenfoot{\evenfoot=\oddfoot}

\def\pagenumstyle#1{\@setnumstyle\pagenum{#1}}

\newif\ifdraft
\def\draft{\drafttrue\leftmargin=.5in \overfullrule=5pt }

\def\outputstyle#1{\global\expandafter\let\expandafter
          \@outputstyle\csname#1output\endcsname
     \usename{#1setup}}

\output={\@outputstyle}

\def\normaloutput{\the\everyoutput
     \global\advance\pagenum by\@ne
     \ifodd\pagenum
          \voffset=\oddtopmargin \hoffset=\oddleftmargin
     \else\voffset=\eventopmargin \hoffset=\evenleftmargin
     \fi
     \advance\voffset by-1in  \advance\hoffset by-1in
     \count0=\pagenum
     \expandafter\shipout\pagebox
     \ifnum\outputpenalty>-\@MM\else\dosupereject\fi}

\newdimen\fullhsize
\newbox\leftpage
\newcount\leftpagenum
\newcount\outputpagenum \outputpagenum=0
\let\leftorright=L

\def\twoupoutput{\the\everyoutput
     \global\advance\pagenum by\@ne
     \if L\leftorright
          \global\setbox\leftpage=\leftline{\pagebox}%
          \global\leftpagenum=\pagenum
          \global\let\leftorright=R%
     \else\global\advance\outputpagenum by\@ne
          \ifodd\outputpagenum
               \voffset=\oddtopmargin \hoffset=\oddleftmargin
          \else\voffset=\eventopmargin \hoffset=\evenleftmargin
          \fi
          \advance\voffset by-1in  \advance\hoffset by-1in
          \count0=\leftpagenum \count1=\pagenum
          \shipout\vbox{\hbox to\fullhsize
               {\box\leftpage\hfil\leftline{\pagebox}}}%
          \global\let\leftorright=L%
     \fi
     \ifnum\outputpenalty>-\@MM
     \else\dosupereject
          \if R\leftorright
               \globaldefs=\@ne\head={\hfil}\foot={\hfil}\globaldefs=\z@
               \null\newpage
          \fi
     \fi}

\def\pagebox{\vbox{\makeheadline\pagebody\makefootline}}

\def\makeheadline{%
     \vbox to\z@{\baselinestretch=\@m
          \vskip\topskip\vskip-.708\baselineskip\vskip-\headskip
          \line{\vbox to\ht\strutbox{}%
               \ifodd\pagenum\the\oddhead\else\the\evenhead\fi}%
          \vss}%
     \nointerlineskip}

\def\pagebody{\vbox to\vsize{%
     \boxmaxdepth\maxdepth
     \ifvoid\topins\else\unvbox\topins\fi
     \depthofpage=\dp255
     \unvbox255
     \ifraggedbottom\kern-\depthofpage\vfil\fi
     \ifvoid\footins
     \else\vskip\skip\footins
          \footnoterule
          \unvbox\footins
          \vskip-\footnoteskip
     \fi}}

\def\makefootline{\baselineskip=\footskip
     \line{\ifodd\pagenum\the\oddfoot\else\the\evenfoot\fi}}

%************** Sectioning commands *************************

\newskip\abovechapterskip
\newskip\belowchapterskip
\newskip\abovesectionskip
\newskip\belowsectionskip
\newskip\abovesubsectionskip
\newskip\belowsubsectionskip

\def\chapterstyle#1{\global\expandafter\let\expandafter\@chapterstyle
     \csname#1text\endcsname}
\def\sectionstyle#1{\global\expandafter\let\expandafter\@sectionstyle
     \csname#1text\endcsname}
\def\subsectionstyle#1{\global\expandafter\let\expandafter\@subsectionstyle
     \csname#1text\endcsname}

\def\chapter#1{%
     \ifdim\lastskip=17sp \else\chapterbreak\vskip\abovechapterskip\fi
     \@chapterstyle{\ifblank\chapternumstyle\then
          \else\newchapternum=\next\chapternumformat\ \fi#1}%
     \nobreak\vskip\belowchapterskip\vskip17sp }

\def\section#1{%
     \ifdim\lastskip=17sp \else\sectionbreak\vskip\abovesectionskip\fi
     \@sectionstyle{\ifblank\sectionnumstyle\then
          \else\newsectionnum=\next\sectionnumformat\ \fi#1}%
     \nobreak\vskip\belowsectionskip\vskip17sp }

\def\subsection#1{%
     \ifdim\lastskip=17sp \else\subsectionbreak\vskip\abovesubsectionskip\fi
     \@subsectionstyle{\ifblank\subsectionnumstyle\then
          \else\newsubsectionnum=\next\subsectionnumformat\ \fi#1}%
     \nobreak\vskip\belowsubsectionskip\vskip17sp }

%************** Text formatting commands ********************

\let\TeXunderline=\underline
\let\TeXoverline=\overline
\def\underline#1{\relax\ifmmode\TeXunderline{#1}\else
     $\TeXunderline{\hbox{#1}}$\fi}
\def\overline#1{\relax\ifmmode\TeXoverline{#1}\else
     $\TeXoverline{\hbox{#1}}$\fi}

\def\baselinestretch{\afterassignment\@baselinestretch\count@}
\def\@baselinestretch{\baselineskip=\normalbaselineskip
     \divide\baselineskip by\@m\baselineskip=\count@\baselineskip
     \setbox\strutbox=\hbox{\vrule
          height.708\baselineskip depth.292\baselineskip width\z@}%
     \bigskipamount=\the\baselineskip
          plus.25\baselineskip minus.25\baselineskip
     \medskipamount=.5\baselineskip
          plus.125\baselineskip minus.125\baselineskip
     \smallskipamount=.25\baselineskip
          plus.0625\baselineskip minus.0625\baselineskip}

\def\\{\ifhmode\ifnum\lastpenalty=-\@M\else\hfil\penalty-\@M\fi\fi
     \ignorespaces}
\def\newpage{\vfil\break}

\def\lefttext#1{\par{\@text\leftskip=\z@\rightskip=\centering
     \noindent#1\par}}
\def\righttext#1{\par{\@text\leftskip=\centering\rightskip=\z@
     \noindent#1\par}}
\def\centertext#1{\par{\@text\leftskip=\centering\rightskip=\centering
     \noindent#1\par}}
\def\@text{\parindent=\z@ \parfillskip=\z@ \everypar={}%
     \spaceskip=.3333em \xspaceskip=.5em
     \def\\{\ifhmode\ifnum\lastpenalty=-\@M\else\penalty-\@M\fi\fi
          \ignorespaces}}

\def\beginleft{\par\@text\leftskip=\z@ \rightskip=\centering}
     
\def\beginright{\par\@text\leftskip=\centering\rightskip=\z@ }
     
\def\begincenter{\par\@text\leftskip=\centering\rightskip=\centering}

\def\beginnarrow{\defaultoption[\parindent]\@beginnarrow}
\def\@beginnarrow[#1]{\par\advance\leftskip by#1\advance\rightskip by#1}

\begingroup
\catcode`\[=1 \catcode`\{=11 \gdef\beginignore[\endgroup\bgroup
     \catcode`\e=0 \catcode`\\=12 \catcode`\{=11 \catcode`\f=12 \let\or=\relax
     \let\nd{ignor=\fi \let\}=\egroup
     \iffalse}
\endgroup

\long\def\marginnote#1{\leavevmode
     \edef\@marginsf{\spacefactor=\the\spacefactor\relax}%
     \ifdraft\strut\vadjust{%
          \hbox to\z@{\hskip\hsize\hskip.1in
               \vbox to\z@{\vskip-\dp\strutbox
                    \marginnoteformat
                    \vskip-\ht\strutbox
                    \noindent\strut#1\par
                    \vss}%
               \hss}}%
     \fi
     \@marginsf}

%************** The \bye command ****************************

\newtoks\everybye \everybye={\par\vfil}
\outer\def\bye{\the\everybye
     \footnotecheck
     \prelabelcheck
     \streamcheck
     \supereject
     \TeXend}

%************************************************************
%*
%*             Footnotes
%*
%************************************************************
\message{footnotes,}

\newcount\footnotenum \footnotenum=0
\newskip\footnoteskip
\let\@footnotelist=\empty

\def\footnotenumstyle#1{\@setnumstyle\footnotenum{#1}%
     \useafter\ifx{@footnotenumstyle}\symbols
          \global\let\@footup=\empty
     \else\global\let\@footup=\markup
     \fi}

\def\footnote{\footnotecheck\defaultoption[]\@footnote}
\def\@footnote[#1]{\@footnotemark[#1]\@footnotetext}

\def\footnotemark{\defaultoption[]\@footnotemark}
\def\@footnotemark[#1]{\let\@footsf=\empty
     \ifhmode\edef\@footsf{\spacefactor=\the\spacefactor\relax}\/\fi
     \ifnoarg#1\then
          \global\advance\footnotenum by\@ne
          \@footup{\footnotenumformat}%
          \edef\@@foota{\footnotenum=\the\footnotenum\relax}%
          \expandafter\additemR\expandafter\@footup\expandafter
               {\@@foota\footnotenumformat}\to\@footnotelist
          \global\let\@footnotelist=\@footnotelist
     \else\markup{#1}%
          \additemR\markup{#1}\to\@footnotelist
          \global\let\@footnotelist=\@footnotelist
     \fi
     \@footsf}

\def\footnotetext{%
     \ifx\@footnotelist\empty\err@extrafootnotetext\else\@footnotetext\fi}
\def\@footnotetext{%
     \getitemL\@footnotelist\to\@@foota
     \global\let\@footnotelist=\@footnotelist
     \insert\footins\bgroup
     \footnoteformat
     \splittopskip=\ht\strutbox\splitmaxdepth=\dp\strutbox
     \interlinepenalty=\interfootnotelinepenalty\floatingpenalty=\@MM
     \noindent\llap{\@@foota}\strut
     \bgroup\aftergroup\@footnoteend
     \let\@@scratcha=}
\def\@footnoteend{\strut\par\vskip\footnoteskip\egroup}

\def\footnoterule{\normalfonts
     \kern-.3em \hrule width2in height.04em \kern .26em }

\def\footnotecheck{%
     \ifx\@footnotelist\empty
     \else\err@extrafootnotemark
          \global\let\@footnotelist=\empty
     \fi}

%************************************************************
%*
%*             Labelling macros
%*
%************************************************************
\message{labels,}

\let\@@labeldef=\xdef
\newif\if@labelfile
\newwrite\@labelfile
\let\@prelabellist=\empty

\def\label#1#2{\trim#1\to\@@labarg\edef\@@labtext{#2}%
     \edef\@@labname{lab@\@@labarg}%
     \useafter\ifundefined\@@labname\then\else\@yeslab\fi
     \useafter\@@labeldef\@@labname{#2}%
     \ifstreaming
          \expandafter\toks@\expandafter\expandafter\expandafter
               {\csname\@@labname\endcsname}%
          \immediate\write\streamout{\noexpand\label{\@@labarg}{\the\toks@}}%
     \fi}
\def\@yeslab{%
     \useafter\ifundefined{if\@@labname}\then
          \err@labelredef\@@labarg
     \else\useif{if\@@labname}\then
               \err@labelredef\@@labarg
          \else\global\usename{\@@labname true}%
               \useafter\ifundefined{pre\@@labname}\then
               \else\useafter\ifx{pre\@@labname}\@@labtext
                    \else\err@badlabelmatch\@@labarg
                    \fi
               \fi
               \if@labelfile
               \else\global\@labelfiletrue
                    \immediate\write\sixt@@n{--> Creating file \jobname.lab}%
                    \immediate\openout\@labelfile=\jobname.lab
               \fi
               \immediate\write\@labelfile
                    {\noexpand\prelabel{\@@labarg}{\@@labtext}}%
          \fi
     \fi}

\def\putlab#1{\trim#1\to\@@labarg\edef\@@labname{lab@\@@labarg}%
     \useafter\ifundefined\@@labname\then\@nolab\else\usename\@@labname\fi}
\def\@nolab{%
     \useafter\ifundefined{pre\@@labname}\then
          \undefinedlabelformat
          \err@needlabel\@@labarg
          \useafter\xdef\@@labname{\undefinedlabelformat}%
     \else\usename{pre\@@labname}%
          \useafter\xdef\@@labname{\usename{pre\@@labname}}%
     \fi
     \useafter\newif{if\@@labname}%
     \expandafter\additemR\@@labarg\to\@prelabellist}

\def\prelabel#1{\useafter\gdef{prelab@#1}}

\def\ifundefinedlabel#1\then{%
     \expandafter\ifx\csname lab@#1\endcsname\relax}
\def\useiflab#1\then{\csname iflab@#1\endcsname}

\def\prelabelcheck{{%
     \def\^^\##1{\useiflab{##1}\then\else\err@undefinedlabel{##1}\fi}%
     \@prelabellist}}

%************************************************************
%*
%*             Equation numbering
%*
%************************************************************
\message{equation numbering,}

\newcount\chapternum
\newcount\sectionnum
\newcount\subsectionnum
\newcount\equationnum
\newcount\subequationnum
\newcount\figurenum
\newcount\subfigurenum
\newcount\tablenum
\newcount\subtablenum

\newif\if@subeqncount
\newif\if@subfigcount
\newif\if@subtblcount

\def\newchapternum{\newsectionnum=\z@\@resetnum\chapternum}
\def\newsectionnum{\newsubsectionnum=\z@\@resetnum\sectionnum}
\def\newsubsectionnum{\newequationnum=\z@\newfigurenum=\z@\newtablenum=\z@
     \@resetnum\subsectionnum}
\def\newequationnum{\newsubequationnum=\z@\@resetnum\equationnum}
\def\newsubequationnum{\@resetnum\subequationnum}
\def\newfigurenum{\newsubfigurenum=\z@\@resetnum\figurenum}
\def\newsubfigurenum{\@resetnum\subfigurenum}
\def\newtablenum{\newsubtablenum=\z@\@resetnum\tablenum}
\def\newsubtablenum{\@resetnum\subtablenum}

\def\@resetnum#1{\global\advance#1by1 \edef\next{\the#1\relax}\global#1}

\newchapternum=0

\def\chapternumstyle#1{\@setnumstyle\chapternum{#1}}
\def\sectionnumstyle#1{\@setnumstyle\sectionnum{#1}}
\def\subsectionnumstyle#1{\@setnumstyle\subsectionnum{#1}}
\def\equationnumstyle#1{\@setnumstyle\equationnum{#1}}
\def\subequationnumstyle#1{\@setnumstyle\subequationnum{#1}%
     \ifblank\subequationnumstyle\then\global\@subeqncountfalse\fi
     \ignorespaces}
\def\figurenumstyle#1{\@setnumstyle\figurenum{#1}}
\def\subfigurenumstyle#1{\@setnumstyle\subfigurenum{#1}%
     \ifblank\subfigurenumstyle\then\global\@subfigcountfalse\fi
     \ignorespaces}
\def\tablenumstyle#1{\@setnumstyle\tablenum{#1}}
\def\subtablenumstyle#1{\@setnumstyle\subtablenum{#1}%
     \ifblank\subtablenumstyle\then\global\@subtblcountfalse\fi
     \ignorespaces}

\def\eqnlabel#1{%
     \if@subeqncount
          \newsubequationnum=\next
     \else\newequationnum=\next
          \ifblank\subequationnumstyle\then
          \else\global\@subeqncounttrue
               \newsubequationnum=\@ne
          \fi
     \fi
     \label{#1}{\puteqnformat}(\puteqn{#1})%
     \ifdraft\rlap{\hskip.1in{\tt#1}}\fi}

\let\puteqn=\putlab

\def\equation#1#2{\useafter\gdef{eqn@#1}{#2\eqno\eqnlabel{#1}}}
\def\Equation#1{\useafter\gdef{eqn@#1}}

\def\putequation#1{\useafter\ifundefined{eqn@#1}\then
     \err@undefinedeqn{#1}\else\usename{eqn@#1}\fi}

\def\eqnseriesstyle#1{\gdef\@eqnseriesstyle{#1}}
\def\begineqnseries{\subequationnumstyle{\@eqnseriesstyle}%
     \defaultoption[]\@begineqnseries}
\def\@begineqnseries[#1]{\edef\@@eqnname{#1}}
\def\endeqnseries{\subequationnumstyle{blank}%
     \expandafter\ifnoarg\@@eqnname\then
     \else\label\@@eqnname{\puteqnformat}%
     \fi
     \aftergroup\ignorespaces}

\def\figlabel#1{%
     \if@subfigcount
          \newsubfigurenum=\next
     \else\newfigurenum=\next
          \ifblank\subfigurenumstyle\then
          \else\global\@subfigcounttrue
               \newsubfigurenum=\@ne
          \fi
     \fi
     \label{#1}{\putfigformat}\putfig{#1}%
     {\def\marginnoteformat{\tt}\marginnote{#1}}}

\let\putfig=\putlab

\def\figseriesstyle#1{\gdef\@figseriesstyle{#1}}
\def\beginfigseries{\subfigurenumstyle{\@figseriesstyle}%
     \defaultoption[]\@beginfigseries}
\def\@beginfigseries[#1]{\edef\@@figname{#1}}
\def\endfigseries{\subfigurenumstyle{blank}%
     \expandafter\ifnoarg\@@figname\then
     \else\label\@@figname{\putfigformat}%
     \fi
     \aftergroup\ignorespaces}

\def\tbllabel#1{%
     \if@subtblcount
          \newsubtablenum=\next
     \else\newtablenum=\next
          \ifblank\subtablenumstyle\then
          \else\global\@subtblcounttrue
               \newsubtablenum=\@ne
          \fi
     \fi
     \label{#1}{\puttblformat}\puttbl{#1}%
     {\def\marginnoteformat{\tt}\marginnote{#1}}}

\let\puttbl=\putlab

\def\tblseriesstyle#1{\gdef\@tblseriesstyle{#1}}
\def\begintblseries{\subtablenumstyle{\@tblseriesstyle}%
     \defaultoption[]\@begintblseries}
\def\@begintblseries[#1]{\edef\@@tblname{#1}}
\def\endtblseries{\subtablenumstyle{blank}%
     \expandafter\ifnoarg\@@tblname\then
     \else\label\@@tblname{\puttblformat}%
     \fi
     \aftergroup\ignorespaces}

%************************************************************
%*
%*             Reference numbering
%*
%************************************************************
\message{reference numbering,}

\newcount\referencenum \referencenum=0
\newcount\@@prerefcount \@@prerefcount=0
\newcount\@@thisref
\newcount\@@lastref
\newcount\@@loopref
\newcount\@@refseq
\newdimen\refnumindent
\let\@undefreflist=\empty

\def\referencenumstyle#1{\@setnumstyle\referencenum{#1}}

\def\referencestyle#1{\usename{@ref#1}}

\def\@refsequential{%
     \gdef\@refpredef##1{\global\advance\referencenum by\@ne
          \let\^^\=0\label{##1}{\^^\{\the\referencenum}}%
          \useafter\gdef{ref@\the\referencenum}{{##1}{\undefinedlabelformat}}}%
     \gdef\@reference##1##2{%
          \ifundefinedlabel##1\then
          \else\def\^^\####1{\global\@@thisref=####1\relax}\putlab{##1}%
               \useafter\gdef{ref@\the\@@thisref}{{##1}{##2}}%
          \fi}%
     \gdef\endputreferences{%
          \loop\ifnum\@@loopref<\referencenum
                    \advance\@@loopref by\@ne
                    \expandafter\expandafter\expandafter\@printreference
                         \csname ref@\the\@@loopref\endcsname
          \repeat
          \par}}

\def\@refpreordered{%
     \gdef\@refpredef##1{\global\advance\referencenum by\@ne
          \additemR##1\to\@undefreflist}%
     \gdef\@reference##1##2{%
          \ifundefinedlabel##1\then
          \else\global\advance\@@loopref by\@ne
               {\let\^^\=0\label{##1}{\^^\{\the\@@loopref}}}%
               \@printreference{##1}{##2}%
          \fi}
     \gdef\endputreferences{%
          \def\^^\####1{\useiflab{####1}\then
               \else\reference{####1}{\undefinedlabelformat}\fi}%
          \@undefreflist
          \par}}

\def\beginprereferences{\par
     \def\reference##1##2{\global\advance\referencenum by1\@ne
          \let\^^\=0\label{##1}{\^^\{\the\referencenum}}%
          \useafter\gdef{ref@\the\referencenum}{{##1}{##2}}}}
\def\endprereferences{\global\@@prerefcount=\the\referencenum\par}

\def\beginputreferences{\par
     \refnumindent=\z@\@@loopref=\z@
     \loop\ifnum\@@loopref<\referencenum
               \advance\@@loopref by\@ne
               \setbox\z@=\hbox{\referencenum=\@@loopref
                    \referencenumformat\enskip}%
               \ifdim\wd\z@>\refnumindent\refnumindent=\wd\z@\fi
     \repeat
     \putreferenceformat
     \@@loopref=\z@
     \loop\ifnum\@@loopref<\@@prerefcount
               \advance\@@loopref by\@ne
               \expandafter\expandafter\expandafter\@printreference
                    \csname ref@\the\@@loopref\endcsname
     \repeat
     \let\reference=\@reference}

\def\@printreference#1#2{\ifx#2\undefinedlabelformat\err@undefinedref{#1}\fi
     \noindent\ifdraft\rlap{\hskip\hsize\hskip.1in \tt#1}\fi
     \llap{\referencenum=\@@loopref\referencenumformat\enskip}#2\par}

\def\reference#1#2{{\par\refnumindent=\z@\putreferenceformat\noindent#2\par}}

\def\putref#1{\trim#1\to\@@refarg
     \expandafter\ifnoarg\@@refarg\then
          \toks@={\relax}%
     \else\@@lastref=-\@m\def\@@refsep{}\def\@more{\@nextref}%
          \toks@={\@nextref#1,,}%
     \fi\the\toks@}
\def\@nextref#1,{\trim#1\to\@@refarg
     \expandafter\ifnoarg\@@refarg\then
          \let\@more=\relax
     \else\ifundefinedlabel\@@refarg\then
               \expandafter\@refpredef\expandafter{\@@refarg}%
          \fi
          \def\^^\##1{\global\@@thisref=##1\relax}%
          \global\@@thisref=\m@ne
          \setbox\z@=\hbox{\putlab\@@refarg}%
     \fi
     \advance\@@lastref by\@ne
     \ifnum\@@lastref=\@@thisref\advance\@@refseq by\@ne\else\@@refseq=\@ne\fi
     \ifnum\@@lastref<\z@
     \else\ifnum\@@refseq<\thr@@
               \@@refsep\def\@@refsep{,}%
               \ifnum\@@lastref>\z@
                    \advance\@@lastref by\m@ne
                    {\referencenum=\@@lastref\putrefformat}%
               \else\undefinedlabelformat
               \fi
          \else\def\@@refsep{--}%
          \fi
     \fi
     \@@lastref=\@@thisref
     \@more}

%************************************************************
%*
%*             Job streaming
%*
%************************************************************
\message{streaming,}

\newif\ifstreaming

\def\streamto{\defaultoption[\jobname]\@streamto}
\def\@streamto[#1]{\global\streamingtrue
     \immediate\write\sixt@@n{--> Streaming to #1.str}%
     \newwrite\streamout\immediate\openout\streamout=#1.str }

\def\streamfrom{\defaultoption[\jobname]\@streamfrom}
\def\@streamfrom[#1]{\newread\streamin\openin\streamin=#1.str
     \ifeof\streamin
          \expandafter\err@nostream\expandafter{#1.str}%
     \else\immediate\write\sixt@@n{--> Streaming from #1.str}%
          \let\@@labeldef=\gdef
          \ifstreaming
               \edef\@elc{\endlinechar=\the\endlinechar}%
               \endlinechar=\m@ne
               \loop\read\streamin to\@@scratcha
                    \ifeof\streamin
                         \streamingfalse
                    \else\toks@=\expandafter{\@@scratcha}%
                         \immediate\write\streamout{\the\toks@}%
                    \fi
                    \ifstreaming
               \repeat
               \@elc
               \input #1.str
               \streamingtrue
          \else\input #1.str
          \fi
          \let\@@labeldef=\xdef
     \fi}

\def\streamcheck{\ifstreaming
     \immediate\write\streamout{\pagenum=\the\pagenum}%
     \immediate\write\streamout{\footnotenum=\the\footnotenum}%
     \immediate\write\streamout{\referencenum=\the\referencenum}%
     \immediate\write\streamout{\chapternum=\the\chapternum}%
     \immediate\write\streamout{\sectionnum=\the\sectionnum}%
     \immediate\write\streamout{\subsectionnum=\the\subsectionnum}%
     \immediate\write\streamout{\equationnum=\the\equationnum}%
     \immediate\write\streamout{\subequationnum=\the\subequationnum}%
     \immediate\write\streamout{\figurenum=\the\figurenum}%
     \immediate\write\streamout{\subfigurenum=\the\subfigurenum}%
     \immediate\write\streamout{\tablenum=\the\tablenum}%
     \immediate\write\streamout{\subtablenum=\the\subtablenum}%
     \immediate\closeout\streamout
     \fi}

%************************************************************
%*
%*             Error messages
%*
%************************************************************

\def\err@badtypesize{%
     \errhelp={The limited availability of certain fonts requires^^J%
          that the base type size be 10pt, 12pt, or 14pt.^^J}%
     \errmessage{--> Illegal base type size}}

\def\err@badsizechange{\immediate\write\sixt@@n
     {--> Size change not allowed in math mode, ignored}}

\def\err@sizetoolarge#1{\immediate\write\sixt@@n
     {--> \noexpand#1 too big, substituting HUGE}}

\def\err@sizenotavailable#1{\immediate\write\sixt@@n
     {--> Size not available, \noexpand#1 ignored}}

\def\err@fontnotavailable#1{\immediate\write\sixt@@n
     {--> Font not available, \noexpand#1 ignored}}

\def\err@sltoit{\immediate\write\sixt@@n
     {--> Style \noexpand\sl not available, substituting \noexpand\it}%
     \it}

\def\err@bfstobf{\immediate\write\sixt@@n
     {--> Style \noexpand\bfs not available, substituting \noexpand\bf}%
     \bf}

\def\err@badgroup#1#2{%
     \errhelp={The block you have just tried to close was not the one^^J%
          most recently opened.^^J}%
     \errmessage{--> \noexpand\end{#1} doesn't match \noexpand\begin{#2}}}

\def\err@badcountervalue#1{\immediate\write\sixt@@n
     {--> Counter (#1) out of bounds}}

\def\err@extrafootnotemark{\immediate\write\sixt@@n
     {--> \noexpand\footnotemark command
          has no corresponding \noexpand\footnotetext}}

\def\err@extrafootnotetext{%
     \errhelp{You have given a \noexpand\footnotetext command without first
          specifying^^Ja \noexpand\footnotemark.^^J}%
     \errmessage{--> \noexpand\footnotetext command has no corresponding
          \noexpand\footnotemark}}

\def\err@labelredef#1{\immediate\write\sixt@@n
     {--> Label "#1" redefined}}

\def\err@badlabelmatch#1{\immediate\write\sixt@@n
     {--> Definition of label "#1" doesn't match value in \jobname.lab}}

\def\err@needlabel#1{\immediate\write\sixt@@n
     {--> Label "#1" cited before its definition}}

\def\err@undefinedlabel#1{\immediate\write\sixt@@n
     {--> Label "#1" cited but never defined}}

\def\err@undefinedeqn#1{\immediate\write\sixt@@n
     {--> Equation "#1" not defined}}

\def\err@undefinedref#1{\immediate\write\sixt@@n
     {--> Reference "#1" not defined}}

\def\err@nostream#1{%
     \errhelp={You have tried to input a stream file that doesn't exist.^^J}%
     \errmessage{--> Stream file #1 not found}}

%************************************************************
%*
%*             Initialization
%*
%************************************************************
\message{jyTeX initialization}

\everyjob{\immediate\write16{--> jyTeX version \fmtversion}%
     \edef\@@jobname{\jobname}%
%     \openin0=\inputpath jysupp
%     \ifeof0
%     \else\closein0
%          \immediate\write16{--> Additional macros loaded from jysupp.tex}%
%          \jyinput jysupp
%     \fi
%     \openin0=\inputpath jylocal
%     \ifeof0
%     \else\closein0
%          \immediate\write16{--> Additional macros loaded from jylocal.tex}%
%          \jyinput jylocal
%     \fi
     \edef\jobname{\@@jobname}%
     \settime
     \openin0=\jobname.lab
     \ifeof0
     \else\closein0
          \immediate\write16{--> Getting labels from file \jobname.lab}%
          \input\jobname.lab
     \fi}

%************** Spacing *************************************

\def\fixedskipslist{%
     \^^\{\topskip}%
     \^^\{\splittopskip}%
     \^^\{\maxdepth}%
     \^^\{\skip\topins}%
     \^^\{\skip\footins}%
     \^^\{\headskip}%
     \^^\{\footskip}}

\def\scalingskipslist{%
     \^^\{\p@renwd}%
     \^^\{\delimitershortfall}%
     \^^\{\nulldelimiterspace}%
     \^^\{\scriptspace}%
     \^^\{\jot}%
     \^^\{\normalbaselineskip}%
     \^^\{\normallineskip}%
     \^^\{\normallineskiplimit}%
     \^^\{\baselineskip}%
     \^^\{\lineskip}%
     \^^\{\lineskiplimit}%
     \^^\{\bigskipamount}%
     \^^\{\medskipamount}%
     \^^\{\smallskipamount}%
     \^^\{\parskip}%
     \^^\{\parindent}%
     \^^\{\abovedisplayskip}%
     \^^\{\belowdisplayskip}%
     \^^\{\abovedisplayshortskip}%
     \^^\{\belowdisplayshortskip}%
     \^^\{\abovechapterskip}%
     \^^\{\belowchapterskip}%
     \^^\{\abovesectionskip}%
     \^^\{\belowsectionskip}%
     \^^\{\abovesubsectionskip}%
     \^^\{\belowsubsectionskip}}

%************** Document layout *****************************

\def\twoupsetup{%                                % setup for twoup style
     \topmargin=.75in
     \leftmargin=.5in
     \vsize=6.9in
     \hsize=4.75in
     \fullhsize=10in
     \let\draft=\relax}

\outputstyle{normal}                             % page style

\def\marginnoteformat{\subscriptsize             % paragraphing of margin notes
     \hsize=1in \baselinestretch=1000 \everypar={}%
     \tolerance=5000 \hbadness=5000 \parskip=0pt \parindent=0pt
     \leftskip=0pt \rightskip=0pt \raggedright}

\head={\ifdraft\normalfonts\it\hfil DRAFT\hfil   % format of headline
     \llap{\number\day\ \monthword\month\ \militarytime}\else\hfil\fi}
\foot={\hfil\normalfonts\numstyle\pagenum\hfil}  % format of footline

\normalbaselineskip=12pt                         % usual \baselineskip
\normallineskip=0pt                              % usual \lineskip
\normallineskiplimit=0pt                         % usual \lineskiplimit
\normalbaselines                                 % set \baselineskip

\topskip=.85\baselineskip \splittopskip=\topskip \headskip=2\baselineskip
\footskip=\headskip

\pagenumstyle{arabic}                            % counter style

\parskip=0pt                                     % no skip between paragraphs
\parindent=20pt                                  % usual \parindent

\baselinestretch=1000                            % set \big-, \med-, \smallskip

%************** Sectioning **********************************

\chapterstyle{left}                              % position of heading
\chapternumstyle{blank}                          % counter style
\def\chapterbreak{\newpage}                      % break before heading
\abovechapterskip=0pt                            % space before heading
\belowchapterskip=1.5\baselineskip               % space after heading
     plus.38\baselineskip minus.38\baselineskip
\def\chapternumformat{\numstyle\chapternum.}     % format of heading counter

\sectionstyle{left}                              % position of heading
\sectionnumstyle{blank}                          % counter style
\def\sectionbreak{\vskip0pt plus4\baselineskip\penalty-100
     \vskip0pt plus-4\baselineskip}              % break before heading
\abovesectionskip=1.5\baselineskip               % space before heading
     plus.38\baselineskip minus.38\baselineskip
\belowsectionskip=\the\baselineskip              % space after heading
     plus.25\baselineskip minus.25\baselineskip
\def\sectionnumformat{%                          % format of heading counter
     \ifblank\chapternumstyle\then\else\numstyle\chapternum.\fi
     \numstyle\sectionnum.}

\subsectionstyle{left}                           % position of heading
\subsectionnumstyle{blank}                       % counter style
\def\subsectionbreak{\vskip0pt plus4\baselineskip\penalty-100
     \vskip0pt plus-4\baselineskip}              % break before heading
\abovesubsectionskip=\the\baselineskip           % space before heading
     plus.25\baselineskip minus.25\baselineskip
\belowsubsectionskip=.75\baselineskip            % space after heading
     plus.19\baselineskip minus.19\baselineskip
\def\subsectionnumformat{%                       % format of heading counter
     \ifblank\chapternumstyle\then\else\numstyle\chapternum.\fi
     \ifblank\sectionnumstyle\then\else\numstyle\sectionnum.\fi
     \numstyle\subsectionnum.}

%************** Footnotes ***********************************

\footnotenumstyle{symbols}                       % counter style
\footnoteskip=0pt                                % jyTeX spacing parameter
\def\footnotenumformat{\numstyle\footnotenum}    % \footnotemark format
\def\footnoteformat{\footnotesize                % paragraphing of text
     \everypar={}\parskip=0pt \parfillskip=0pt plus1fil
     \leftskip=1em \rightskip=0pt
     \spaceskip=0pt \xspaceskip=0pt
     \def\\{\ifhmode\ifnum\lastpenalty=-10000
          \else\hfil\penalty-10000 \fi\fi\ignorespaces}}

%************** Labels **************************************

\def\undefinedlabelformat{$\bullet$}             % mark for undefined label

%************** Equation numbering **************************

\equationnumstyle{arabic}                        % counter style
\subequationnumstyle{blank}                      % counter style
\figurenumstyle{arabic}                          % counter style
\subfigurenumstyle{blank}                        % counter style
\tablenumstyle{arabic}                           % counter style
\subtablenumstyle{blank}                         % counter style

\eqnseriesstyle{alphabetic}                      % sub-counter style for series
\figseriesstyle{alphabetic}                      % sub-counter style for series
\tblseriesstyle{alphabetic}                      % sub-counter style for series

\def\puteqnformat{\hbox{%                        % equation number format
     \ifblank\chapternumstyle\then\else\numstyle\chapternum.\fi
     \ifblank\sectionnumstyle\then\else\numstyle\sectionnum.\fi
     \ifblank\subsectionnumstyle\then\else\numstyle\subsectionnum.\fi
     \numstyle\equationnum
     \numstyle\subequationnum}}
\def\putfigformat{\hbox{%                        % figure number format
     \ifblank\chapternumstyle\then\else\numstyle\chapternum.\fi
     \ifblank\sectionnumstyle\then\else\numstyle\sectionnum.\fi
     \ifblank\subsectionnumstyle\then\else\numstyle\subsectionnum.\fi
     \numstyle\figurenum
     \numstyle\subfigurenum}}
\def\puttblformat{\hbox{%                        % table number format
     \ifblank\chapternumstyle\then\else\numstyle\chapternum.\fi
     \ifblank\sectionnumstyle\then\else\numstyle\sectionnum.\fi
     \ifblank\subsectionnumstyle\then\else\numstyle\subsectionnum.\fi
     \numstyle\tablenum
     \numstyle\subtablenum}}

%************** Reference numbering *************************

\referencestyle{sequential}                      % referencing method
\referencenumstyle{arabic}                       % counter style
\def\putrefformat{\numstyle\referencenum}        % format of reference citation
\def\referencenumformat{\numstyle\referencenum.} % format of number in list
\def\putreferenceformat{%                        % paragraphing of list
     \everypar={\hangindent=1em \hangafter=1 }%
     \def\\{\hfil\break\null\hskip-1em \ignorespaces}%
     \leftskip=\refnumindent\parindent=0pt \interlinepenalty=1000 }

%************** Font initialization *************************

\normalsize

%*****************************************************************************

\def\fmtversion{2.6M (June 1992)}

\catcode`\@=12
% ------------------ End of jytex.tex -----------------

%\input jytex.tex   % available from hep-th
\typesize=10pt \magnification=1200 \baselineskip17truept
%\baselineskip25truept
\footnotenumstyle{arabic} \hsize=6truein\vsize=8.5truein
\input epsf
%\draft
%\leftmargin=1.25in
%\oddleftmargin=.5in
%\evenleftmargin=1.5in
\sectionnumstyle{blank}
\chapternumstyle{blank}
\chapternum=1
\sectionnum=1
\pagenum=0
%\referencestyle{preordered}
% title style follows

\def\begintitle{\pagenumstyle{blank}\parindent=0pt
\begin{narrow}[0.4in]}
\def\endtitle{\end{narrow}\newpage\pagenumstyle{arabic}}

% exercise style follows

\def\beginexercise{\vskip 20truept\parindent=0pt\begin{narrow}[10
truept]}
\def\endexercise{\vskip 10truept\end{narrow}}

% **************    my jyTeX abbreviations   *****************

\def\eql#1{\eqno\eqnlabel{#1}}
\def\ref{\reference}
\def\peq{\puteqn}
\def\pref{\putref}

\def\mgn{\marginnote}
\def\bex{\begin{exercise}}
\def\eex{\end{exercise}}

% *********************** My definitions ************************

\font\open=msbm10 %scaled\magstep1 % For VAX. Borde p195.

 %scaled\magstep1 % For VAX. Borde p195.
%\font\open=msym10 %scaled\magstep1 % For Arbortxt on PC
%\font\opens=msym8 %scaled\magstep1 % For Arbortxt on PC
  % For Arbortxt on PC, and VAX. Borde p199
 
\def\StretchRtArr#1{{\count255=0\loop\relbar\joinrel\advance\count255 by1
\ifnum\count255<#1\repeat\rightarrow}}
\def\StretchLtArr#1{\,{\leftarrow\!\!\count255=0\loop\relbar
\joinrel\advance\count255 by1\ifnum\count255<#1\repeat}}

\def\StretchLRtArr#1{\,{\leftarrow\!\!\count255=0\loop\relbar\joinrel\advance
\count255 by1\ifnum\count255<#1\repeat\rightarrow\,\,}}

\def\mbox#1{{\leavevmode\hbox{#1}}}

\def\hspace#1{{\phantom{\mbox#1}}}
\def\oR{\mbox{\open\char82}}

\def\oZ{\mbox{\open\char90}}

\def\al{\alpha}

 %in jyTeX
 %in jyTeX
 %in jyTeX
 %in jyTeX
 %in jyTeX
 %in jyTeX
 %in jyTeX
 %in jyTeX
 %in jyTeX
 %in jyTeX
 %in jyTeX
 %in jyTeX
 %in jyTeX
% in jyTeX
% in jyTeX
% in jyTeX
% in jyTeX
% in jyTeX

\def\ga{\gamma}
\def\de{\delta}

\def\la{\lambda}

\def\si{\sigma}

\def\ze{\zeta}

\def\De{\Delta}

\def\caC{{\cal C}}

\def\caP{{\cal P}}

\def\det{{\rm det\,}}

\def\Real{{\rm Re\,}}

\def\sc{{\rm sc }}

\def\zf{$\zeta$--function}

     % Newline

\def\frac#1/#2{\leavevmode\kern.1em
\raise.5ex\hbox{\the\scriptfont0 #1}\kern-.1em/\kern-.15em
\lower.25ex\hbox{\the\scriptfont0 #2}}
\def\sfrac#1/#2{\leavevmode\kern.1em
\raise.5ex\hbox{\the\scriptscriptfont0 #1}\kern-.1em/\kern-.15em
\lower.25ex\hbox{\the\scriptscriptfont0 #2}}

\def\gtorder{\mathrel{\raise.3ex\hbox{$>$}\mkern-14mu
             \lower0.6ex\hbox{$\sim$}}}
\def\ltorder{\mathrel{\raise.3ex\hbox{$<$}\mkern-14mu
             \lower0.6ex\hbox{$\sim$}}}

\def\semidirprod{\rlap{\ss C}\raise1pt\hbox{$\mkern.75mu\times$}}
\def\for{\lower6pt\hbox{$\Big|$}}
\def\fish{\kern-.25em{\phantom{abcde}\over \phantom{abcde}}\kern-.25em}

 %triple
%dot
 %double
%dot
 %double dot
%for small #1

\def\boxit#1{\vbox{\hrule\hbox{\vrule\kern3pt
        \vbox{\kern3pt#1\kern3pt}\kern3pt\vrule}\hrule}}
\def\dalemb#1#2{{\vbox{\hrule height .#2pt
        \hbox{\vrule width.#2pt height#1pt \kern#1pt \vrule
                width.#2pt} \hrule height.#2pt}}}

\def\ol{\overline}
        %double stroke
\def\frac#1#2{{{#1}\over{#2}}}
 %lower covariant deriv.
 %upper covariant deriv.
 %lower covariant deriv semicolon.
    %lower ordinary  deriv.
    %lower ordinary  deriv comma.

\def\noin{\noindent}

      %Connection
    %Connection'
\def\comb#1#2{{\left(#1\atop#2\right)}}

\def\cosec{{\rm cosec\,}}

\def\eg{{\it e.g.}}
\def\ie{{\it i.e. }}
\def\cf{{\it cf }}
\def\pa{\partial}

 %gives average <#1>
 %gives thermal average <<#1>>
   %gives bracket <#1|#2>
   %gives comma bracket <#1,#2>
 %gives round bracket (#1,#2)
 %gives round bracket (#1,|#2)
 %gives big bracket <#1|#2>
\def\me#1#2#3{\langle{#1}\mid\!{#2}\!\mid{#3}\rangle}  %gives
%matrix element <#1|#2|#3>

%gives reduced matrix element
%<#1||#2||#3>

\def\tr{{\rm tr\,}}

\def\3j#1#2#3#4#5#6{\left\lgroup\matrix{#1&#2&#3\cr#4&#5&#6\cr}
\right\rgroup}

\def\m?{\mgn{?}}
% KK's defs

\def\pa{\partial}

\def\beq{\begin{eqnarray}}
\def\eeq{\end{eqnarray}}

%  *******************  Journal refs **********************

\def\aop#1#2#3{{\it Ann. Phys.} {\bf {#1}} ({#2}) #3}
\def\cjp#1#2#3{{\it Can. J. Phys.} {\bf {#1}} ({#2}) #3}
\def\cmp#1#2#3{{\it Comm. Math. Phys.} {\bf {#1}} ({#2}) #3}
\def\cqg#1#2#3{{\it Class. Quant. Grav.} {\bf {#1}} ({#2}) #3}

\def\ijmp#1#2#3{{\it Int. J. Mod. Phys.} {\bf {#1}} ({#2}) #3}

\def\jmp#1#2#3{{\it J. Math. Phys.} {\bf {#1}} ({#2}) #3}
\def\jpa#1#2#3{{\it J. Phys.} {\bf A{#1}} ({#2}) #3}
\def\jpc#1#2#3{{\it J. Phys.} {\bf C{#1}} ({#2}) #3}
\def\lnm#1#2#3{{\it Lect. Notes Math.} {\bf {#1}} ({#2}) #3}

\def\np#1#2#3{{\it Nucl. Phys.} {\bf B{#1}} ({#2}) #3}
\def\npa#1#2#3{{\it Nucl. Phys.} {\bf A{#1}} ({#2}) #3}
\def\pl#1#2#3{{\it Phys. Lett.} {\bf {#1}} ({#2}) #3}
\def\phm#1#2#3{{\it Phil.Mag.} {\bf {#1}} ({#2}) #3}
\def\prp#1#2#3{{\it Phys. Rep.} {\bf {#1}} ({#2}) #3}
\def\pr#1#2#3{{\it Phys. Rev.} {\bf {#1}} ({#2}) #3}
\def\prA#1#2#3{{\it Phys. Rev.} {\bf A{#1}} ({#2}) #3}

\def\prD#1#2#3{{\it Phys. Rev.} {\bf D{#1}} ({#2}) #3}
\def\prE#1#2#3{{\it Phys. Rev.} {\bf E{#1}} ({#2}) #3}
\def\prl#1#2#3{{\it Phys. Rev. Lett.} {\bf #1} ({#2}) #3}

\def\rmp#1#2#3{{\it Rev. Mod. Phys.} {\bf {#1}} ({#2}) #3}

\def\zfp#1#2#3{{\it Z. f. Phys.} {\bf {#1}} ({#2}) #3}

\def\cras#1#2#3{{\it Comptes Rend. Acad. Sci. (Paris)} {\bf{#1}} (#2) #3}
\def\prs#1#2#3{{\it Proc. Roy. Soc.} {\bf A{#1}} ({#2}) #3}
\def\pcps#1#2#3{{\it Proc. Camb. Phil. Soc.} {\bf{#1}} ({#2}) #3}
\def\mpcps#1#2#3{{\it Math. Proc. Camb. Phil. Soc.} {\bf{#1}} ({#2}) #3}

\def\amsh#1#2#3{{\it Abh. Math. Sem. Ham.} {\bf {#1}} ({#2}) #3}
\def\am#1#2#3{{\it Acta Mathematica} {\bf {#1}} ({#2}) #3}
\def\aim#1#2#3{{\it Adv. in Math.} {\bf {#1}} ({#2}) #3}
\def\ajm#1#2#3{{\it Am. J. Math.} {\bf {#1}} ({#2}) #3}
\def\amm#1#2#3{{\it Am. Math. Mon.} {\bf {#1}} ({#2}) #3}

\def\aom#1#2#3{{\it Ann. of Math.} {\bf {#1}} ({#2}) #3}
\def\cjm#1#2#3{{\it Can. J. Math.} {\bf {#1}} ({#2}) #3}
\def\bams#1#2#3{{\it Bull.Am.Math.Soc.} {\bf {#1}} ({#2}) #3}

\def\cmh#1#2#3{{\it Comm. Math. Helv.} {\bf {#1}} ({#2}) #3}

\def\dmj#1#2#3{{\it Duke Math. J.} {\bf {#1}} ({#2}) #3}
\def\invm#1#2#3{{\it Invent. Math.} {\bf {#1}} ({#2}) #3}

\def\jdg#1#2#3{{\it J. Diff. Geom.} {\bf {#1}} ({#2}) #3}

\def\joa#1#2#3{{\it J. of Algebra} {\bf {#1}} ({#2}) #3}
\def\jram#1#2#3{{\it J. f. Reine u. Angew. Math.} {\bf {#1}} ({#2}) #3}
\def\jims#1#2#3{{\it J. Indian. Math. Soc.} {\bf {#1}} ({#2}) #3}
\def\jlms#1#2#3{{\it J. Lond. Math. Soc.} {\bf {#1}} ({#2}) #3}
\def\jmpa#1#2#3{{\it J. Math. Pures. Appl.} {\bf {#1}} ({#2}) #3}
\def\ma#1#2#3{{\it Math. Ann.} {\bf {#1}} ({#2}) #3}

\def\mz#1#2#3{{\it Math. Zeit.} {\bf {#1}} ({#2}) #3}
\def\ojm#1#2#3{{\it Osaka J.Math.} {\bf {#1}} ({#2}) #3}
\def\pams#1#2#3{{\it Proc. Am. Math. Soc.} {\bf {#1}} ({#2}) #3}
\def\pems#1#2#3{{\it Proc. Edin. Math. Soc.} {\bf {#1}} ({#2}) #3}

\def\plb#1#2#3{{\it Phys. Letts.} {\bf {B#1}} ({#2}) #3}
\def\pla#1#2#3{{\it Phys. Letts.} {\bf {A#1}} ({#2}) #3}
\def\plms#1#2#3{{\it Proc. Lond. Math. Soc.} {\bf {#1}} ({#2}) #3}
\def\pgma#1#2#3{{\it Proc. Glasgow Math. Ass.} {\bf {#1}} ({#2}) #3}
\def\qjm#1#2#3{{\it Quart. J. Math.} {\bf {#1}} ({#2}) #3}
\def\qjpam#1#2#3{{\it Quart. J. Pure and Appl. Math.} {\bf {#1}} ({#2}) #3}

\def\rmjm#1#2#3{{\it Rocky Mountain J. Math.} {\bf {#1}} ({#2}) #3}

\def\tams#1#2#3{{\it Trans.Am.Math.Soc.} {\bf {#1}} ({#2}) #3}

% *******************   Main text *********************
\begin{title}
\vglue 0.5truein
%\righttext {MUTP/96/23}
%\righttext{hep-th/96}
\vskip15truept
%\leftline{\today}
%\vskip 30truept
\centertext {\Bigfonts \bf Heat--kernels on the discrete circle and interval} \vskip7truept
\vskip10truept\centertext{\Bigfonts \bf }
 \vskip 20truept
\centertext{J.S.Dowker\footnote{dowker@man.ac.uk}} \vskip 7truept \centertext{\it
Theory Group,} \centertext{\it School of Physics and Astronomy,} \centertext{\it The
University of Manchester,} \centertext{\it Manchester, England} \vskip 7truept
\centertext{}

\vskip 7truept

\vskip40truept
\begin{narrow}
As is known, the free heat--kernel on the integers (a modified Bessel function) is turned into
the periodic free heat--kernel on the discrete circle by factoring, giving a pre--image sum. I
generalise existing treatments by making the functions periodic up to a phase, thus
introducing an extra parameter into the analysis.  Identifying the classical paths form with
the conventional eigenfunction expression, I find a combinatorial trace identity which allows
various Bessel identities to be extracted, such as a generalisation of the Jacobi--Anger
expansion.

The free Dirichlet, Neumann and hybrid Dirichlet--Neumann heat--kernels on a discrete
interval are constructed using both modes and images. The Neumann imaging mirror has to
be placed at a half--integer.

The corresponding lattice Green functions are expressed in terms of Chebyshev polynomials
and the Laplacian matrices extracted. The generating functions for circuits with bumps are
evaluated.

\end{narrow}
\vskip 5truept \vskip 60truept
%\righttext{Typeset in \jyTeX}
\vfil
\end{title}
\pagenum=0
\newpage

\section{\bf 1. Introduction}

In the next two sections I summarise and rework some known material, [\pref{CandY2}],
[\pref{KandN}], on the free heat--kernels on the integers and on the discrete circle with $p$
vertices (or $p$--cycle). In section 3, I extend the discussion to complex functions on the
circle periodic up to a phase. This is mathematically possible and gives extra room for
manoeuvre in the analysis and allows for a more general trace identity. Other sections
contain some algebraic consequences, in particular Bessel function identities, which might be
novel. I also construct the Dirichlet, Neumann and Dirichlet--Neumann kernels on a discrete
interval using both modes and images and then, by Laplace transformation, Chebyshev
polynomial forms for the lattice Green functions.

I am only concerned here with one dimension but higher dimensions can be reached by
cartesian products and, as physical motivation for revisiting this topic, I remark that the
spectral problem, \eg\ determinants, on higher dimensional lattices, particularly two
dimensional ones, has relevance in statistical mechanics, [\pref{Kasteleyn}],
[\pref{DuandD}], string theory, [\pref{GandT}], [\pref{Thorn}], network theory and many
other areas.

The two--dimensional discrete (periodic) lattice also appears in the description of the
generators of SU(N), [\pref{FFZ}], [\pref{FandZ}], and can have the interpretation of a
phase space. It occurs thus in a finite model of two--dimensional ideal hydrodynamics on the
torus, [\pref{DandWo}], [\pref{Zeitlin}]. In [\pref{DandWo}], the discrete \zf\ appears
when constructing the stream function of a lattice vortex by analogy with the continuum
version.

\section{\bf 2. Heat--Kernel on the integers}

The free heat--kernel, $K_{\oZ}(j,j';t)$, on the integers, $j,j'\in\oZ$, has been given by
Chung and Yau, [\pref{CandY2}], for example. The defining equation is the discrete heat, or
diffusion, equation\footnote{ Some authors have a factor of $1/2$ in the first term.}
  $$
  \bigg(-\nabla\Delta +{\pa\over\pa t}\bigg)K_{\oZ}(j,j';t)=\de(t)
  \eql{heqn}
  $$
with initial condition,
  $$
  K_{\oZ}(j,j';0)=\de_{jj'}
  \eql{incond}
  $$
and the forward propagation condition, $K_{\oZ}(j,j';t)=0 $ for $t<0$.

The discrete Laplacian is
  $$
 \nabla\Delta\, y(j) \equiv y(j+1)-2 y(j)+y(j-1)\,.
 \eql{disclap}
  $$

The explicit expression for $K$ is derived in [\pref{CandY2}] and [\pref{KandN}]. I give an
algebraically slightly different, and more rapid, development.

By translation invariance, it is sufficient to set $j'=0$ and consider $K_{\oZ}(j;t)\equiv
K_{\oZ}(j,0;t)$. It is also expedient to remove the middle term in the Laplacian,
(\peq{disclap}), at the outset by setting,
  $$
  K_{\oZ}(j;t)=e^{-2t}\,\ol K_{\oZ}(j;t)\,,
  $$
so that the operator solution of (\peq{heqn}) is effectively expressed in,
  $$
  \ol K_{\oZ}(j;t)=e^{t(E+E^{-1})}\,\de(j)=e^{tE}\, e^{tE^{-1}}\,\de(j)\,,
  \eql{hk}
  $$
where $E$ is the usual stepping operator, $E\,y(j)=y(j+1)$, and $\de(j)=\de_{j0}$, in
terms of the Kronecker delta.

Expanding the exponentials in (\peq{hk}) it is immediately seen that,
  $$\eqalign{
  e^{tE^{-1}}\de(j)&={t^j\over j!}\cr
  \ol K_{\oZ}(j;t)&=e^{tE}\,{t^j\over j!}=
  \sum_{n=0}^\infty {t^{2n+j}\over n!(n+j)!}\,\quad j\ge0\,,
  }
  $$
which is the standard series expression for the Bessel function, $I_j(2t)$, and I obtain the
published formula for the free heat--kernel on (all)  the integers,\footnote {Equivalently,
one requires the term $E^{-j}$ in $(E+E^{-1})^n$ in order to pull the index $j$ in
$\de_{j0}$ back to zero, \cf\ [\pref{CandY2}].}
  $$
   K_{\oZ}(j;t)=e^{-2t}\,I_j(2t)\,,\quad t\ge0.
  $$
This result is not surprising, \cf\ [\pref{KandN}], in view of the fact that Basset's modified
Bessel function, $I_\nu$, satisfies the recursion,
  $$
( E+E^{-1})I_\nu(z)={\pa\over\pa z}\,I_\nu(z)\,.
  $$
Further, the initial condition, (\peq{incond}), is validated using $I_j(0)=\de(j)$.

By discrete translation invariance,
  $$
   K_{\oZ}(j,j';t)=e^{-2t}\,I_{j-j'}(2t)\,.
   \eql{ZK}
  $$
Karlsson and Neuhauser [\pref{KandN}], note that a probabilistic derivation of this result is
given by Feller, [\pref{Feller}], and also remark on the relevance of the composition
rule,\footnote{ By an infinite folding, one could obtain an analogue of a path integral.}
  $$
  I_{j-j''}(z_1+z_2)=\sum_{j'=-\infty}^\infty I_{j-j'}(z_1)\,I_{j'-j''}(z_2)\,.
  $$

As is very well known in the continuum case, the heat--kernel on the real line, $K_\oR$, can
be found in many ways, the standard one being eigenfunction expansion. Equating these
various expressions gives identities. I will not do this for the integers because I wish to pass
immediately to the factor, $ \oZ/p\oZ$, which is the discrete circle  with $p$ vertices (a
$p$--cycle, $\caC$). obtained from the covering, $\oZ$, by identification, exactly as for the
continuum.

\section{\bf 3. Heat--Kernel on the discrete circle}

A standard, geometric, way of obtaining the heat--kernel on the covered space is by a
pre--image sum of that on the covering space to give a periodic function.\footnote{This
procedure has occurred  in numerous places at various times.} Hence,
  $$\eqalign{
   K_{\oZ/p\oZ}(j,j';t)&= \sum_{m=-\infty}^\infty K_{\oZ}(j,j'+mp\,;t)\cr
  &=e^{-2t}\,\sum_{m=-\infty}^\infty I_{j-j'+mp}(2t)\,,
  }
  \eql{socp}
  $$
which can be called the sum over classical paths form, by analogy with the more common
continuum situation. I do not distinguish, notationally, between points on $\oZ$ and those
on $\oZ/p\oZ$. Quantities on the covering space are the thermodynamic limits of those on
the covered space and correspond to the direct term, $m=0$, in the sum (\peq{socp}).

A dual representation arises from the eigenfunction  expression. The Laplacian eigenproblem
on the discrete circle is old, \eg\ [\pref{Fort}]. I therefore just write down the
eigenfunctions, in complex form, and the eigenvalues.

The normalised modes on the discrete circle are,
    $$
   \psi_n(j)={1\over\sqrt p}\,e^{2\pi inj /p}\,,\quad n=0,\ldots,p-1\,,
   \eql{eigf}
    $$
with corresponding eigenvalues,
  $$
  \la_n=4\sin^2{\pi n\over p}\,,
  \eql{Peigen}
  $$
so tha
  $$
   K_{\oZ/p\oZ}(j;t)={1\over p} \sum_{n=0}^{p-1}
   e^{-4\sin^2(\pi n/p)\,t}\,e^{2\pi inj /p}\,.
   \eql{eigform}
  $$

Equating the two representations of the heat--kernel yields the identity, [\pref{KandN}],
  $$
  e^{-2t}\,\sum_{m=-\infty}^\infty I_{j+mp}(2t)=
  {1\over p} \sum_{n=0}^{p-1}e^{-4\sin^2(\pi n/p)\,t}\,e^{2\pi inj /p}\,.
  \eql{id}
  $$

Cosmetically re-expressed, this simplifies to,
  $$
  \sum_{m=-\infty}^\infty I_{j+mp}(z)=
  {1\over p} \sum_{n=0}^{p-1}e^{\cos(2\pi n/p)\,z}\,e^{2\pi inj /p}\,.
  \eql{ident1}
  $$
For $j=0$, this result is to be found in Al--Jarrah, Dempsey and Glasser, [\pref{ADG}],
together with numerous Bessel series. Their method involves a generalised Poisson formula
due to Titchmarsh.

Since the left--hand side of (\peq{id}) is real (if $2t$ is) one can conclude,
  $$\eqalign{
  e^{-2t}\,\sum_{m=-\infty}^\infty I_{j+mp}(2t)&=
  {1\over p} \sum_{n=0}^{p-1}e^{-4\sin^2(\pi n/p)\,t}\,\cos{2\pi nj /p}\cr
  0&=\sum_{n=0}^{p-1}e^{-4\sin^2(\pi n/p)\,t}\,\sin2\pi nj /p\,.
  }
  \eql{idr}
  $$

\section{\bf 4. Heat--kernel on the circle for twisted fields.}

So far, I have reproduced published results. I wish now to generalise the analysis a little. In
the continuum case, according to the results of Schulman, [\pref{Schulman}], Laidlaw and
DeWitt, [\pref{LandD}], and Dowker, [\pref{Dow}], in the quantum mechanical context the
wave functions on the covered space can be be twisted by a unitary representation of the
character group of the fundamental group, $\pi_1$, of the covered manifold.\footnote{ For
line bundles, $\pi_1$ can be replaced by the more tractable first homology group, $H_1$ as
discussed in [\pref{dowaustin}], [\pref{dowstat}].} For the circle, $\pi_1=\oZ$, and the
unitary representations are generated by the phase, $\exp 2\pi i\al$, where $\al$ is a real
parameter. The wavefunction picks up this phase as its angle argument increases by $2\pi$,
\ie as the circle is completed. A possible interpretation of this involves an Aharonov--Bohm
flux.

In terms of the propagator, this twisting translates into the statement that each term of the
pre--image sum enters with a multiplying factor of a power of this phase.

Here I want to use this flexibility in the discrete case, and for diffusion. That is, I consider a
(now complex) function, $\psi(j)$, defined on the points, $j$, and satisfying the twisted
periodicity condition,
  $$
  \psi(p)=e^{2\pi i\al}\psi(0)\,,\quad\psi(p-1)=e^{2\pi i\al}\psi(-1)\,,
  \eql{twbc}
  $$
which generalises the usual one.\footnote{ Fort, [\pref{Fort}]. discusses the case $\al=1/2$
which corresponds to (real) anti--periodic functions.} At the moment, I present this simply
as a mathematical generalisation.\footnote{ The use of such twisted boundary conditions is
commonplace in statistical physics and string theory, both continuous and discrete. As a
typical example I cite [\pref{IIH}].}

Firstly, the classical paths expression is modified to the twisted form,
  $$\eqalign{
   K_{\oZ/p\oZ}(j,j';t)&= \sum_{m=-\infty}^\infty e^{-2\pi i m\al}\,K_{\oZ}(j,j'+mp\,;t)\cr
  &=e^{-2t}\,\sum_{m=-\infty}^\infty  e^{-2\pi i m\al}\,I_{j-j'+mp}(2t)\,.
  }
  $$

Then, instead of (\peq{eigf}) and (\peq{Peigen})  the eigenfunctions are,\mgn{check sign
of $\al$}
    $$
   \psi^{\al}_n(j)={1\over\sqrt p}\,e^{2\pi i(n+\al)j /p}\,,\quad n=0,
   \ldots,p-1\,,\quad 0\le\al<1\,,
    $$
with corresponding eigenvalues,
  $$
  \la_n(\al)=4\sin^2{\pi (n+\al)\over p}\,.
  \eql{Peigen2}
  $$

As before, identifying the paths and eigenfunction expressions yields, instead of (\peq{id}),
the twisted identity, ($z=2t$),
  $$
  e^{-z}\,\sum_{m=-\infty}^\infty e^{-2\pi im\al}\,I_{j+mp}(z)=
  {1\over p} \sum_{n=0}^{p-1}e^{-2z\sin^2\big(\pi (n+\al)/p\big)}\,
  e^{2\pi i(n+\al)j /p}\,.
  \eql{id2}
  $$
  
I record the special case $\al=1/2$, (anti--periodic functions),
    $$\eqalign{
  e^{-z}\,\sum_{m=-\infty}^\infty (-1)^m\,I_{j+mp}(z)&=
  {1\over p} \sum_{n=0}^{p-1}e^{-2z\sin^2\big(\pi (2n+1)/2p\big)}\,
  e^{\pi i(2n+1)j /p}\cr
  &={1\over p} \sum_{n=0}^{p-1}e^{-2z\sin^2\big(\pi (2n+1)/2p\big)}\,
  \cos\big(\pi (2n+1)j /p\big)\,,
  }
  \eql{id4}
  $$
used later.
 
When $j=0$ see also the computations of Cojocaru, [\pref{Cojocaru}].

\section{\bf 5. Further identities}

Before discussing the above results, I outline some known things. The textbook identity,
  $$
  \prod_{m=0}^{p-1}\bigg(4\sin^2\big(\pi(m+\al)/p\big)+4\sinh^2\ga\bigg)=2\big(
  \cosh 2p\ga-\cos 2\pi\al\big)\,,
  \eql{ident3}
  $$
can be proved by elementary trigonometry, \eg\ Hobson, [\pref{Hobson2}], Loney,
[\pref{Loney}], \S368, \ Bromwich, [\pref{Bromwich}],p.211, and so can reasonably be
considered as given.

It can be understood as equating two expressions for the determinant of the operator
$-\nabla\De +4\sinh^2\ga$. The left--hand side is the product of the eigenvalues (the
definition) and the right--hand side results from a non--eigenvalue evaluation using, say, the
Gel'fand--Yaglom technique. This could be considered an alternative derivation to the
trigonometric one.

Taking logs and differentiating with respect to $\ga$ produces the known,
  $$\eqalign{
  {1\over4p}\sum_{m=0}^{p-1} {1\over \sin^2\big(\pi(m+\al)/p\big)+\sinh^2\ga}
  &={1\over2}{\sinh 2p\ga\over\sinh 2\ga}{1\over  \cosh 2p\ga-\cos 2\pi\al}\cr
  &={U_{p-1}(\cosh2\ga)\over 2\big(T_p(\cosh2\ga)-\cos2\pi \al\big)}\,,
  }
  \eql{ident4}
  $$
where $T$ and $U$ are Chebyshev polynomials of the first and second kind.

In [\pref{dowverlinde}] I gave a fragmentary history and some discussion of these
expressions in connection with Verlinde's formula for the dimensions of vector bundles on
moduli spaces. In particular it was noted that the values of the twisted \zf,
  $$
  \ze_{\oZ/p\oZ}(s;\al)\equiv \sum _{m=0}^{p-1} {1\over \big(4\sin^2
  \pi(m+\al)/p\big)^s}\,,
  \eql{twzeta}
  $$
on the discrete circle, at positive integers, \ie,
  $$
 2^{-2n} \sum_{m=0}^{p-1}\cosec^{2n}\big( \pi(m+\al)/p\big)\,,\quad \al>0\,,
 \eql{cosecsums}
  $$
can be obtained by expanding (\peq{ident4}) in powers of $4\sinh^2\ga$.This is just an
example of the Euler-Rayleigh technique. See also Dikii, [\pref{Dikii}].

The quantity $4\sinh^2\ga$ can be thought of as a mass squared, $m^2$, a Laplace
transform parameter, $s$, a fugacity, $z$, or as a spectral parameter, $\la$. One
interpretation of the left--hand side of (\peq{ident4}) is a resolvent which is the derivative
of logdet with respect to $m^2$.

Alternatively, setting $\ga=0$ in (\peq{ident3}) and taking logs reduces to  the ancient, and
directly derivable,  Kubert identity,
  $$
  \sum_{m=0}^{p-1}\log2\sin \pi(m+\al)/p=\log2\sin \pi\al\,,
  $$
which can be differentiated, this time with respect to $\al$, also to give the $\cosec$ sums
(\peq{cosecsums}). For general interest, I give two examples, \eg\ Fisher, [\pref{Fisher}],
  $$\eqalign{
   \ze_{\oZ/p\oZ}(1;\al)&={1\over4}\bigg(\cosec^2\pi\al/p+p^2\cosec^2\al-
   \cosec^2 \al/p\bigg)\cr
    \ze_{\oZ/p\oZ}(2;\al)&=\!{1\over16}\bigg(\cosec^4\pi\al/p+p^4\cosec^4\al-
   {2\over3}p^2(p^2-1)\,\cosec^2\al-   \cosec^4 \al/p\bigg).
   }
  $$
The first term is to be dropped if $\al=0$ as it corresponds to a zero mode and is not
included in the \zf.

Other relevant    formulae are the Fourier series,
  $$\eqalign{
   {\sin2\pi\al\over\cosh2\ga-\cos2\pi\al}
   &=2\sum_{n=1}^\infty e^{-2\ga n}\,\sin 2\pi n\al\cr
   {\sinh2\ga\over\cosh2\ga-\cos2\pi\al}
   &=1+2\sum_{n=1}^\infty e^{-2\ga n}\,\cos 2\pi n\al\,,\quad \ga>0\,,
   }
   \eql{fs}
  $$
again derivable by school trigonometry, \eg\ Loney, [\pref{Loney}] \S\S293--294.

After this recapitulation, I can now return to the results of the earlier sections. Karlsson and
Neuhauser, [\pref{KandN}], derive the identity, (\peq{ident4}), for $\al=0$, by taking the
Laplace transform of the heat--kernel equation, (\peq{id}), evaluated at $j=0$ (which
amounts to taking the trace\footnote{ The factor of $p$ is the volume}). The right--hand
side gives the resolvent, corresponding to the left--hand side of (\peq{ident4}). The
transform of the summand in the left--hand side of (\peq{id}) can be given as a standard
closed form and, on summation, effectively gives the right--hand side of
(\peq{ident4}).\footnote{ The relation between variables is that the $s$ of [\pref{KandN}]
equals $\sqrt2\sinh\ga$ here.}

It is no harder to show that the more general identity, (\peq{ident4}), follows on Laplace
transforming the heat--kernel relation, (\peq{id2}), at $j=0$. As used in [\pref{KandN}], in
a different notation, the tabulated Laplace transform,
  $$
  \int_0^\infty dz\, e^{-z}I_n(z)\,e^{-2z\sinh^2\ga}={1\over\sinh 2\ga}\,e^{-2\ga\,n}\,,\quad
  n\ge0\,,
  \eql{lt}
  $$
allows the summation over $m$ to be performed immediately, after this has been written
over non--negative integers using $I_{-n}(z)=I_n(z)$. Simple algebra then quickly delivers
(\peq{ident4}). Depending on one's starting point, this can viewed either as  a derivation or
as a confirmation of (\peq{ident4}).

Continuing with this theme, starting with the basic integral form of the modified Bessel,
 $$
   I_n(z)={1\over\pi}\int_0^\pi d\phi\, e^{z\cos\phi}\cos n\phi\,,
 $$
the Laplace transform, (\peq{lt}), can be {\it derived} purely trigonometrically on applying
the elementary Fourier series (\peq{fs}). Alternative proofs can be more complicated and
may involve contour manipulation, or series expansion. (Consult Gray and Mathews,
[\pref{GrandM}] Chap.VI and Ex.14,p.76.)

I finally note that the heat--kernel relation (\peq{id2}) can be more neatly written as in
(\peq{ident1}),
  $$
  \sum_{m=-\infty}^\infty e^{-2\pi im\al}\,I_{j+mp}(z)=
  {1\over p} \sum_{n=0}^{p-1}e^{z\cos\big(2\pi (n+\al)/p\big)}\,e^{2\pi i(n+\al)j /p}\,.
  \eql{id3}
  $$
Some consequences of this are derived in the next section.

\section{\bf 6. Bessel relations}

Taking the trace of (\peq{id3}) (\ie\ setting $j=0$) gives the Fourier cosine series,
  $$
  {1\over2}\,I_0(z)+\sum_{m=1}^\infty\cos 2\pi m\al\, I_{mp}(z)
  ={1\over2p}\sum_{n=0}^{p-1}e^{z\cos\big(2\pi (n+\al)/p\big)}\,.
  \eql{flc}
  $$
(See also the analysis of Cojocaru [\pref{Cojocaru}].)

For example, for $p=1$ and $p=2$, I find,
  $$\eqalign{
  {1\over2}\,I_0(z)+\sum_{m=1}^\infty\cos 2\pi m\al\, I_{m}(z)
  &={1\over2}\,e^{z\,\cos2\pi\al}\cr
  {1\over2}\,I_0(z)+\sum_{m=1}^\infty\cos 2\pi m\al\, I_{2m}(z)
  &={1\over2}\cosh (z\cos\pi\al)\,.
  }
  \eql{flc2}
  $$

Next, setting $\al=1/2$ in these produces the anti--periodic expressions,
  $$\eqalign{
  {1\over2}\,I_0(z)+\sum_{m=1}^\infty(-1)^m\, I_{m}(z)&={1\over2}\,e^{-z}\cr
  {1\over2}\,I_0(z)+\sum_{m=1}^\infty(-1)^m\, I_{2m}(z)&={1\over2}\,.
  }
  \eql{flc3}
  $$

Adding the first of this last set to the  $p=1$, $\al=0$ equation in (\peq{flc2}), gives the
$p=2$, $\al=0$ relation in (\peq{flc2}). Subtracting, on the other hand, yields,
  $$
  2\sum_{\mu=0}^\infty I_{2\mu+1}(z)=\sinh z\,.
  \eql{sodds}
  $$

These relations can be obtained from those in [\pref{ADG}], which are actually expressed in
terms of the $J_n$  Bessel function. They are listed results, [\pref{ADG}] providing the
references.

Furthermore, the first equation in (\peq{flc2}) is recognised as a real form of the
Jacobi--Anger expansion which, as is well known, yields a generating function for $I_m$ if
one puts $t=e^{2\pi i\al}$ when it turns into the basic,
  $$
  \sum_{m=-\infty}^\infty I_m(z)\, t^m=e^{z(t+t^{-1})/2}\,.
  $$

The other interesting value is $\al=1/4$ for which (\peq{flc2}) becomes,\footnote{ The
value $\al=1/4$ is a turning point between the `boson' value, $\al=0$, and the `fermion',
$\al=1/2.$}
  $$\eqalign{
 1&= I_0(z)+2\sum_{m=1}^\infty (-1)^m \,I_{2m}(z)\cr
  \cosh (z/\sqrt2)&=I_0(z)+2\sum_{m=1}^\infty (-1)^m\, I_{4m}(z)\,.
  }
  \eql{flc4}
  $$
The first of these agrees with the second of (\peq{flc3}).

For comparison, I present an elementary derivation of the first equation in (\peq{flc4}).
First I note by simple cancellation and recursion  that, (all arguments are $z$),
  $$\eqalign{
  I_1&=\sum_{m=0}^\infty (-1)^{m+1}I_{2m+1}
  +\sum_{m=1}^\infty(-1)^m I_{2m+1}\cr
  &=\sum_{m=1}^\infty (-1)^{m}\big(I_{2m-1}+ I_{2m+1}\big)\cr
  &=\sum_{m=1}^\infty (-1)^{m}I'_{2m}\,.
  }
  $$
Then, using $I_1=I'_0/2$, integration yields the required relation after setting $z=0$ to fix
the unknown constant.

Finally in this section, I give two further examples of (\peq{flc}), this time for $p=3$,
  $$\eqalign{
    {1\over2}I_0(z)+\sum_{m=1}^\infty (-1)^m\,I_{3m}(z)&=
    {1\over6}\big(e^{z/2}+2e^{-z/4}\cosh 3z/4\big)\cr
    {1\over2}I_0(z)+\sum_{m=1}^\infty (-1)^m\,I_{6m}(z)&=
    {1\over6}\big(1+2\cosh \sqrt3z/2\big)\,.
   }
  $$
For other uses of similar Bessel relations see [\pref{CoandB}].
  
\section{\bf 7. Heat--Kernel on the discrete Dirichlet interval}

It is straightforward to find the Dirichlet (D) heat--kernel on the discrete interval,or path,
$\caP$, defined to be a set of $p+1$ vertices, labelled by $j\in\oZ$. The end points, $j=0$
and $j=p$, are boundary vertices at which the function, and hence the heat--kernel,
vanishes. The remaining, internal vertices, $j=1,\ldots,p-1$, are the dynamical ones. Then
$\caP$ is a $(p-1)$-path.

Translation invariance is lost and one has to consider $K^D_\caP(j,j';t)$ which, just as in the
continuum case, can be constructed from $K_\oZ(j,j';t)$ by images. The positioning of the
continuum images is classic and given \eg\ in Thomson, [\pref{Thomson}], and has been
used frequently in heat conduction and hydrodynamical problems. (See Hicks,
[\pref{Hicks}], Carslaw and Jaeger, [\pref{CandJ}], Basset, [\pref{Basset}] \S57.)  In the
discrete situation, the mirrors are also placed at $j=0$ and $j=p$, enclosing the $p-1$
dynamical, free vertices and it is easy to see that the discrete Dirichlet conditions hold at the
end points.

This construction results in the image summation,
  $$\eqalign{
  K^D_\caP(j,j';t)&=\sum_{m=-\infty}^\infty
  \big(K_\oZ(j,j'-2mp;t)-K_\oZ(j,-j'-2mp;t)\big)\cr
  &=e^{-2t}\sum_{m=-\infty}^\infty\big(I_{j-j'+2mp}(2t)-I_{j+j'+2mp}(2t)\big)\,.
  }
  \eql{dimage}
  $$

Dual to this image form is  the eigenfunction expression. The eigenvalues and modes here
are genuinely historic, being,
  $$
  \la_n=4\sin^2 \!{\pi n\over 2p}\,,\quad n=1,\ldots,p-1\,,
  \eql{Deigen}
  $$
and
   $$
  y_n(j)=\sqrt{2\over p}\sin {j\pi n\over p}\,,
  $$
respectively, \eg\ Fort, [\pref{Fort}], so that the heat--kernel is,
  $$\eqalign{
  K^D_\caP(j,j';t)&={2\over p}\sum_{n=1}^{p-1}e^{-4t\sin^2 (\pi n/ 2p)}
  \sin {n\pi j\over p}\,\sin {n\pi j'\over p}    \cr
  &={1\over p}\sum_{n=1}^{p-1}e^{-4t\sin^2 (\pi n/ 2p)}
  \bigg(\cos {n\pi(j-j')\over p}-\cos {n\pi(j+j')\over p}\bigg)\,,
  }
  \eql{dhk}
  $$
which can be seen to be equivalent to the image form, (\peq{dimage}), by virtue of the
identity, (\peq{idr}), given earlier, after setting $p\to2p$. I give a few details, as they are
not quite obvious. I first note that the $n=0$ term in (\peq{idr}) cancels by the subtraction
in (\peq{dhk}). Then I  remark on the identity,
  $$\eqalign{
  &\sum_{n=1}^{p-1}\cos (\pi nj/p)\big(e^{-4t\sin^2(\pi n/2p)}
  -(-1)^je^{-4t\cos^2(\pi n/2p)}\big)\cr
  &=e^{-2t}\sum_{n=1}^{p-1}\cos (\pi nj/p)\bigg(e^{2t\cos \pi n/p}
  -(-1)^je^{-2t\cos \pi n/p}\bigg)=0\,.
  }
  $$
which comes into play when the sum $\sum_{n=1}^{2p-1}$ in (\peq{idr}) is split into one
from $n=1$ to $p-1$  plus one from $p$ to $2p-1$ and shows that the two sums are equal
so that (\peq{idr}) becomes,
 $$\eqalign{
  e^{-2t}\,\sum_{m=-\infty}^\infty I_{j+2mp}(2t)&={1\over2p}+
  {1\over p} \sum_{n=1}^{p-1}e^{-4\sin^2(\pi n/p)\,t}\,\cos{2\pi nj /p}\,.\cr
  }
  \eql{idr2}
  $$
The required equivalence of (\peq{dhk}) and (\peq{dimage}) then follows easily.

In the continuum case, this equivalence is effectively Poisson summation, as has been
known for about 150 years, in various guises, and the identity (\peq{id}) is a discrete
analogue, [\pref{CandY2}],  [\pref{KandN}].

Taking the trace, \ie setting $j=j'$ and summing over $j$ from $1$ to $p-1$, yields the
image form,
  $$\eqalign{
  \sum_{j=1}^{p-1} K^D_\caP(j,j;t)
  &=e^{-2t}(p-1)\sum_{m=-\infty}^\infty I_{2mp}(2t)-e^{-2t}
 \sum_{m=-\infty}^\infty\sum_{j=1}^{p-1} I_{2j+2mp}(2t)\cr
 &=e^{-2t}p\sum_{m=-\infty}^\infty I_{2mp}(2t)-e^{-2t}
 \sum_{m=-\infty}^\infty\sum_{j=0}^{p-1} I_{2j+2mp}(2t)\cr
&=e^{-2t}p\sum_{m=-\infty}^\infty I_{2mp}(2t)-e^{-2t}
 \sum_{m=-\infty}^\infty I_{2m}(2t)\cr
&=e^{-2t}p\sum_{m=-\infty}^\infty I_{2mp}(2t)-e^{-2t}\cosh2t
  }
  \eql{trace}
  $$
where I have used the listed (\peq{flc2}), for $\al=0$ and have split all the integers into
residue classes mod p, $N=j+mp$, $m\in\oZ$ and $0\le j<p$.

Equating this to the trace obtained from the eigenfunction form yields an identity that
contains nothing new over the periodic case, as the equivalence, proved just above, shows.

\section{\bf 8. Heat--Kernel on the discrete Neumann interval}

Again I approach the construction from the image side and now offset the mirrors by $1/2$
to the {\it half}--integer points $1/2$ and $p+1/2$  so as, this time, to enclose a $p$--path,
$\caP$ ($j=1$ to $j=p$) of $p$ free vertices. The end points, $j=0$ and $j=p+1$, which are
sometimes referred to as fictional, or ghost, points, are now to be regarded as images and
tied to the free vertices at $j=1$ and $j=p$, respectively, thus reproducing the standard
discrete Neumann conditions. In the continuum limit the $1/2$ offset becomes irrelevant.

It is easily concluded that the even images of the point $j$ are at $j-2mp$ and the odd ones
at $-j-2mp+1$ ($m=-\infty \to\infty$), so as to yield the image form,
  $$\eqalign{
  K^N_\caP(j,j';t)
  &=\sum_{m=-\infty}^\infty \big(K_\oZ(j,j'-2mp;t)+K_\oZ(j,-j'+1-2mp;t)\big)\cr
  &=\sum_{m=-\infty}^\infty \big(K_\oZ(j-j'+2mp;t)+K_\oZ(j+j'-1+2mp;t)\big)\,,
  }
  \eql{nimage3}
  $$
both sets of images coming in with the same sign.

I now look at the eigenfunction expression. The eigenvalues are as in (\peq{Deigen})
except that the mode label extends to $n=0$, corresponding to a zero mode.

The $N$--eigenfunctions on the $p$--path are again standard,
  $$\eqalign{
  y_n^N(j)&=\sqrt{2\over p}\cos {n\pi(2j-1)\over2p}\,,\quad n=1,\ldots p-1\cr
  &={1\over\sqrt p}\,,\quad n=0\,,
  }
  \eql{neigenf}
  $$
and the free, off--diagonal heat--kernel can be written,
  $$\eqalign{
  K^N_\caP(j,j';t)&={1\over p}+{2\over p}\sum_{n=1}^{p-1}e^{-4t\sin^2 (\pi n/ 2p)}
  \cos {n\pi(2j-1)\over2p}\,\cos {n\pi(2j'-1)\over2p}    \cr
  &={1\over p}+{1\over p}\sum_{n=1}^{p-1}e^{-4t\sin^2 (\pi n/ 2p)}
  \bigg(\cos {n\pi(j+j'-1)\over p}+\cos {n\pi(j-j')\over  p}\bigg)\,.
  }
  \eql{nhk}
  $$

To transform this into an image form one proceeds as in the last section by using
(\peq{idr2}) to obtain,
  $$\eqalign{
  K^N_\caP&(j,j';t)=e^{-2t}\sum_{m=-\infty}^\infty \big(I_{j-j'+2mp}(2t)+
  I_{j+j'-1+2mp}(2t)\big)\cr
  &=\sum_{m=-\infty}^\infty \big(K_\oZ(j-j'+2mp;t)+K_\oZ(j+j'-1+2mp;t)\big)\,,\cr
  }
  \eql{nimage2}
  $$
in agreement with (\peq{nimage3}).

The trace can be constructed, but, again, yields nothing new and so I will not write it out.

\section{\bf 9. The D--N interval}

Having treated the D--D and the N--N intervals, I now complete the set by tackling the
hybrid, D--N case, although nothing especially new is expected to emerge.

I choose $j=0$ as the D end point, \ie\ $y(0)=0$, and set the mirrors at $j=0$ and at
$j=p+1/2$ enclosing $p$ free vertices, as in the N--N case. The images are located at
$j-2m(2p+1)$ for even reflections and at the opposite values, $-j-2m(2p+1)$, for odd,
($-\infty \le m\le \infty$). The even terms enter with the sign, $(-1)^m$, while the odd
ones have $(-1)^{m+1}$. Hence the heat--kernel is,
  $$\eqalign{
  K^{ND}_\caP(j,j';t)
  &=\!\!\sum_{m=-\infty}^\infty (-1)^m\bigg(K_\oZ(j,j'-m(2p+1);t)\!-
  \!K_\oZ(j,-j'-m(2p+1);t)\!\bigg)\cr
  &=\!\!\!\sum_{m=-\infty}^\infty \!\!(-1)^m\! \bigg(K_\oZ(j-j'\!+m(2p+1);t)\!-
  \!K_\oZ(j+j'\!+m(2p+1);t)\!\!\bigg).\cr
  &=e^{-2t}\!\!\!\sum_{m=-\infty}^\infty \!\!(-1)^m\!\bigg(I_{j-j'\!+m(2p+1)}(2t)-
  I_{j+j'\!+m(2p+1)}(2t)\bigg)\,.
  }
  \eql{dnimage}
  $$

The transformation of this into the eigenfunction form provides a use for the particular
twisted identity, (\peq{id4}), corresponding to anti--periodic functions. I repeat it here, with
$p\to2p+1$,
 $$\eqalign{
  e^{-z}\sum_{m=-\infty}^\infty (-1)^m\,I_{j+m(2p+1)}(z)=
  {1\over 2p+1} \sum_{n=0}^{2p}&e^{-2z\sin^2\big(\pi (2n+1)/2(2p+1)\big)}\cr
  &\hspace{******}\times\cos{\pi(2 n+1)j \over(2p+1)}\,.
  }
  \eql{id5}
  $$
  
Substitution into (\peq{dnimage}) produces, at first,
 
 $$\eqalign{
  K^{ND}_\caP(j,j';t)
  ={1\over 2p+1} \bigg[\sum_{n=0}^{p-1}&+\sum_{n=p+1}^{2p}\bigg]
  e^{-4t\sin^2\big(\pi (2n+1)/2(2p+1)\big)}\cr
  &\times\bigg(\cos{\pi (2n+1)(j-j') \over2p+1}
  -\cos{\pi(2n+1)(j+j') \over2p+1}\bigg)\,,
  }
  \eql{dn}
  $$
after noting that the $n=p$ summand is zero. As earlier, the two sums are equal and so,
   $$\eqalign{
  K^{ND}_\caP(j,j';t)
  ={4\over 2p+1} \sum_{n=0}^{p-1}&
  e^{-4t\sin^2\big(\pi (2n+1)/2(2p+1)\big)}\cr
  &\times\sin{\pi (2n+1)j \over2p+1}\,\sin{\pi(2n+1)j' \over2p+1}\,,
  }
  \eql{dn2}
  $$
from which the eigenvalues and eigenfunctions can be read off, in a standard way, as,
  $$
    \la^{DN}_n=4\sin^2{\pi (2n+1)\over2(2p+1)}\,,\quad n=0,\ldots p-1
  $$
and
   $$
     y_n^{DN}(j)={2\over\sqrt{2p+1}}\sin{\pi (2n+1)j \over2p+1}\,.
   $$
These agree with available expressions \eg\ [\pref{DuandD}], [\pref{Thorn}].

\section{\bf 10. Lattice Green functions}

In the continuum, the Laplace transform of the heat--kernel is the Green function, and this
holds in the discrete case too.

The transform of the right--hand side of (\peq{id2}) (times $1/2$) is the Green function for
the $p$--cycle, $\caC$,
  $$\eqalign{
    G^\caC_p(j,j',\al,\ga)&={1\over 4p}\sum_{n=0}^{p-1}{e^{2\pi i(n+\al)(j-j')/p}\over
    \sin^2\big(\pi(n+\al)/p\big)+\sinh^2\ga}\cr
    &=G_p(j-j',0,\al,\ga)\cr
    &\equiv G_p(j-j',\al,\ga)\,.
    }
    \eql{ms}
  $$
  
In order to apply the transform to the left--hand side of (\peq{id2}) the sum has to be
arranged so that the order of the Bessel function is always non--negative \ie $m$ is such
that $j+mp\ge0$. Those terms for which $j+mp$ is negative can be converted to positive
$j+mp$ using $I_{-j-mp}=I_{j+mp}$.

It is best to use residues mod $p$ and set $j=Jp+r$ with $J\in\oZ$ and $0\le r<p$. The
positive terms then correspond to $m\ge -J+1$ and the negative ones to $m\le -J-1$. The
zero term, $m=-J$, gives an $I_r$ and the sum therefore can be rearranged to,
  $$
\sum_{m=-J+1}^\infty \big(   e^{-2\pi i m\al}\,I_{Jp+r+mp}
+e^{2\pi i m\al+4\pi i J\al}\,I_{Jp-r+mp}\big)+e^{2\pi iJ\al}\,I_r(z)\,,
  $$
allowing the Laplace transform to be taken giving,
  $$
{1\over2\sinh2\ga}\!\!\sum_{m=-J+1}^\infty\! e^{-2\ga mp}\bigg(\!
e^{-2\ga Jp}\big( e^{-2\ga r}  e^{-2\pi i m\al}
+e^{2\ga r}\,e^{2\pi i m\al+4\pi i J\al}\big)+e^{2\pi iJ\al}\,e^{-2\ga r}\!\!\bigg)\,.
\eql{lt2}
  $$
  
The $m$-summation can be done as before and, taking the real part, yields the identity,
  $$\eqalign{
  {1\over 4p}\sum_{n=0}^{p-1}&{\cos{2\pi (n+\al)j/p}\over
    \sin^2\big(\pi(n+\al)/p\big)+\sinh^2\ga}
  = {\cos{2\pi\al\! J}\over2\sinh2\ga}\bigg(
   {\cosh2\ga r\sinh2p\ga\over\cosh2p\ga
   -\cos2\pi\al}-\sinh 2\ga\,r\bigg)\cr
   &\hspace{***************************}-{\sin2\pi\al J\over\sinh2\ga}
   {\sin2\pi\al \sinh2\ga r\over\cosh2p\ga-\cos2\pi\al}\cr
   &\hspace{*****}={\cos(2\pi\al J)\,U_{p-1-r}(\cosh2\ga)
   +\cos\big(2\pi\al(J+1)\big)\,U_{r-1}(\cosh2\ga)\over
   2\big(T_p(\cosh2\ga)-\cos2\pi\al\big)}
   }
   \eql{lgf}
  $$
again in terms of the Chebyshev polynomials, which are the basic building blocks for the
Green functions of the free equation.
  
As a check, when $j=0$ ($J=0=r$) equating this to the real part of (\peq{ms}) gives the
identity, (\peq{ident4}). Also, when $\al=0$, it is periodic in $J$ of period 1 (\ie\ in $j$  of
period $p$) as it should be. Actually it is then independent of $J$ the expression
being\footnote{ This is the standard result found in Feller, [\pref{Feller}]. See Montroll and
Weiss, [\pref{MandW}], equn.II.19.}
  $$\eqalign{
   \Real G^\caC_p(Jp+r,0,\ga)&={\cosh(2r-p)\ga\over2\sinh2\ga\sinh p\ga}\cr
   &={U_{p-r-1}(\cosh2\ga)+U_{r-1}(\cosh2\ga)
   \over 2\big(T_p(\cosh2\ga)-1\big)}\,.
   }
   \eql{reg}
  $$
  
I also write out the anti--periodic Green function, ($\al=1/2$),
   $$\eqalign{
   \Real G^\caC_p(Jp+r,1/2,\ga)&=(-1)^J\,{\sinh(p-2r)\ga\over2\sinh2\ga\cosh p\ga}\cr
   &=(-1)^J{U_{p-r-1}(\cosh2\ga)-U_{r-1}(\cosh2\ga)
   \over 2\big(T_p(\cosh2\ga)+1\big)}\,.
   }
   \eql{reg2}
  $$
  
When $\al\ne0$, the twisted periodicity appears in its SO(2) aspect and, as a by--product,
yields the imaginary part of (\peq{ms}) which can be confimed by performing the
summation, (\peq{lt2}). I will not expose the easily found expressions.

This lattice Green function has also been evaluated using Laplace transforms by Cojocaru,
[\pref{Cojocaru}], equn.(20).

The formula (\peq{lgf}) generalises that of Bendito, Encinas and Carmona, [\pref{BEC}],
who give the untwisted (\peq{reg}).

The interval Green functions, which are real, can be obtained by combining the cycle
expressions, (\peq{reg}) and (\peq{reg2}), according to the image forms, (\peq{dimage}),
(\peq{nimage2}) and (\peq{dnimage}).

The Dirichlet Green function for the $p$--path, $\caP$, is found to be, after some simple
trigonometric rearrangement,\footnote{ On the interval $j=r$, $j'=r'$. As a tactical point, I
find it easier to use trigonometric formulae than relations between Chebyshev polynomials.}
  $$
   G^D_p(r,r',\ga)={U_{p-r}(\cosh2\ga)\,U_{r'-1}(\cosh2\ga)\over U_p(\cosh2\ga)}
   \eql{regd}
  $$
where $r\ge r'$. If $r<r'$, the symmetry of the Green function in $r$ and $r'$ can be
employed. This result agrees with that of Chung and Yau, [\pref{CandY2}], derived in a
more involved way, and that of Bendito, Encinas and Carmona, [\pref{BEC}], obtained via a
Sturm--Liouville approach which illuminates the structure of (\peq{regd}) and avoids
mention of the eigenproblem.

The Neumann Green function is found to be, again after simple trigonometric manipulation,
  $$\eqalign{
   G^N_p(r,r',\ga)&={V_{p-r}(\cosh2\ga)\,V_{r'-1}(\cosh2\ga)
   \over2\,(\cosh2\ga-1)\, U_{p-1}(\cosh2\ga)}\,,\quad r\ge r'\,,
   }
   \eql{reg3}
  $$
which agrees with [\pref{BEC}]. The third kind Chebyshev polynomial, $V$, arises through
the Neumann offset $r\to r-1/2$, $r'\to r'-1/2$.

Finally I find the N--D Green function to be,
  $$
   G^{ND}_p(r,r',\ga)={V_{p-r}(\cosh2\ga)\,U_{r'-1}(\cosh2\ga)
   \over V_{p}(\cosh2\ga)}\,,\quad r\ge r'\,.
   \eql{reg4}
  $$
To attain this form, one sees from (\peq{dnimage}) that  it is the {\it anti}--periodic cycle
Green function, (\peq{reg2}), that is required here.
  
Bass, [\pref{Bass}], gives the Green functions for all three interval types, but has a more
complicated structure. He obtains the form for the Dirichlet and Neumann cases in terms of
the periodic one by algebra.

\section{\bf 11. Graph considerations}

The stepping operator $E$ is the coordinate (or vertex)  space representation of the abstract
stepping operator ${\bf E}$, and  ${\bf E+E^{-1}}$ is the adjacency operator, ${\bf A}$,
having the usual adjacency matrix representation on the integer lattice, $\oZ$, coordinates.

One sees from the definition (\peq{hk}) that $\ol {\bf K}(t)$ is a generating function for the
powers of the adjacency operator,
  $$
    \ol {\bf K}_\oZ(t)=e^{{\bf A}_{\!\!Z }t}\,,
  $$
 with
   $$
    \ol K_\oZ(j,j';t)=\me j{{\bf K}_\oZ(t)}{j'}\,.
   $$

A similar relation holds on the $p$--cycle projection, $\caC_p=\oZ/p\oZ$. According to
(\peq{socp}),
  $$
    \ol {\bf K}_\caC(t)=e^{{\bf A}_{\caC }t}\,,
    \eql{cyck}
  $$
which are operators in the $p$--dimensional vector space describing $\caC_p$. ${\bf
A}_{\caC }$ is represented by the usual periodic adjacency matrix.

The trace of the expansion of (\peq{cyck}) yields,
  $$
   \tr \ol {\bf K}_\caC(t)=\sum_{k=0}^\infty{t^k\over k!}\,\tr {\bf A}_{\caC }^k\,,
   \eql{trkc}
  $$
and, from the definition of adjacency matrix, $\tr{\bf A}^k_\caC$ is the number of closed
paths (circuits) of length $k$ on the $p$--cycle, which equals, by homogeneity (or
transitivity) $p$ times the number of circuits starting and finishing on any one vertex.
Denoting this latter number by $g_k$, one can derive identities from the various forms of
the left--hand side of (\peq{trkc}) which allow the computation of $g_k$, or, equivalently,
of the series generating function defined by,
   $$
     g(\si)=\sum_{k=0}^\infty g_k\,\si^k\,.
   $$
   
The Laplace transform of (\peq{trkc}) produces,
  $$
  g(\si)={1\over\si}\int_0^\infty dt\, e^{-t/\si}\,\tr \ol {\bf K}_\caC(t)\,,
  $$
which is the trace of the cycle lattice Green function, already computed, in (\peq{ident4}) or
(\peq{reg}), the relation between variables being $2\cosh2\ga=1/\si\ge2$. Hence one has
the explicit rational formula for the generating function,\mgn{check 2s}
  $$
   g(\si)={1\over2\si}{\sinh2p\ga\over\sinh2\ga}{1\over \cosh 2p\ga-1}
   ={1\over2\si}{U_{p-1}({1/2\si})\over\big(T_p({1/2\si})-1\big)}\,.
   \eql{genfun}
  $$

A more general quantity in graph theory is the number of circuits of length $k$ with $l$
bumps (or immediate backtrackings), attached to a vertex. Denoting this number by
$f_{kl}$, the corresponding generating function is,
  $$
   f(u,\si)=\sum_{k,l=0}^\infty f_{l\,k}\,u^l\,\si^k\,,
  $$
so that, as a special case, $g(\si)=f(1,\si)$ and for the number of circuits with no
backtrackings, $f(\si)\equiv f(0,\si)$.

If the graph is $d$--regular (for the cycle $d=2$) there is the relation, [\pref{Bartholdi}],
  $$
   {f(1-u,\si)\over1-u^2\si^2}=g\bigg({\si\over1+u(d-u)\si^2)}\bigg)\,,
   \eql{Barth}
  $$
which allows the more interesting $f(u,\si)$ to be obtained from the simpler $g(\si)$.
Trigonometric expansion  (\eg\ Loney, [\pref{Loney}] \S\S 293--294, can be used to
evaluate this but it is simpler to employ Chebyshev polynomials as rapid computation is
available through,
  $$\eqalign{
   U_n(x)=\tr \big({ C}^n\,(x)\,{Q}\big)\,,\quad
   T_n(x)={1\over2}\,\tr {C}^n(x)\,,\quad V_n(x)=
   \tr \big({C}^n\,(x)\,{R}\big)\,,
   }
  $$
where ${C}$, ${Q}$, ${R}$ and ${S}$ are given by,\footnote{ I introduce ${S}$ because
the denominator of the Neumann Green function, (\peq{reg3}), can be written
$U_{p}+U_{p-2}-2U_{p-1}=\tr ({C}^p\,{S})$.}
  $$
   {C}(x)=\left(\matrix{0&1\cr-1&2x}\right)\,,\quad {Q}=
   \left(\matrix{0&0\cr0&1}\right)\,,\quad {R}=
   \left(\matrix{0&0\cr-1&1}\right)\,,
   \quad {S}=
   \left(\matrix{-1&1\cr-1&1}\right).
  $$

Hand evaluation quickly produces the first two examples,
  $$
   f_1(u,\si)={1+(1-u)\si\over1-(1+u)\si}\,,\quad f_2(u,\si)=
   {1+(1-u^2)\si^2\over1-(1+u)^2\si^2}\,,
  $$
where the subscript refers to the number of vertices, \ie $p$.

For larger $p$, it is sufficient to give the $g_p$ from which the $f_p(u,\si)$ can readily be
found on applying (\peq{Barth}). I list the results for $p=1$ to $p=12$,
  $$\eqalign{
&-{\frac{1}{2\,\sigma-1}},-\frac{1}{4\,{\sigma}^{2}-1},\frac{\sigma-1}
{2\,{\sigma}^{2}+\sigma-1},\frac{2\,{\sigma}^{2}-1}{4\,{\sigma}^{2}-1},
-\frac{{\sigma}^{2}+\sigma-1}{2\,{\sigma}^{3}-3\,{\sigma}^{2}-\sigma+1},\cr
& -\frac{3\,{\sigma}^{2}-1}{4\,{\sigma}^{4}-5\,{\sigma}^{2}+1},
\frac{{\sigma}^{3}-2\,{\sigma}^{2}-\sigma+1}{2\,{\sigma}^{4}+3
\,{\sigma}^{3}-4\,{\sigma}^{2}-\sigma+1},\frac{2\,{\sigma}^{4}-
4\,{\sigma}^{2}+1}{8\,{\sigma}^{4}-6\,{\sigma}^{2}+1},\cr
&-\frac{{\sigma}^{4}+2\,{\sigma}^{3}-3\,{\sigma}^{2}-\sigma+1}
{2\,{\sigma}^{5}-5\,{\sigma}^{4}-4\,{\sigma}^{3}+5\,{\sigma}^{2}
+\sigma-1},-\frac{5\,{\sigma}^{4}-5\,{\sigma}^{2}+1}{4\,{\sigma}^{6}
-13\,{\sigma}^{4}+7\,{\sigma}^{2}-1}\,.\cr
  }
  \eql{genfuns}
  $$

The ensuing $f_p(u,\si)$ agree with Bartholdi, [\pref{Bartholdi}], (who uses more involved
algebra) except that his even expressions appear to have misprints.

Taylor expansion allows the coefficients $f_{lk}$ to be read off and the correctness of
$f_4(u,\si)$, say, can be checked. For interest, I write it out,
  $$\eqalign{
   f_4(u,\si)=
   1&+2\,u\,{\sigma}^{2}+2\left( {u}^{3}+{u}^{2}+u+1\right)
   \,{\sigma}^{4}\cr
   &+2\left( {u}^{5}+2\,{u}^{4}+4\,{u}^{3}+6
   \,{u}^{2}+6\,u\right) \,{\sigma}^{6}+\ldots\,.\cr
   }
     $$
     
It is possible to confirm numerically the determinantal identity (\peq{ident3}) by computing
the Laplacian determinant using the explicit forms of the generating functions
(\peq{genfuns}). This is no more than a check of algebra of course, but it makes a
comforting exercise.

The Laplacian graph eigenvalues, denoted $\la_n$, ($n=1\to p$), are related to the
adjacency eigenvalues, $\mu_n$ by $\la_n=2-\mu_n$, {\it in this cycle case}. The first
$\la_n$, \ie\ $\la_1$,  is zero and so $\mu_1=2$. The log Laplacian determinant, defined
omitting the zero eigenvalue, is, therefore,
  $$\eqalign{
   \log\det'_{\!L}(p)&=\sum_{n=2}^p\log\la_n=\sum_{n=2}^p\log(2-\mu_n)\cr
   &=(p-1)\log2-\sum_{k=1}^{\infty}\sum_{n=2}^p{1\over k\,2^k}\,\mu_n^k\cr
   &=(p-1)\log2-\sum_{k=1}^{\infty}{1\over k\,2^k}\,(\tr{\bf A}_\caC^k-2^k)\cr
   &=(p-1)\log2-\sum_{k=1}^{\infty}{1\over k\,2^k}\,(p\,g_k-2^k)\,.\cr
   }
   \eql{logdet}
  $$
  
Now define the subtracted generating function, $g^*$, by
  $$
   p\,g^*_p(\si)=p\,g_p(\si)-{1\over1-2\si}
  $$
which is finite at $\si=1/2$, then (\peq{logdet}) becomes,
  $$\eqalign{
   \det'_{\!L}(p)&=2^{p-1}\exp\bigg(-\int_0^{1/2}{d\si\over\si}\,
    \big(p\,g^*_p(\si)-p+1\big)\bigg)\cr
   }
   \eql{logdet2}
  $$
which can be evaluated and gives the right answer\footnote{ According to one version of
Kirchhoff's matrix-tree theorem, it gives the number of spanning trees, $\det'_{\!L}(p)/p$,
correctly as $p$ for the cycle.}, $\det'_{\!L}(p)=p^2$, case by case, using
(\peq{genfuns}). The integration is the reverse of the step taking (\peq{ident3}) into
(\peq{ident4}) for arbitrary $p$.

A further check of the expressions, if one were needed, is provided by computing the graph
adjacency matrix from the expansion of the Green function in powers of
$\rho\equiv\si/(1-2\si)$. The coefficient of $-\rho^2 $ is the combinatorial Laplacian matrix,
$L$, from which the adjacency matrix can be found. The definition of Laplacian I use is the
operator relation, ${\bf G}=\rho/({\bf I}+\rho{\bf L})$.

I do not give the details since the answers are as expected. However, for the $p$--path, I
do write out the Laplacian matrices for the D--D, N--N and D--N  boundary conditions
computed from (\peq{regd}), (\peq{reg3}) and (\peq{reg4}), using $p=3$ as an
illustration,
  $$
    L_D=\left(\matrix {2&-1&0\cr
                                -1&2&-1\cr
                                0&-1&2}\right)\!,
      L_{N}=\left(\matrix {1&-1&0\cr
                                -1&2&-1\cr
                                0&-1&1\cr
                                }\right)\!,
        L_{DN}=\left(\matrix {2&-1&0\cr
                                -1&2&-1\cr
                                0&-1&1\cr
                                }\right).
   \eql{laplm}
  $$
The adjacency matrix is obtained by replacing the diagonal elements by zeros.

The Neumann diagonal elements correspond directly to the degrees  of the free vertices
whereas the first and last Dirichlet elements include the effects of the two fixed end points,
which are not part of the free graph and serve only to make $L_D$ non--singular.\footnote{
In the terminology of Chung and Yau, [\pref{CandY}], the graph is an induced subgraph of a
{\it host} graph, which includes the end, or boundary, points. This can be seen from the
form of the matrices in (\peq{laplm}). $L_D$ is obtained from $L_N$, in two higher
dimensions, by striking out the rows and columns that refer to the end points. } This is
confirmed by the D--N matrix.
  
A combinatorial expression for the number of closed circuits, $g_k$, on the $p$--cycle also
follows directly from the operator definition,
  $$\eqalign{
   \,g_k&={1\over p}\tr {\bf A}_\caC^k={1\over p}\tr({\bf E}
   +{\bf E}^{-1})^k\,,\quad {\bf E}^p={\bf I}\cr
   &=\sum_{{m=-[k/p]}\atop k-mp\,{\rm even}}^{[k/p]}\comb{k}{(k-mp)/2}\,.
   }
  $$

\section{\bf 12. The thermodynamical limit}

As mentioned before, the covering space (an infinite lattice) corresponds to the
thermodynamic limit of the covered (a finite lattice) (\eg\ [\pref{DuandD}]). For example,
very easily, the Laplace transform of the heat--kernel on $\oZ$, (\peq{ZK}), gives the
thermodynamic limit of the cycle free Green function on $\oZ/p\oZ$ directly as,
  $$
    G(j,j',\ga)={e^{-2\ga(j-j')}\over2\sinh2\ga}\,,\quad j\ge j'.
  $$
  
The parameter, $\ga$, can be extended into the complex plane so long as it retains a
positive real part (assuming $j\ge j')$ and then one can compare with the formulae in
Katsura and Inawashiro, [\pref{KandI}], Appendix, (although I have problems with equation
(A.4)).

\section{\bf 13. Remarks}

Most of the preceding expressions and relations are rather elementary examples of more
general structures that have been investigated for many years, in many different areas, as
mentioned in the introduction. For example, (bulk) lattice Green functions, in all dimensions,
have been {\it rewritten} in terms of Bessel functions both for numerical and formal
reasons, an early reference being van der Pol and Bremmer, [\pref{PandB}], p.368, in
connection with electrical networks. Later works include [\pref{KIA}] and [\pref{KandI}]
where Mellin--Barnes type integrals for Bessel products are utilised to give more explicit
representations of the lattice Green functions, a topic of some considerable activity and
manipulative complexity. See the reviews, Guttmann, [\pref{Guttman}], and Zucker,
[\pref{Zucker}].

Incidentally, van der Pol and Bremmer also take the infinite square lattice expression to an
elliptic integral, see [\pref{PandB}], equn.(91). A number of forms for this quantity have
been obtained over the years, the earliest being, perhaps, by McCrea and Whipple,
[\pref{MandW2}], for a random walk problem. More mathematical are the computations by
St\"ohr, [\pref{Stohr}], who touches on the  resistor network realisation which itself has
also seen a goodly amount of development. Flanders, [\pref{Flanders}], lists some square
expressions.

\section{\bf 14. Potentials, higher dimensions and continuum limit}

I wish to mention possible extensions to the preceding calculations which I will not be
carrying out here.

The analysis in this work has all been concerned with free propagation. An obvious
development would be to consider the case with a potential described by the recurrence,
  $$
  \bigg(-\nabla\Delta +V(j)+{\pa\over\pa t}\bigg)K_{\oZ}(j,j';t)=\de(t)\,,
  \eql{heqn2}
  $$
instead of (\peq{heqn}). In order to project down to $\oZ/p\oZ$, for example, $V(j)$ would
have to be periodic, $V(j+p)=V(j)$. Working on the interval, $\caP$, would also be possible
directly and related to a Sturm--Liouville problem.

Unless exact solutions were available, one might be interested in the discrete analogue of
the short--time behaviour of the heat--kernel, and the resulting expansion coefficients. One
possible structure has been analysed by Iliev, [\pref{Iliev}].

I have also restricted myself to one dimension.  Higher (free) dimensions can be assembled
in the usual way by taking the cartesian products of  one--dimensional structures, for
example a torus from the circle, a rectangle from the interval and and cubic lattices from
the integers. Some basic notions are outlined in [\pref{CandY2}] and a more extensive
discussion of tori in Chinta, Jorgenson and Karlsson, [\pref{CJK}], [\pref{CJK2}] (see also
Chaumard, [\pref{Chaumard}]), who are particularly concerned with the continuum limit,
another important topic I have not addressed.

\section{\bf 15. Conclusion}

I have made a slight extension to the computation of the free heat--kernel on a $p$--cycle
and have used the resulting flexibility to derive a bigger variety of Bessel relations, for
example.

An organisational point is that, if one is interested just in the discrete resolvent (the
left--hand side of (\peq{ident4})) then an appeal can be made to the identity (\peq{ident3})
and the Bessel form can be bypassed.

I have also calculated the heat--kernel on a discrete interval for various boundary conditions
using both modes and images. As usual, the Neumann images are all positive while the
Dirichlet ones alternate in sign (positive for even reflections and negative for odd). In the
continuum, when adding Neumann to Dirichlet the odd D images cancel corresponding N
ones leaving just the images for a periodic circle of circumference twice the size of the
interval, as is well known. This cancellation does not happen in the discrete case due to the
offset of 1/2 to the Neumann mirrors.

Expressions for the free Green functions in terms of Chebyshev polynomials are derived.
That on the D--N  path is perhaps novel. As a check, the combinatorial Laplacian matrices
are extracted.

Relatedly, rational generating functions  for circuits of specific length and number of bumps
on the cycle graph are readily computed and compared with those of Bartholdi,
[\pref{Bartholdi}].

Paths and cycles are graphs and these calculations illustrate very general graph--theoretic
procedures. I intend to expand on this aspect elsewhere.

   \newpage
 \vglue 20truept

\noin{\bf References.} \vskip5truept

\begin{putreferences}
  \ref{Hicks}{Hicks,W.M. \qjm{15}{1878}{274}.}
  \ref{Bartholdi}{Bartholdi,L. {\it L'Enseignement Math.} {\bf 45} (1999) 83.}
  \ref{PandB}{van der Pol,B. and Bremmer,H.   {\it Operational calculus}. (CUP,Cambridge
  1950).}
  \ref{MandW2}{McCrea,W.H. and Whipple,F.J.W. {\it Proc.Roy.Soc.Edin.}
   {\bf 60} (1940) 281.}
  \ref{Grone}{Grone}
  \ref{Guttman}{Guttmann,A.J. {\it Lattice Green functions in all dimensions}
   ArXiv:math-ph/ 1004. 1435.}
  \ref{Zucker}{Zucker,I.J. {\it J.Stat.Phys.} {\bf 145} (2011) 591.}
  \ref{Flanders}{Flanders,H. {\it J.Math.Anal.Appl.} {\bf 40} (1972) 30.}
  \ref{Stohr}{St\"ohr,A. {\it Math.Nach.} {\bf 3} (1950) 208, 295, 330.}
  \ref{MandW}{Montroll,E.W. and Weiss,G.H. \jmp{6}{1965}{167}.}
  \ref{KIA}{Katsura,S., Inawashiro,S. and Abe, Y. \jmp{12}{1971}{895}.}
  \ref{KandI}{Katsura,S. and Inawashiro,S. \jmp{12}{1971}{1622}.}
  \ref{KandI2}{Katsura,S. and Inawashiro,S. \jmp{12}{1971}{}.}
  \ref{Cojocaru}{Cojocaru,S. {\it Int.J.Mod.Phys.} B{\bf 20}(2006)593.}
  \ref{CoandB}{B\^{a} rsan,V. and Cojocaru,S. {\it Bessel functions of integer order in terms of
  hyperbolic functions} ArXiv:math-ph/0703010.}
  \ref{CandJ}{Carslaw,H.S. and Jaeger,J.C. {\it Conduction of heat in solids} (Oxford,
  Clarendon, 1959).}
  \ref{IIH}{Ivashkevich,E.V. Izmailian,N.Sh. and Hu,C--K, \jpa{35}{2002}{5543}.}
  \ref{GandT}{Giles,R. and Thorn,C.B. \prD{16}{1977}{366}.}
  \ref{Thorn}{Thorn,C.B. {\it Determinants for the Lightcone Worldsheet}  ArXiv:1205.5815.}
  \ref{Zeitlin}{Zeitlin,V. {\it Physica} {\bf D49}(1991)353.}
  \ref{FFZ}{Fairlie,D.B.,Fletcher,P. and Zachos,C.K. \plb{218}{1989}{203}.}
  \ref{FandZ}{Fairlie,D.B. and Zachos,C.K. \plb{224}{1989}{101}.}
  \ref{Kasteleyn}{Kasteleyn,P.W. {\it Physica} {\bf 27}(1961)1209.}
  \ref{DuandD}{Duplantier,B. and David.F. {\it J.Stat.Phys.} {\bf 51} (1988) 327.}
  \ref{CJK}{Chinta,G., Jorgenson,J. and Karlsson,A. {\it Zeta functions, heat kernels and
  spectral asymptotics on degenerating families of discrete tori} ArXiv:0806.2014.}
  \ref{CJK2}{Chinta,G., Jorgenson,J. and Karlsson,A. {\it Complexity and the heights
  of tori} ArXiv:1110.6841.}
  \ref{Chaumard}{Chaumard,L. {\it Bull.Soc.Math.France} {\bf134} (2006) 327.}
  \ref{Iliev}{Iliev,P. {\it Selecta Math.} (N.S.) {\bf13} (2007) 497.}
  \ref{Loney}{Loney,S.L. {\it Plane trigonometry} (CUP, Cambridge, 1893).}
  \ref{Hobson2}{Hobson,E.W. {\it A treatise on plane trigonometry}
  (CUP, Cambridge, 1891).}
  \ref{Basset}{Basset,A.B. {\it A Treatise on Hydrodynamics} (Deighton Bell,
  Cambridge,1888).}
   \ref{Dow}{Dowker,J.S. \jpa{5}{1972}{936}.}
   \ref{Dikii}{Dikii,L.A. {\it Usp.Mat.Nauk.} {\bf 13} (1958) 111.}
  \ref{Thomson}{Thomson, J.J. {\it Elements of the Mathematical Theory of Electricity and
  Magnetism}
  (CUP,Cambridge,1895).}
  \ref{dowstat}{Dowker,J.S. \jpa{18}{1985}{3521}.}
  \ref{dowaustin}{Dowker,J.S. 1979 {\it Selected topics in topology and quantum
  field theory}
  (Lectures at Center for Relativity, University of Texas, Austin).}
  \ref{LandD}{Laidlaw,M. and De Witt, C. \prD{3}{1971}{1375}.}
  \ref{Fisher}{Fisher,M.E. {\it SIAM Rev.} {\bf 13} 116 (Solution of problem 69-14).}
  \ref{dowverlinde}{Dowker,J.S. \jpa{25}{1992}{2641}.}
  \ref{CandY2}{Chung, F. and Yau, S.T. {\it A combinatorial trace identity},
  Tsing Hua lectures on geometry and analysis (S.T.Yau,ed.) (International Press,
  Cambridge, MA, 1987, p.107).}
  \ref{KandN}{Karlsson,A. and Neuhauser,M. {\it Contemporary Mathematics}
  {\bf 394}(2006)177.}
  \ref{Feller}{Feller, W. {\it An Introduction to Probability Theory} Vol.2 (Wiley,
  New York,1971).}
  \ref{ADG}{Al--Jarrah,A., Dempsey,M. and Glasser,M.I. {\it J.Comput.Appl.Math.}
  {\bf 143}(2002)1.}
  \ref{Annaby}{Annaby,M.H. {\it Analysis} {\bf 18} (1998) 55.}
  \ref{Watson2}{Watson,G.N. {\it A Treatise on the theory of Bessel Functions} (CUP,
  Cambridge, 1944).}
  \ref{Dowgelyag}{Dowker,J.S. {\it Discrete Determinants and the Gel'fand--Yaglom
  formula}, \jpa{}{}{}. }
  \ref{Flugge}{Fl\"ugge,S.   {\it Practical Quantum Mechanics} (Springer,Berlin,1971).}
  \ref{GandG}{Graf,J.H. and Gubler,E. {\it Einleitung in die Theorie der Bessel'schen
  Funktionen} (Wyss, Bern,1898-1900).}
  \ref{Lommel}{Lommel, E. {\it Studien \"uber die Bessel'schen Functionen}
  (Teubner, Leipzig, 1868).}
  \ref{Nielsen}{Nielsen,N. {\it Handbuch der Theorie der Cylinderfunktionen}
  (Teubner, Leipzig, 1904.).}
  \ref{LandT}{Lakshmikantham,V. and Trigiante,D. {\it Theory of Difference Equations}
  (Academic Press, San Diego, 1988).}
  \ref{Agarwal}{Agarwal,R.P. {\it Difference Equations and Inequalities} (Dekker, New York,
  1992).}
  \ref{Kleinert}{Kleinert, H. {\it Path Integrals in Quantum mechanics} (World Scientific,
   Singapore, 2004).}
  \ref{Miller2}{Miller,K.S. {\it Linear Difference Equations} (Benjamin, New York, 1968).}
  \ref{GrandM}{Gray,A. and Matthews, G.B. {\it A treatise on Bessel functions}, (MacMillan,
  London, 1922).}
  \ref{Birkhoff}{Birkhoff,G.D. \tams{12}{1911}{243}.}
  \ref{LeandL}{Levy,H. and Lessman,F. {\it Finite Difference Equations} (Pitman, London,
  1959).}
  \ref{EandM}{Erhardt and Mickens .}
  \ref{Bohmer}{B\"ohmer,E. {\it Differenzengleichungen und bestimmte Integrale} (Koehler,
  Leipzig, 1939).}
  \ref{Barnesrec}{Barnes,E.W. {\it Mess. Math.} {\bf34}(1904) 52.}
   \ref{Bass}{Bass,R. \jmp{26}{1985}{3068}.}
   \ref{GandK}{Gantmacher,F.R. and Krein,M.G. {\it Oszillazionsmatrizen}
   (Akad.-Verlag.Berlin, 1960).}
   \ref{Hald}{Hald,O.H. {\it Numer.Math.} {\bf27} (1977) 249.}
   \ref{CandY}{Chung,F. and Yau S.--T. {\it J.Comb.Theory} {\bf A91} (2000) 191.}
   \ref{BEC}{Bendito,E, Encinas,A.M. and Carmona,A. {\it Appl.Anal.Disc.Math.}{\bf 3}
   (2009) 282.}
   \ref{AandA}{Annaby,M.H. and Asharabi,R.M. {\it Acta.Math.Scientia} {\bf 31B}
   (2011)408.}
   \ref{MandH}{Mason,J.C. and Handscomb,D.C. {\it Chebyshev Polynomials}
   (Chapman and Hall,   Boca Raton, 2002).}
   \ref{LandB}{Levy,H. and Baggott,E.A. \phm{18}{1934}{177}.}
   \ref{BlandM}{Bleich,Fr. and Melan,E. {\it Die Gew\"onlichen und Partiellen
   Differenzengleichungen der Baustatik}, (Springer, Berlin, 1927).}
   \ref{Spiegel}{Spiegel,M.R. {\it Schaum's Outline of Calculus of Finite Differences},
   (McGraw-Hill, New York, 1971).}
   \ref{Porter}{Porter,M.B. \aom{3}{1901}{55}.}
    \ref{Dunne}{Dunne,G. \jpa{41}{2008}{304006}.}
    \ref{MandF}{Morse,P.M. and  Feshbach,H. {\it Methods of Theoretical Physics},
    (McGraw--Hill, New York, 1953).}
    \ref{KandM}{Kirsten, K. and McKane,A. \aop{308}{2003}{502}.}
    \ref{LandS}{Levit,S. and Smilansky,U. \pams{65}{1977}{299}.}
    \ref{GeandY}{Gel'fand, I.M. and Yaglom,A.M. \jmp {1}{1960}{48}.}
    \ref{ABR}{Actor,A., Bender,C. and Reingruber,J., {\it Fortschr.Phys.} {\bf48} (2000) 4.}
    \ref{Rayleigh}{Rayleigh, Lord, {\it Theory of Sound}, (MacMillan, London, 1894).}
    \ref{Floratos}{Floratos,E.G. \plb{228}{1989}{335}.}
    \ref{deverdiere}{de Verdi\`ere, C. {\it Ann. Inst. Fourier} {\bf 49} (1999) 861.}
    \ref{Goldberg}{Goldberg,S. {\it Introduction to Difference Equations},
    (Wiley, New York, 1958).}
    \ref{Jordan}{Jordan,C. {\it Calculus of Finite Differences}, (Budapest, 1939).}
    \ref{Fort}{Fort,T. {\it Finite Differences}, (Clarendon Press, Oxford,1948). }
    \ref{Atkinson}{Atkinson,F.V. {\it Discrete and Continuous Boundary Problems},
    (Academic Press, New York,1964).}
   \ref{Elaydi}{Elaydi,S.N. {\it An Introduction to Difference Equations}, (Springer, New York,
   1999.}
   \ref{Forman}{Forman,R. \cmp{147}{1992}{485}.}
    \ref{Boole}{Boole, G. {\it Calculus of Finite Differences}, (MacMillan, Cambridge,
1860).}
     \ref{Cayley5}{Cayley,A.  {\it Phil. Trans. Roy. Soc. Lond.} {\bf 148} (1858) 47.}
     \ref{Milne-Thomson}{Milne-Thomson, L.M. {\it The Calculus of Finite Differences},
     (MacMillan, London, 1933).}
    \ref{Herschel}{Herschel, J.F.W.  {\it Phil. Trans. Roy. Soc. Lond.} {\bf 106} (1816) 25.}
    \ref{Littlewood2}{Littlewood,D.E. {\it The Theory of Group Characters}
    (Clarendon Press, Oxford, 1950).}
    \ref{Wright}{Wright,E.M. \amm{68}{1961}{144}.}
\ref{Carlitz}{Carlitz,L. \dmj{27}{1960}{401}.}
     \ref{Netto}{Netto,E. {\it Lehrbuch der Combinatorik} 2nd Edn.
     (Teubner, Leipzig, 1927).}
    \ref{FdeB}{Fa\`{a} de Bruno, F. {\it Th\'eorie des Formes Binaires}
    (Brero, Turin,1876).}
    \ref{Ehrhart}{Ehrhart,E. \jram{227}{25}{1967}.}
    \ref{Bell}{Bell,E.T.\ajm{65}{1943}{382}.}
    \ref{BandR}{Beck,M. and Robins,S. {\it Computing the Continuous Discretely,}
    (Springer, New York, 2007).}
    \ref{BandR2}{Beck, M. and Robins,S. {\it Discrete and Comp. Geom.}
    {\bf 27}(2002) 443.}
    \ref{Harmer}{Harmer,M. {\it J.Australian Math.Soc.} {\bf 84}(2008)217.}
    \ref{RandF}{Rubinstein, B.Y. and Fel,L.G., {\it Ramanujan J.} {\bf11}(2006)331.}
    \ref{BGK}{Beck,M., Gessel, I.M. and Komatsu,T. {\it Electronic Journal of Combinatorics}
    {\bf8}(2001) 1.}
    \ref{Sylvester}{Sylvester,J.J. \qjpam{1}{1858}{81}.}
    \ref{Sylvester2}{Sylvester,J.J. \qjpam{1}{1858}{142}.}
    \ref{Sylvester3}{Sylvester,J.J. \ajm{5}{1882}{79}.}
    \ref{Sylvester4}{Sylvester,J.J. \plms{28}{1896}{33}.}
    \ref{Dowgta}{J.S.Dowker, {\it Group theory aspects of spectral problems on
    spherical factors},
   ArXiv.Math.DG: 0907.1309.}
    \ref{BDR}{Beck,M., Diaz and Robins,S. {\it J.Numb.Theory} {\bf 96} (2002) 1.}
    \ref{PandS}{P\'{o}lya, G. and Szeg\H{o},G. {\it Aufgaben und Lehrs\"atze
    aus der Analysis} (Springer--Verlag, Berlin, 1925).}
    \ref{EOS}{Elizalde,E., Odintsov, S.D. and Saharian, A.A. \prD{79}{2009}{065023}.}
    \ref{Cavalcanti}{Cavalcanti,R.M. \prD{69}{2004}{065015}.}
    \ref{MWK}{Milton, K.A., Wagner,J. and Kirsten,K. \prD{80}{2009}{125028}.}
    \ref{EBM2}{Ellingsen,S.A., Brevik,I. and Milton,K.A. \prE{81}{2010}{065031}.}
    \ref{EBM}{Ellingsen,S.A., Brevik,I. and Milton,K.A. \prE{80}{2009}{021125}.}
    \ref{BEM}{Brevik,I., Ellingsen,S.A. and Milton,K.A. \prE{79}{2009}{041120}.}
    \ref{FKW}{Fulling,S.A, Kaplan L. and Wilson,J.H. \prA{76}{2007}{012118}.}
    \ref{Lukosz}{Lukosz,W, {\it Physica} {\bf 56} (1971) 109; \zfp{258}{1973}{99}
    ;\zfp{262}{1973}{327}.}
    \ref{Gromes}{Gromes, D. \mz{94}{1966}{110}.}
    \ref{FandK1}{Kirsten,K. and Fulling,S.A. \prD{79}{2009}{065019} .}
    \ref{FandK2}{Fucci,G. and Kirsten,K, JHEP (2011), 1103:016.}
    \ref{dowgjms}{Dowker,J.S. {\it Determinants and conformal anomalies
    of GJMS operators on spheres}, ArXiv: 1007.3865.}
    \ref{Dowcascone}{Dowker,J.S. \prD{36}{1987}{3095}.}
    \ref{Dowcos}{Dowker,J.S. \prD{36}{1987}{3742}.}
    \ref{Dowtherm}{Dowker,J.S. \prD{18}{1978}{1856}.}
    \ref{Dowgeo}{Dowker,J.S. \cqg{11}{1994}{L55}.}
    \ref{ApandD2}{Dowker,J.S. and Apps,J.S. \cqg{12}{1995}{1363}.}
   \ref{HandW}{Hertzberg,M.P. and Wilczek,F. {\it Some calculable contributions to
   Entanglement Entropy}, ArXiv:1007.0993.}
   \ref{KandB}{Kamela,M. and Burgess,C.P. \cjp{77}{1999}{85}.}
   \ref{Dowhyp}{Dowker,J.S. \jpa{43}{2010}{445402}; ArXiv:1007.3865.}
   \ref{LNST}{Lohmayer,R., Neuberger,H, Schwimmer,A. and Theisen,S.
   \plb{685}{2010}{222}.}
   \ref{Allen2}{Allen,B. PhD Thesis, University of Cambridge, 1984.}
   \ref{MyandS}{Myers,R.C. and Sinha,A. {\it Seeing a c-theorem with
   holography}, ArXiv:1006.1263}
   \ref{MyandS2}{Myers,R.C. and Sinha,A. {\it Holographic c-theorems in
   arbitrary dimensions},\break ArXiv: 1011.5819.}
   \ref{RyandT}{Ryu,S. and Takayanagi,T. JHEP {\bf 0608}(2006)045.}
   \ref{CaandH}{Casini,H. and Huerta,M. {\it Entanglement entropy
   for the n--sphere},\break arXiv:1007.1813.}
   \ref{CaandH3}{Casini,H. and Huerta,M. \jpa {42}{2009}{504007}.}
   \ref{Solodukhin}{Solodukhin,S.N. \plb{665}{2008}{305}.}
   \ref{Solodukhin2}{Solodukhin,S.N. \plb{693}{2010}{605}.}
   \ref{CaandW}{Callan,C.G. and Wilczek,F. \plb{333}{1994}{55}.}
   \ref{FandS1}{Fursaev,D.V. and Solodukhin,S.N. \plb{365}{1996}{51}.}
   \ref{FandS2}{Fursaev,D.V. and Solodukhin,S.N. \prD{52}{1995}{2133}.}
   \ref{Fursaev}{Fursaev,D.V. \plb{334}{1994}{53}.}
   \ref{Donnelly2}{Donnelly,H. \ma{224}{1976}{161}.}
   \ref{ApandD}{Apps,J.S. and Dowker,J.S. \cqg{15}{1998}{1121}.}
   \ref{FandM}{Fursaev,D.V. and Miele,G. \prD{49}{1994}{987}.}
   \ref{Dowker2}{Dowker,J.S.\cqg{11}{1994}{L137}.}
   \ref{Dowker1}{Dowker,J.S.\prD{50}{1994}{6369}.}
   \ref{FNT}{Fujita,M.,Nishioka,T. and Takayanagi,T. JHEP {\bf 0809}
   (2008) 016.}
   \ref{Hund}{Hund,F. \zfp{51}{1928}{1}.}
   \ref{Elert}{Elert,W. \zfp {51}{1928}{8}.}
   \ref{Poole2}{Poole,E.G.C. \qjm{3}{1932}{183}.}
   \ref{Bellon}{Bellon,M.P. {\it On the icosahedron: from two to three
   dimensions}, arXiv:0705.3241.}
   \ref{Bellon2}{Bellon,M.P. \cqg{23}{2006}{7029}.}
   \ref{McLellan}{McLellan,A,G. \jpc{7}{1974}{3326}.}
   \ref{Boiteaux}{Boiteaux, M. \jmp{23}{1982}{1311}.}
   \ref{HHandK}{Hage Hassan,M. and Kibler,M. {\it On Hurwitz
   transformations} in {Le probl\`eme de factorisation de Hurwitz}, Eds.,
   A.Ronveaux and D.Lambert (Fac.Univ.N.D. de la Paix, Namur, 1991),
   pp.1-29.}
   \ref{Weeks2}{Weeks,Jeffrey \cqg{23}{2006}{6971}.}
   \ref{LandW}{Lachi\`eze-Rey,M. and Weeks,Jeffrey, {\it Orbifold construction of
   the modes on the Poincar\'e dodecahedral space}, arXiv:0801.4232.}
   \ref{Cayley4}{Cayley,A. \qjpam{58}{1879}{280}.}
   \ref{JMS}{Jari\'c,M.V., Michel,L. and Sharp,R.T. {\it J.Physique}
   {\bf 45} (1984) 1. }
   \ref{AandB}{Altmann,S.L. and Bradley,C.J.  {\it Phil. Trans. Roy. Soc. Lond.}
   {\bf 255} (1963) 199.}
   \ref{CandP}{Cummins,C.J. and Patera,J. \jmp{29}{1988}{1736}.}
   \ref{Sloane}{Sloane,N.J.A. \amm{84}{1977}{82}.}
   \ref{Gordan2}{Gordan,P. \ma{12}{1877}{147}.}
   \ref{DandSh}{Desmier,P.E. and Sharp,R.T. \jmp{20}{1979}{74}.}
   \ref{Kramer}{Kramer,P., \jpa{38}{2005}{3517}.}
   \ref{Klein2}{Klein, F.\ma{9}{1875}{183}.}
   \ref{Hodgkinson}{Hodgkinson,J. \jlms{10}{1935}{221}.}
   \ref{ZandD}{Zheng,Y. and Doerschuk, P.C. {\it Acta Cryst.} {\bf A52}
   (1996) 221.}
   \ref{EPM}{Elcoro,L., Perez--Mato,J.M. and Madariaga,G.
   {\it Acta Cryst.} {\bf A50} (1994) 182.}
    \ref{PSW2}{Prandl,W., Schiebel,P. and Wulf,K.
   {\it Acta Cryst.} {\bf A52} (1999) 171.}
    \ref{FCD}{Fan,P--D., Chen,J--Q. and Draayer,J.P.
   {\it Acta Cryst.} {\bf A55} (1999) 871.}
   \ref{FCD2}{Fan,P--D., Chen,J--Q. and Draayer,J.P.
   {\it Acta Cryst.} {\bf A55} (1999) 1049.}
   \ref{Honl}{H\"onl,H. \zfp{89}{1934}{244}.}
   \ref{PSW}{Patera,J., Sharp,R.T. and Winternitz,P. \jmp{19}{1978}{2362}.}
   \ref{LandH}{Lohe,M.A. and Hurst,C.A. \jmp{12}{1971}{1882}.}
   \ref{RandSA}{Ronveaux,A. and Saint-Aubin,Y. \jmp{24}{1983}{1037}.}
   \ref{JandDeV}{Jonker,J.E. and De Vries,E. \npa{105}{1967}{621}.}
   \ref{Rowe}{Rowe, E.G.Peter. \jmp{19}{1978}{1962}.}
   \ref{KNR}{Kibler,M., N\'egadi,T. and Ronveaux,A. {\it The Kustaanheimo-Stiefel
   transformation and certain special functions} \lnm{1171}{1985}{497}.}
   \ref{GLP}{Gilkey,P.B., Leahy,J.V. and Park,J-H, \jpa{29}{1996}{5645}.}
   \ref{Kohler}{K\"ohler,K.: Equivariant Reidemeister torsion on
   symmetric spaces. Math.Ann. {\bf 307}, 57-69 (1997)}
   \ref{Kohler2}{K\"ohler,K.: Equivariant analytic torsion on ${\bf P^nC}$.
   Math.Ann.{\bf 297}, 553-565 (1993) }
   \ref{Kohler3}{K\"ohler,K.: Holomorphic analytic torsion on Hermitian
   symmetric spaces. J.Reine Angew.Math. {\bf 460}, 93-116 (1995)}
   \ref{Zagierzf}{Zagier,D. {\it Zetafunktionen und Quadratische
   K\"orper}, (Springer--Verlag, Berlin, 1981).}
   \ref{Stek}{Stekholschkik,R. {\it Notes on Coxeter transformations and the McKay
   correspondence.} (Springer, Berlin, 2008).}
   \ref{Pesce}{Pesce,H. \cmh {71}{1996}{243}.}
   \ref{Pesce2}{Pesce,H. {\it Contemp. Math} {\bf 173} (1994) 231.}
   \ref{Sutton}{Sutton,C.J. {\it Equivariant isospectrality
   and isospectral deformations on spherical orbifolds}, ArXiv:math/0608567.}
   \ref{Sunada}{Sunada,T. \aom{121}{1985}{169}.}
   \ref{GoandM}{Gornet,R, and McGowan,J. {\it J.Comp. and Math.}
   {\bf 9} (2006) 270.}
   \ref{Suter}{Suter,R. {\it Manusc.Math.} {\bf 122} (2007) 1-21.}
   \ref{Lomont}{Lomont,J.S. {\it Applications of finite groups} (Academic
   Press, New York, 1959).}
   \ref{DandCh2}{Dowker,J.S. and Chang,Peter {\it Analytic torsion on
   spherical factors and tessellations}, arXiv:math.DG/0904.0744 .}
   \ref{Mackey}{Mackey,G. {\it Induced representations}
   (Benjamin, New York, 1968).}
   \ref{Koca}{Koca, {\it Turkish J.Physics}.}
   \ref{Brylinski}{Brylinski, J-L., {\it A correspondence dual to McKay's}
    ArXiv alg-geom/9612003.}
   \ref{Rossmann}{Rossman,W. {\it McKay's correspondence
   and characters of finite subgroups of\break SU(2)} {\it Progress in Math.}
      Birkhauser  (to appear) .}
   \ref{JandL}{James, G. and Liebeck, M. {\it Representations and
   characters of groups} (CUP, Cambridge, 2001).}
   \ref{IandR}{Ito,Y. and Reid,M. {\it The Mckay correspondence for finite
   subgroups of SL(3,C)} Higher dimensional varieties, (Trento 1994),
   221-240, (Berlin, de Gruyter 1996).}
   \ref{BandF}{Bauer,W. and Furutani, K. {\it J.Geom. and Phys.} {\bf
   58} (2008) 64.}
   \ref{Luck}{L\"uck,W. \jdg{37}{1993}{263}.}
   \ref{LandR}{Lott,J. and Rothenberg,M. \jdg{34}{1991}{431}.}
   \ref{DoandKi} {Dowker.J.S. and Kirsten, K. {\it Analysis and Appl.}
   {\bf 3} (2005) 45.}
   \ref{dowtess1}{Dowker,J.S. \cqg{23}{2006}{1}.}
   \ref{dowtess2}{Dowker,J.S. {\it J.Geom. and Phys.} {\bf 57} (2007) 1505.}
   \ref{MHS}{De Melo,T., Hartmann,L. and Spreafico,M. {\it Reidemeister
   Torsion and analytic torsion of discs}, ArXiv:0811.3196.}
   \ref{Vertman}{Vertman, B. {\it Analytic Torsion of a  bounded
   generalized cone}, ArXiv:0808.0449.}
   \ref{WandY} {Weng,L. and You,Y., {\it Int.J. of Math.}{\bf 7} (1996)
   109.}
   \ref{ScandT}{Schwartz, A.S. and Tyupkin,Yu.S. \np{242}{1984}{436}.}
   \ref{AAR}{Andrews, G.E., Askey,R. and Roy,R. {\it Special functions}
   (CUP, Cambridge, 1999).}
   \ref{Tsuchiya}{Tsuchiya, N.: R-torsion and analytic torsion for spherical
   Clifford-Klein manifolds.: J. Fac.Sci., Tokyo Univ. Sect.1 A, Math.
   {\bf 23}, 289-295 (1976).}
   \ref{Tsuchiya2}{Tsuchiya, N. J. Fac.Sci., Tokyo Univ. Sect.1 A, Math.
   {\bf 23}, 289-295 (1976).}
  \ref{Lerch}{Lerch,M. \am{11}{1887}{19}.}
  \ref{Lerch2}{Lerch,M. \am{29}{1905}{333}.}
  \ref{TandS}{Threlfall, W. and Seifert, H. \ma{104}{1930}{1}.}
  \ref{RandS}{Ray, D.B., and Singer, I. \aim{7}{1971}{145}.}
  \ref{RandS2}{Ray, D.B., and Singer, I. {\it Proc.Symp.Pure Math.}
  {\bf 23} (1973) 167.}
  \ref{Jensen}{Jensen,J.L.W.V. \aom{17}{1915-1916}{124}.}
  \ref{Rosenberg}{Rosenberg, S. {\it The Laplacian on a Riemannian Manifold}
  (CUP, Cambridge, 1997).}
  \ref{Nando2}{Nash, C. and O'Connor, D-J. {\it Int.J.Mod.Phys.}
  {\bf A10} (1995) 1779.}
  \ref{Fock}{Fock,V. \zfp{98}{1935}{145}.}
  \ref{Levy}{Levy,M. \prs {204}{1950}{145}.}
  \ref{Schwinger2}{Schwinger,J. \jmp{5}{1964}{1606}.}
  \ref{Muller}{M\"uller, \lnm{}{}{}.}
  \ref{VMK}{Varshalovich.}
  \ref{DandWo}{Dowker,J.S. and Wolski, A. \prA{46}{1992}{6417}.}
  \ref{Zeitlin1}{Zeitlin,V. {\it Physica D} {\bf 49} (1991).  }
  \ref{Zeitlin0}{Zeitlin,V. {\it Nonlinear World} Ed by
   V.Baryakhtar {\it et al},  Vol.I p.717,  (World Scientific, Singapore, 1989).}
  \ref{Zeitlin2}{Zeitlin,V. \prl{93}{2004}{264501}. }
  \ref{Zeitlin3}{Zeitlin,V. \pla{339}{2005}{316}. }
  \ref{Groenewold}{Groenewold, H.J. {\it Physica} {\bf 12} (1946) 405.}
  \ref{Cohen}{Cohen, L. \jmp{7}{1966}{781}.}
  \ref{AandW}{Argawal G.S. and Wolf, E. \prD{2}{1970}{2161,2187,2206}.}
  \ref{Jantzen}{Jantzen,R.T. \jmp{19}{1978}{1163}.}
  \ref{Moses2}{Moses,H.E. \aop{42}{1967}{343}.}
  \ref{Carmeli}{Carmeli,M. \jmp{9}{1968}{1987}.}
  \ref{SHS}{Siemans,M., Hancock,J. and Siminovitch,D. {\it Solid State
  Nuclear Magnetic Resonance} {\bf 31}(2007)35.}
 \ref{Dowk}{Dowker,J.S. \prD{28}{1983}{3013}.}
 \ref{Heine}{Heine, E. {\it Handbuch der Kugelfunctionen}
  (G.Reimer, Berlin. 1878, 1881).}
  \ref{Pockels}{Pockels, F. {\it \"Uber die Differentialgleichung $\De
  u+k^2u=0$} (Teubner, Leipzig. 1891).}
  \ref{Hamermesh}{Hamermesh, M., {\it Group Theory} (Addison--Wesley,
  Reading. 1962).}
  \ref{Racah}{Racah, G. {\it Group Theory and Spectroscopy}
  (Princeton Lecture Notes, 1951). }
  \ref{Gourdin}{Gourdin, M. {\it Basics of Lie Groups} (Editions
  Fronti\'eres, Gif sur Yvette. 1982.)}
  \ref{Clifford}{Clifford, W.K. \plms{2}{1866}{116}.}
  \ref{Story2}{Story, W.E. \plms{23}{1892}{265}.}
  \ref{Story}{Story, W.E. \ma{41}{1893}{469}.}
  \ref{Poole}{Poole, E.G.C. \plms{33}{1932}{435}.}
  \ref{Dickson}{Dickson, L.E. {\it Algebraic Invariants} (Wiley, N.Y.
  1915).}
  \ref{Dickson2}{Dickson, L.E. {\it Modern Algebraic Theories}
  (Sanborn and Co., Boston. 1926).}
  \ref{Hilbert2}{Hilbert, D. {\it Theory of algebraic invariants} (C.U.P.,
  Cambridge. 1993).}
  \ref{Olver}{Olver, P.J. {\it Classical Invariant Theory} (C.U.P., Cambridge.
  1999.)}
  \ref{AST}{A\v{s}erova, R.M., Smirnov, J.F. and Tolsto\v{i}, V.N. {\it
  Teoret. Mat. Fyz.} {\bf 8} (1971) 255.}
  \ref{AandS}{A\v{s}erova, R.M., Smirnov, J.F. \np{4}{1968}{399}.}
  \ref{Shapiro}{Shapiro, J. \jmp{6}{1965}{1680}.}
  \ref{Shapiro2}{Shapiro, J.Y. \jmp{14}{1973}{1262}.}
  \ref{NandS}{Noz, M.E. and Shapiro, J.Y. \np{51}{1973}{309}.}
  \ref{Cayley2}{Cayley, A. {\it Phil. Trans. Roy. Soc. Lond.}
  {\bf 144} (1854) 244.}
  \ref{Cayley3}{Cayley, A. {\it Phil. Trans. Roy. Soc. Lond.}
  {\bf 146} (1856) 101.}
  \ref{Wigner}{Wigner, E.P. {\it Gruppentheorie} (Vieweg, Braunschweig. 1931).}
  \ref{Sharp}{Sharp, R.T. \ajop{28}{1960}{116}.}
  \ref{Laporte}{Laporte, O. {\it Z. f. Naturf.} {\bf 3a} (1948) 447.}
  \ref{Lowdin}{L\"owdin, P-O. \rmp{36}{1964}{966}.}
  \ref{Ansari}{Ansari, S.M.R. {\it Fort. d. Phys.} {\bf 15} (1967) 707.}
  \ref{SSJR}{Samal, P.K., Saha, R., Jain, P. and Ralston, J.P. {\it
  Testing Isotropy of Cosmic Microwave Background Radiation},
  astro-ph/0708.2816.}
  \ref{Lachieze}{Lachi\'eze-Rey, M. {\it Harmonic projection and
  multipole Vectors}. astro- \break ph/0409081.}
  \ref{CHS}{Copi, C.J., Huterer, D. and Starkman, G.D.
  \prD{70}{2003}{043515}.}
  \ref{Jaric}{Jari\'c, J.P. {\it Int. J. Eng. Sci.} {\bf 41} (2003) 2123.}
  \ref{RandD}{Roche, J.A. and Dowker, J.S. \jpa{1}{1968}{527}.}
  \ref{KandW}{Katz, G. and Weeks, J.R. \prD{70}{2004}{063527}.}
  \ref{Waerden}{van der Waerden, B.L. {\it Die Gruppen-theoretische
  Methode in der Quantenmechanik} (Springer, Berlin. 1932).}
  \ref{EMOT}{Erdelyi, A., Magnus, W., Oberhettinger, F. and Tricomi, F.G. {
  \it Higher Transcendental Functions} Vol.1 (McGraw-Hill, N.Y. 1953).}
   \ref{EMOT2}{Erdelyi, A., Magnus, W., Oberhettinger, F. and Tricomi, F.G. {
  \it Higher Transcendental Functions} Vol.2 (McGraw-Hill, N.Y. 1953).}
  \ref{Dowzilch}{Dowker, J.S. {\it Proc. Phys. Soc.} {\bf 91} (1967) 28.}
  \ref{DandD}{Dowker, J.S. and Dowker, Y.P. {\it Proc. Phys. Soc.}
  {\bf 87} (1966) 65.}
  \ref{DandD2}{Dowker, J.S. and Dowker, Y.P. \prs{}{}{}.}
  \ref{Dowk3}{Dowker,J.S. \cqg{7}{1990}{1241}.}
  \ref{Dowk5}{Dowker,J.S. \cqg{7}{1990}{2353}.}
  \ref{CoandH}{Courant, R. and Hilbert, D. {\it Methoden der
  Mathematischen Physik} vol.1 \break (Springer, Berlin. 1931).}
  \ref{Applequist}{Applequist, J. \jpa{22}{1989}{4303}.}
  \ref{Torruella}{Torruella, \jmp{16}{1975}{1637}.}
  \ref{Weinberg}{Weinberg, S.W. \pr{133}{1964}{B1318}.}
  \ref{Meyerw}{Meyer, W.F. {\it Apolarit\"at und rationale Curven}
  (Fues, T\"ubingen. 1883.) }
  \ref{Ostrowski}{Ostrowski, A. {\it Jahrsb. Deutsch. Math. Verein.} {\bf
  33} (1923) 245.}
  \ref{Kramers}{Kramers, H.A. {\it Grundlagen der Quantenmechanik}, (Akad.
  Verlag., Leipzig, 1938).}
  \ref{ZandZ}{Zou, W.-N. and Zheng, Q.-S. \prs{459}{2003}{527}.}
  \ref{Weeks1}{Weeks, J.R. {\it Maxwell's multipole vectors
  and the CMB}.  astro-ph/0412231.}
  \ref{Corson}{Corson, E.M. {\it Tensors, Spinors and Relativistic Wave
  Equations} (Blackie, London. 1950).}
  \ref{Rosanes}{Rosanes, J. \jram{76}{1873}{312}.}
  \ref{Salmon}{Salmon, G. {\it Lessons Introductory to the Modern Higher
  Algebra} 3rd. edn. \break (Hodges,  Dublin. 1876.)}
  \ref{Milnew}{Milne, W.P. {\it Homogeneous Coordinates} (Arnold. London. 1910).}
  \ref{Niven}{Niven, W.D. {\it Phil. Trans. Roy. Soc.} {\bf 170} (1879) 393.}
  \ref{Scott}{Scott, C.A. {\it An Introductory Account of
  Certain Modern Ideas and Methods in Plane Analytical Geometry,}
  (MacMillan, N.Y. 1896).}
  \ref{Bargmann}{Bargmann, V. \rmp{34}{1962}{300}.}
  \ref{Maxwell}{Maxwell, J.C. {\it A Treatise on Electricity and
  Magnetism} 2nd. edn. (Clarendon Press, Oxford. 1882).}
  \ref{BandL}{Biedenharn, L.C. and Louck, J.D.
  {\it Angular Momentum in Quantum Physics} (Addison-Wesley, Reading. 1981).}
  \ref{Weylqm}{Weyl, H. {\it The Theory of Groups and Quantum Mechanics}
  (Methuen, London. 1931).}
  \ref{Robson}{Robson, A. {\it An Introduction to Analytical Geometry} Vol I
  (C.U.P., Cambridge. 1940.)}
  \ref{Sommerville}{Sommerville, D.M.Y. {\it Analytical Conics} 3rd. edn.
   (Bell, London. 1933).}
  \ref{Coolidge}{Coolidge, J.L. {\it A Treatise on Algebraic Plane Curves}
  (Clarendon Press, Oxford. 1931).}
  \ref{SandK}{Semple, G. and Kneebone. G.T. {\it Algebraic Projective
  Geometry} (Clarendon Press, Oxford. 1952).}
  \ref{AandC}{Abdesselam A., and Chipalkatti, J. {\it The Higher
  Transvectants are redundant}, arXiv:0801.1533 [math.AG] 2008.}
  \ref{Elliott}{Elliott, E.B. {\it The Algebra of Quantics} 2nd edn.
  (Clarendon Press, Oxford. 1913).}
  \ref{Elliott2}{Elliott, E.B. \qjpam{48}{1917}{372}.}
  \ref{Howe}{Howe, R. \tams{313}{1989}{539}.}
  \ref{Clebsch}{Clebsch, A. \jram{60}{1862}{343}.}
  \ref{Prasad}{Prasad, G. \ma{72}{1912}{136}.}
  \ref{Dougall}{Dougall, J. \pems{32}{1913}{30}.}
  \ref{Penrose}{Penrose, R. \aop{10}{1960}{171}.}
  \ref{Penrose2}{Penrose, R. \prs{273}{1965}{171}.}
  \ref{Burnside}{Burnside, W.S. \qjm{10}{1870}{211}. }
  \ref{Lindemann}{Lindemann, F. \ma{23} {1884}{111}.}
  \ref{Backus}{Backus, G. {\it Rev. Geophys. Space Phys.} {\bf 8} (1970) 633.}
  \ref{Baerheim}{Baerheim, R. {\it Q.J. Mech. appl. Math.} {\bf 51} (1998) 73.}
  \ref{Lense}{Lense, J. {\it Kugelfunktionen} (Akad.Verlag, Leipzig. 1950).}
  \ref{Littlewood}{Littlewood, D.E. \plms{50}{1948}{349}.}
  \ref{Fierz}{Fierz, M. {\it Helv. Phys. Acta} {\bf 12} (1938) 3.}
  \ref{Williams}{Williams, D.N. {\it Lectures in Theoretical Physics} Vol. VII,
  (Univ.Colorado Press, Boulder. 1965).}
  \ref{Dennis}{Dennis, M. \jpa{37}{2004}{9487}.}
  \ref{Pirani}{Pirani, F. {\it Brandeis Lecture Notes on
  General Relativity,} edited by S. Deser and K. Ford. (Brandeis, Mass. 1964).}
  \ref{Sturm}{Sturm, R. \jram{86}{1878}{116}.}
  \ref{Schlesinger}{Schlesinger, O. \ma{22}{1883}{521}.}
  \ref{Askwith}{Askwith, E.H. {\it Analytical Geometry of the Conic
  Sections} (A.\&C. Black, London. 1908).}
  \ref{Todd}{Todd, J.A. {\it Projective and Analytical Geometry}.
  (Pitman, London. 1946).}
  \ref{Glenn}{Glenn. O.E. {\it Theory of Invariants} (Ginn \& Co, N.Y. 1915).}
  \ref{DowkandG}{Dowker, J.S. and Goldstone, M. \prs{303}{1968}{381}.}
  \ref{Turnbull}{Turnbull, H.A. {\it The Theory of Determinants,
  Matrices and Invariants} 3rd. edn. (Dover, N.Y. 1960).}
  \ref{MacMillan}{MacMillan, W.D. {\it The Theory of the Potential}
  (McGraw-Hill, N.Y. 1930).}
   \ref{Hobson}{Hobson, E.W. {\it The Theory of Spherical
   and Ellipsoidal Harmonics} (C.U.P., Cambridge. 1931).}
  \ref{Hobson1}{Hobson, E.W. \plms {24}{1892}{55}.}
  \ref{GandY}{Grace, J.H. and Young, A. {\it The Algebra of Invariants}
  (C.U.P., Cambridge, 1903).}
  \ref{FandR}{Fano, U. and Racah, G. {\it Irreducible Tensorial Sets}
  (Academic Press, N.Y. 1959).}
  \ref{TandT}{Thomson, W. and Tait, P.G. {\it Treatise on Natural Philosophy}
   (Clarendon Press, Oxford. 1867).}
  \ref{Brinkman}{Brinkman, H.C. {\it Applications of spinor invariants in
atomic physics}, North Holland, Amsterdam 1956.}
  \ref{Kramers1}{Kramers, H.A. {\it Proc. Roy. Soc. Amst.} {\bf 33} (1930) 953.}
  \ref{DandP2}{Dowker,J.S. and Pettengill,D.F. \jpa{7}{1974}{1527}}
  \ref{Dowk1}{Dowker,J.S. \jpa{}{}{45}.}
  \ref{Dowk2}{Dowker,J.S. \aop{71}{1972}{577}}
  \ref{DandA}{Dowker,J.S. and Apps, J.S. \cqg{15}{1998}{1121}.}
  \ref{Weil}{Weil,A., {\it Elliptic functions according to Eisenstein
  and Kronecker}, Springer, Berlin, 1976.}
  \ref{Ling}{Ling,C-H. {\it SIAM J.Math.Anal.} {\bf5} (1974) 551.}
  \ref{Ling2}{Ling,C-H. {\it J.Math.Anal.Appl.}(1988).}
 \ref{BMO}{Brevik,I., Milton,K.A. and Odintsov, S.D. \aop{302}{2002}{120}.}
 \ref{KandL}{Kutasov,D. and Larsen,F. {\it JHEP} 0101 (2001) 1.}
 \ref{KPS}{Klemm,D., Petkou,A.C. and Siopsis {\it Entropy
 bounds, monoticity properties and scaling in CFT's}. hep-th/0101076.}
 \ref{DandC}{Dowker,J.S. and Critchley,R. \prD{15}{1976}{1484}.}
 \ref{AandD}{Al'taie, M.B. and Dowker, J.S. \prD{18}{1978}{3557}.}
 \ref{Dow1}{Dowker,J.S. \prD{37}{1988}{558}.}
 \ref{Dowrob}{Dowker,J.S. \cqg{13}{1996}{585}.}
 \ref{Dow30}{Dowker,J.S. \prD{28}{1983}{3013}.}
 \ref{DandK}{Dowker,J.S. and Kennedy,G. \jpa{}{1978}{895}.}
 \ref{Dow2}{Dowker,J.S. \cqg{1}{1984}{359}.}
 \ref{DandKi}{Dowker,J.S. and Kirsten, K. {\it Comm. in Anal. and Geom.
 }{\bf7} (1999) 641.}
 \ref{DandKe}{Dowker,J.S. and Kennedy,G.\jpa{11}{1978}{895}.}
 \ref{Gibbons}{Gibbons,G.W. \pl{60A}{1977}{385}.}
 \ref{Cardy}{Cardy,J.L. \np{366}{1991}{403}.}
 \ref{ChandD}{Chang,P. and Dowker,J.S. \np{395}{1993}{407}.}
 \ref{DandC2}{Dowker,J.S. and Critchley,R. \prD{13}{1976}{224}.}
 \ref{Camporesi}{Camporesi,R. \prp{196}{1990}{1}.}
 \ref{BandM}{Brown,L.S. and Maclay,G.J. \pr{184}{1969}{1272}.}
 \ref{CandD}{Candelas,P. and Dowker,J.S. \prD{19}{1979}{2902}.}
 \ref{Unwin1}{Unwin,S.D. Thesis. University of Manchester. 1979.}
 \ref{Unwin2}{Unwin,S.D. \jpa{13}{1980}{313}.}
 \ref{DandB}{Dowker,J.S. and Banach,R. \jpa{11}{1978}{2255}.}
 \ref{Obhukov}{Obhukov,Yu.N. \pl{109B}{1982}{195}.}
 \ref{Kennedy}{Kennedy,G. \prD{23}{1981}{2884}.}
 \ref{CandT}{Copeland,E. and Toms,D.J. \np {255}{1985}{201}.}
  \ref{CandT2}{Copeland,E. and Toms,D.J. \cqg {3}{1986}{431}.}
 \ref{ELV}{Elizalde,E., Lygren, M. and Vassilevich,
 D.V. \jmp {37}{1996}{3105}.}
 \ref{Malurkar}{Malurkar,S.L. {\it J.Ind.Math.Soc} {\bf16} (1925/26) 130.}
 \ref{Glaisher}{Glaisher,J.W.L. {\it Messenger of Math.} {\bf18}
(1889) 1.} \ref{Anderson}{Anderson,A. \prD{37}{1988}{536}.}
 \ref{CandA}{Cappelli,A. and D'Appollonio,\pl{487B}{2000}{87}.}
 \ref{Wot}{Wotzasek,C. \jpa{23}{1990}{1627}.}
 \ref{RandT}{Ravndal,F. and Tollesen,D. \prD{40}{1989}{4191}.}
 \ref{SandT}{Santos,F.C. and Tort,A.C. \pl{482B}{2000}{323}.}
 \ref{FandO}{Fukushima,K. and Ohta,K. {\it Physica} {\bf A299} (2001) 455.}
 \ref{GandP}{Gibbons,G.W. and Perry,M. \prs{358}{1978}{467}.}
 \ref{Dow4}{Dowker,J.S..}
  \ref{Rad}{Rademacher,H. {\it Topics in analytic number theory,}
Springer-Verlag,  Berlin,1973.}
  \ref{Halphen}{Halphen,G.-H. {\it Trait\'e des Fonctions Elliptiques},
  Vol 1, Gauthier-Villars, Paris, 1886.}
  \ref{CandW}{Cahn,R.S. and Wolf,J.A. {\it Comm.Mat.Helv.} {\bf 51}
  (1976) 1.}
  \ref{Berndt}{Berndt,B.C. \rmjm{7}{1977}{147}.}
  \ref{Hurwitz}{Hurwitz,A. \ma{18}{1881}{528}.}
  \ref{Hurwitz2}{Hurwitz,A. {\it Mathematische Werke} Vol.I. Basel,
  Birkhauser, 1932.}
  \ref{Berndt2}{Berndt,B.C. \jram{303/304}{1978}{332}.}
  \ref{RandA}{Rao,M.B. and Ayyar,M.V. \jims{15}{1923/24}{150}.}
  \ref{Hardy}{Hardy,G.H. \jlms{3}{1928}{238}.}
  \ref{TandM}{Tannery,J. and Molk,J. {\it Fonctions Elliptiques},
   Gauthier-Villars, Paris, 1893--1902.}
  \ref{schwarz}{Schwarz,H.-A. {\it Formeln und
  Lehrs\"atzen zum Gebrauche..},Springer 1893.(The first edition was 1885.)
  The French translation by Henri Pad\'e is {\it Formules et Propositions
  pour L'Emploi...},Gauthier-Villars, Paris, 1894}
  \ref{Hancock}{Hancock,H. {\it Theory of elliptic functions}, Vol I.
   Wiley, New York 1910.}
  \ref{watson}{Watson,G.N. \jlms{3}{1928}{216}.}
  \ref{MandO}{Magnus,W. and Oberhettinger,F. {\it Formeln und S\"atze},
  Springer-Verlag, Berlin 1948.}
  \ref{Klein}{Klein,F. {\it Lectures on the Icosohedron}
  (Methuen, London. 1913).}
  \ref{AandL}{Appell,P. and Lacour,E. {\it Fonctions Elliptiques},
  Gauthier-Villars,
  Paris. 1897.}
  \ref{HandC}{Hurwitz,A. and Courant,C. {\it Allgemeine Funktionentheorie},
  Springer,
  Berlin. 1922.}
  \ref{WandW}{Whittaker,E.T. and Watson,G.N. {\it Modern analysis},
  Cambridge. 1927.}
  \ref{SandC}{Selberg,A. and Chowla,S. \jram{227}{1967}{86}. }
  \ref{zucker}{Zucker,I.J. {\it Math.Proc.Camb.Phil.Soc} {\bf 82 }(1977)
  111.}
  \ref{glasser}{Glasser,M.L. {\it Maths.of Comp.} {\bf 25} (1971) 533.}
  \ref{GandW}{Glasser, M.L. and Wood,V.E. {\it Maths of Comp.} {\bf 25}
  (1971)
  535.}
  \ref{greenhill}{Greenhill,A,G. {\it The Applications of Elliptic
  Functions}, MacMillan. London, 1892.}
  \ref{Weierstrass}{Weierstrass,K. {\it J.f.Mathematik (Crelle)}
{\bf 52} (1856) 346.}
  \ref{Weierstrass2}{Weierstrass,K. {\it Mathematische Werke} Vol.I,p.1,
  Mayer u. M\"uller, Berlin, 1894.}
  \ref{Fricke}{Fricke,R. {\it Die Elliptische Funktionen und Ihre Anwendungen},
    Teubner, Leipzig. 1915, 1922.}
  \ref{Konig}{K\"onigsberger,L. {\it Vorlesungen \"uber die Theorie der
 Elliptischen Funktionen},  \break Teubner, Leipzig, 1874.}
  \ref{Milne}{Milne,S.C. {\it The Ramanujan Journal} {\bf 6} (2002) 7-149.}
  \ref{Schlomilch}{Schl\"omilch,O. {\it Ber. Verh. K. Sachs. Gesell. Wiss.
  Leipzig}  {\bf 29} (1877) 101-105; {\it Compendium der h\"oheren
  Analysis}, Bd.II, 3rd Edn, Vieweg, Brunswick, 1878.}
  \ref{BandB}{Briot,C. and Bouquet,C. {\it Th\`eorie des Fonctions
  Elliptiques}, Gauthier-Villars, Paris, 1875.}
  \ref{Dumont}{Dumont,D. \aim {41}{1981}{1}.}
  \ref{Andre}{Andr\'e,D. {\it Ann.\'Ecole Normale Superior} {\bf 6} (1877)
  265;
  {\it J.Math.Pures et Appl.} {\bf 5} (1878) 31.}
  \ref{Raman}{Ramanujan,S. {\it Trans.Camb.Phil.Soc.} {\bf 22} (1916) 159;
 {\it Collected Papers}, Cambridge, 1927}
  \ref{Weber}{Weber,H.M. {\it Lehrbuch der Algebra} Bd.III, Vieweg,
  Brunswick 190  3.}
  \ref{Weber2}{Weber,H.M. {\it Elliptische Funktionen und algebraische
  Zahlen},
  Vieweg, Brunswick 1891.}
  \ref{ZandR}{Zucker,I.J. and Robertson,M.M.
  {\it Math.Proc.Camb.Phil.Soc} {\bf 95 }(1984) 5.}
  \ref{JandZ1}{Joyce,G.S. and Zucker,I.J.
  {\it Math.Proc.Camb.Phil.Soc} {\bf 109 }(1991) 257.}
  \ref{JandZ2}{Zucker,I.J. and Joyce.G.S.
  {\it Math.Proc.Camb.Phil.Soc} {\bf 131 }(2001) 309.}
  \ref{zucker2}{Zucker,I.J. {\it SIAM J.Math.Anal.} {\bf 10} (1979) 192,}
  \ref{BandZ}{Borwein,J.M. and Zucker,I.J. {\it IMA J.Math.Anal.} {\bf 12}
  (1992) 519.}
  \ref{Cox}{Cox,D.A. {\it Primes of the form $x^2+n\,y^2$}, Wiley,
  New York, 1989.}
  \ref{BandCh}{Berndt,B.C. and Chan,H.H. {\it Mathematika} {\bf42} (1995)
  278.}
  \ref{EandT}{Elizalde,R. and Tort.hep-th/}
  \ref{KandS}{Kiyek,K. and Schmidt,H. {\it Arch.Math.} {\bf 18} (1967) 438.}
  \ref{Oshima}{Oshima,K. \prD{46}{1992}{4765}.}
  \ref{greenhill2}{Greenhill,A.G. \plms{19} {1888} {301}.}
  \ref{Russell}{Russell,R. \plms{19} {1888} {91}.}
  \ref{BandB}{Borwein,J.M. and Borwein,P.B. {\it Pi and the AGM}, Wiley,
  New York, 1998.}
  \ref{Resnikoff}{Resnikoff,H.L. \tams{124}{1966}{334}.}
  \ref{vandp}{Van der Pol, B. {\it Indag.Math.} {\bf18} (1951) 261,272.}
  \ref{Rankin}{Rankin,R.A. {\it Modular forms} C.U.P. Cambridge}
  \ref{Rankin2}{Rankin,R.A. {\it Proc. Roy.Soc. Edin.} {\bf76 A} (1976) 107.}
  \ref{Skoruppa}{Skoruppa,N-P. {\it J.of Number Th.} {\bf43} (1993) 68 .}
  \ref{Down}{Dowker.J.S. {\it Nucl.Phys.}B (Proc.Suppl) ({\bf 104})(2002)153;
  also Dowker,J.S. hep-th/ 0007129.}
  \ref{Eichler}{Eichler,M. \mz {67}{1957}{267}.}
  \ref{Zagier}{Zagier,D. \invm{104}{1991}{449}.}
  \ref{Lang}{Lang,S. {\it Modular Forms}, Springer, Berlin, 1976.}
  \ref{Kosh}{Koshliakov,N.S. {\it Mess.of Math.} {\bf 58} (1928) 1.}
  \ref{BandH}{Bodendiek, R. and Halbritter,U. \amsh{38}{1972}{147}.}
  \ref{Smart}{Smart,L.R., \pgma{14}{1973}{1}.}
  \ref{Grosswald}{Grosswald,E. {\it Acta. Arith.} {\bf 21} (1972) 25.}
  \ref{Kata}{Katayama,K. {\it Acta Arith.} {\bf 22} (1973) 149.}
  \ref{Ogg}{Ogg,A. {\it Modular forms and Dirichlet series} (Benjamin,
  New York,
   1969).}
  \ref{Bol}{Bol,G. \amsh{16}{1949}{1}.}
  \ref{Epstein}{Epstein,P. \ma{56}{1903}{615}.}
  \ref{Petersson}{Petersson.}
  \ref{Serre}{Serre,J-P. {\it A Course in Arithmetic}, Springer,
  New York, 1973.}
  \ref{Schoenberg}{Schoenberg,B., {\it Elliptic Modular Functions},
  Springer, Berlin, 1974.}
  \ref{Apostol}{Apostol,T.M. \dmj {17}{1950}{147}.}
  \ref{Ogg2}{Ogg,A. {\it Lecture Notes in Math.} {\bf 320} (1973) 1.}
  \ref{Knopp}{Knopp,M.I. \dmj {45}{1978}{47}.}
  \ref{Knopp2}{Knopp,M.I. \invm {}{1994}{361}.}
  \ref{LandZ}{Lewis,J. and Zagier,D. \aom{153}{2001}{191}.}
  \ref{DandK1}{Dowker,J.S. and Kirsten,K. {\it Elliptic functions and
  temperature inversion symmetry on spheres} hep-th/.}
  \ref{HandK}{Husseini and Knopp.}
  \ref{Kober}{Kober,H. \mz{39}{1934-5}{609}.}
  \ref{HandL}{Hardy,G.H. and Littlewood, \am{41}{1917}{119}.}
  \ref{Watson}{Watson,G.N. \qjm{2}{1931}{300}.}
  \ref{SandC2}{Chowla,S. and Selberg,A. {\it Proc.Nat.Acad.} {\bf 35}
  (1949) 371.}
  \ref{Landau}{Landau, E. {\it Lehre von der Verteilung der Primzahlen},
  (Teubner, Leipzig, 1909).}
  \ref{Berndt4}{Berndt,B.C. \tams {146}{1969}{323}.}
  \ref{Berndt3}{Berndt,B.C. \tams {}{}{}.}
  \ref{Bochner}{Bochner,S. \aom{53}{1951}{332}.}
  \ref{Weil2}{Weil,A.\ma{168}{1967}{}.}
  \ref{CandN}{Chandrasekharan,K. and Narasimhan,R. \aom{74}{1961}{1}.}
  \ref{Rankin3}{Rankin,R.A. {} {} ().}
  \ref{Berndt6}{Berndt,B.C. {\it Trans.Edin.Math.Soc}.}
  \ref{Elizalde}{Elizalde,E. {\it Ten Physical Applications of Spectral
  Zeta Function Theory}, \break (Springer, Berlin, 1995).}
  \ref{Allen}{Allen,B., Folacci,A. and Gibbons,G.W. \pl{189}{1987}{304}.}
  \ref{Krazer}{Krazer}
  \ref{Elizalde3}{Elizalde,E. {\it J.Comp.and Appl. Math.} {\bf 118}
  (2000) 125.}
  \ref{Elizalde2}{Elizalde,E., Odintsov.S.D, Romeo, A. and Bytsenko,
  A.A and
  Zerbini,S.
  {\it Zeta function regularisation}, (World Scientific, Singapore,
  1994).}
  \ref{Eisenstein}{Eisenstein}
  \ref{Hecke}{Hecke,E. \ma{112}{1936}{664}.}
  \ref{Hecke2}{Hecke,E. \ma{112}{1918}{398}.}
  \ref{Terras}{Terras,A. {\it Harmonic analysis on Symmetric Spaces} (Springer,
  New York, 1985).}
  \ref{BandG}{Bateman,P.T. and Grosswald,E. {\it Acta Arith.} {\bf 9}
  (1964) 365.}
  \ref{Deuring}{Deuring,M. \aom{38}{1937}{585}.}
  \ref{Mordell}{Mordell,J. \prs{}{}{}.}
  \ref{GandZ}{Glasser,M.L. and Zucker, {}.}
  \ref{Landau2}{Landau,E. \jram{}{1903}{64}.}
  \ref{Kirsten1}{Kirsten,K. \jmp{35}{1994}{459}.}
  \ref{Sommer}{Sommer,J. {\it Vorlesungen \"uber Zahlentheorie}
  (1907,Teubner,Leipzig).
  French edition 1913 .}
  \ref{Reid}{Reid,L.W. {\it Theory of Algebraic Numbers},
  (1910,MacMillan,New York).}
  \ref{Milnor}{Milnor, J. {\it Is the Universe simply--connected?},
  IAS, Princeton, 1978.}
  \ref{Milnor2}{Milnor, J. \ajm{79}{1957}{623}.}
  \ref{Opechowski}{Opechowski,W. {\it Physica} {\bf 7} (1940) 552.}
  \ref{Bethe}{Bethe, H.A. \zfp{3}{1929}{133}.}
  \ref{LandL}{Landau, L.D. and Lishitz, E.M. {\it Quantum
  Mechanics} (Pergamon Press, London, 1958).}
  \ref{GPR}{Gibbons, G.W., Pope, C. and R\"omer, H., \np{157}{1979}{377}.}
  \ref{Jadhav}{Jadhav,S.P. PhD Thesis, University of Manchester 1990.}
  \ref{DandJ}{Dowker,J.S. and Jadhav, S. \prD{39}{1989}{1196}.}
  \ref{CandM}{Coxeter, H.S.M. and Moser, W.O.J. {\it Generators and
  relations of finite groups} (Springer. Berlin. 1957).}
  \ref{Coxeter2}{Coxeter, H.S.M. {\it Regular Complex Polytopes},
   (Cambridge University Press, \break Cambridge, 1975).}
  \ref{Coxeter}{Coxeter, H.S.M. {\it Regular Polytopes}.}
  \ref{Stiefel}{Stiefel, E., J.Research NBS {\bf 48} (1952) 424.}
  \ref{BandS}{Brink, D.M. and Satchler, G.R. {\it Angular momentum theory}.
  (Clarendon Press, Oxford. 1962.).}
  %\ref{Racah1}
  \ref{Rose}{Rose}
  \ref{Schwinger}{Schwinger, J. {\it On Angular Momentum}
  in {\it Quantum Theory of Angular Momentum} edited by
  Biedenharn,L.C. and van Dam, H. (Academic Press, N.Y. 1965).}
  \ref{Bromwich}{Bromwich, T.J.I'A. {\it Infinite Series},
  (Macmillan, London, 1926).}
  \ref{Ray}{Ray,D.B. \aim{4}{1970}{109}.}
  \ref{Ikeda}{Ikeda,A. {\it Kodai Math.J.} {\bf 18} (1995) 57.}
  \ref{Kennedy}{Kennedy,G. \prD{23}{1981}{2884}.}
  \ref{Ellis}{Ellis,G.F.R. {\it General Relativity} {\bf2} (1971) 7.}
  \ref{Dow8}{Dowker,J.S. \cqg{20}{2003}{L105}.}
  \ref{IandY}{Ikeda, A and Yamamoto, Y. \ojm {16}{1979}{447}.}
  \ref{BandI}{Bander,M. and Itzykson,C. \rmp{18}{1966}{2}.}
  \ref{Schulman}{Schulman, L.S. \pr{176}{1968}{1558}.}
  \ref{Bar1}{B\"ar,C. {\it Arch.d.Math.}{\bf 59} (1992) 65.}
  \ref{Bar2}{B\"ar,C. {\it Geom. and Func. Anal.} {\bf 6} (1996) 899.}
  \ref{Vilenkin}{Vilenkin, N.J. {\it Special functions},
  (Am.Math.Soc., Providence, 1968).}
  \ref{Talman}{Talman, J.D. {\it Special functions} (Benjamin,N.Y.,1968).}
  \ref{Miller}{Miller, W. {\it Symmetry groups and their applications}
  (Wiley, N.Y., 1972).}
  \ref{Dow3}{Dowker,J.S. \cmp{162}{1994}{633}.}
  \ref{Cheeger}{Cheeger, J. \jdg {18}{1983}{575}.}
  \ref{Cheeger2}{Cheeger, J. \aom {109}{1979}{259}.}
  \ref{Dow6}{Dowker,J.S. \jmp{30}{1989}{770}.}
  \ref{Dow20}{Dowker,J.S. \jmp{35}{1994}{6076}.}
  \ref{Dowjmp}{Dowker,J.S. \jmp{35}{1994}{4989}.}
  \ref{Dow21}{Dowker,J.S. {\it Heat kernels and polytopes} in {\it
   Heat Kernel Techniques and Quantum Gravity}, ed. by S.A.Fulling,
   Discourses in Mathematics and its Applications, No.4, Dept.
   Maths., Texas A\&M University, College Station, Texas, 1995.}
  \ref{Dow9}{Dowker,J.S. \jmp{42}{2001}{1501}.}
  \ref{Dow7}{Dowker,J.S. \jpa{25}{1992}{2641}.}
  \ref{Warner}{Warner.N.P. \prs{383}{1982}{379}.}
  \ref{Wolf}{Wolf, J.A. {\it Spaces of constant curvature},
  (McGraw--Hill,N.Y., 1967).}
  \ref{Meyer}{Meyer,B. \cjm{6}{1954}{135}.}
  \ref{BandB}{B\'erard,P. and Besson,G. {\it Ann. Inst. Four.} {\bf 30}
  (1980) 237.}
  \ref{PandM}{P\'{o}lya,G. and Meyer,B. \cras{228}{1948}{28}.}
  \ref{Springer}{Springer, T.A. Lecture Notes in Math. vol 585 (Springer,
  Berlin,1977).}
  \ref{SeandT}{Threlfall, H. and Seifert, W. \ma{104}{1930}{1}.}
  \ref{Hopf}{Hopf,H. \ma{95}{1925}{313}. }

  \ref{LLL}{Lehoucq,R., Lachi\'eze-Rey,M. and Luminet, J.--P. {\it
  Astron.Astrophys.} {\bf 313} (1996) 339.}
  \ref{LaandL}{Lachi\'eze-Rey,M. and Luminet, J.--P.
  \prp{254}{1995}{135}.}
  \ref{Schwarzschild}{Schwarzschild, K., {\it Vierteljahrschrift der
  Ast.Ges.} {\bf 35} (1900) 337.}
  \ref{Starkman}{Starkman,G.D. \cqg{15}{1998}{2529}.}
  \ref{LWUGL}{Lehoucq,R., Weeks,J.R., Uzan,J.P., Gausman, E. and
  Luminet, J.--P. \cqg{19}{2002}{4683}.}
  \ref{Dow10}{Dowker,J.S. \prD{28}{1983}{3013}.}
  \ref{BandD}{Banach, R. and Dowker, J.S. \jpa{12}{1979}{2527}.}
  \ref{Jadhav2}{Jadhav,S. \prD{43}{1991}{2656}.}
  \ref{Gilkey}{Gilkey,P.B. {\it Invariance theory,the heat equation and
  the Atiyah--Singer Index theorem} (CRC Press, Boca Raton, 1994).}
  \ref{BandY}{Berndt,B.C. and Yeap,B.P. {\it Adv. Appl. Math.}
  {\bf29} (2002) 358.}
  \ref{HandR}{Hanson,A.J. and R\"omer,H. \pl{80B}{1978}{58}.}
  \ref{Hill}{Hill,M.J.M. {\it Trans.Camb.Phil.Soc.} {\bf 13} (1883) 36.}
  \ref{Cayley}{Cayley,A. {\it Quart.Math.J.} {\bf 7} (1866) 304.}
  \ref{Seade}{Seade,J.A. {\it Anal.Inst.Mat.Univ.Nac.Aut\'on
  M\'exico} {\bf 21} (1981) 129.}
  \ref{CM}{Cisneros--Molina,J.L. {\it Geom.Dedicata} {\bf84} (2001)
  \ref{Goette1}{Goette,S. \jram {526} {2000} 181.}
  207.}
  \ref{NandO}{Nash,C. and O'Connor,D--J, \jmp {36}{1995}{1462}.}
  \ref{Dows}{Dowker,J.S. \aop{71}{1972}{577}; Dowker,J.S. and Pettengill,D.F.
  \jpa{7}{1974}{1527}; J.S.Dowker in {\it Quantum Gravity}, edited by
  S. C. Christensen (Hilger,Bristol,1984)}
  \ref{Jadhav2}{Jadhav,S.P. \prD{43}{1991}{2656}.}
  \ref{Dow11}{Dowker,J.S. \cqg{21}{2004}4247.}
  \ref{Dow12}{Dowker,J.S. \cqg{21}{2004}4977.}
  \ref{Dow13}{Dowker,J.S. \jpa{38}{2005}1049.}
  \ref{Zagier}{Zagier,D. \ma{202}{1973}{149}}
  \ref{RandG}{Rademacher, H. and Grosswald,E. {\it Dedekind Sums},
  (Carus, MAA, 1972).}
  \ref{Berndt7}{Berndt,B, \aim{23}{1977}{285}.}
  \ref{HKMM}{Harvey,J.A., Kutasov,D., Martinec,E.J. and Moore,G.
  {\it Localised Tachyons and RG Flows}, hep-th/0111154.}
  \ref{Beck}{Beck,M., {\it Dedekind Cotangent Sums}, {\it Acta Arithmetica}
  {\bf 109} (2003) 109-139 ; math.NT/0112077.}
  \ref{McInnes}{McInnes,B. {\it APS instability and the topology of the brane
  world}, hep-th/0401035.}
  \ref{BHS}{Brevik,I, Herikstad,R. and Skriudalen,S. {\it Entropy Bound for the
  TM Electromagnetic Field in the Half Einstein Universe}; hep-th/0508123.}
  \ref{BandO}{Brevik,I. and Owe,C.  \prD{55}{4689}{1997}.}
  \ref{Kenn}{Kennedy,G. Thesis. University of Manchester 1978.}
  \ref{KandU}{Kennedy,G. and Unwin S. \jpa{12}{L253}{1980}.}
  \ref{BandO1}{Bayin,S.S.and Ozcan,M.
  \prD{48}{2806}{1993}; \prD{49}{5313}{1994}.}
  \ref{Chang}{Chang, P., {\it Quantum Field Theory on Regular Polytopes}.
   Thesis. University of Manchester, 1993.}
  \ref{Barnesa}{Barnes,E.W. {\it Trans. Camb. Phil. Soc.} {\bf 19} (1903) 374.}
  \ref{Barnesb}{Barnes,E.W. {\it Trans. Camb. Phil. Soc.}
  {\bf 19} (1903) 426.}
  \ref{Stanley1}{Stanley,R.P. \joa {49Hilf}{1977}{134}.}
  \ref{Stanley}{Stanley,R.P. \bams {1}{1979}{475}.}
  \ref{Hurley}{Hurley,A.C. \pcps {47}{1951}{51}.}
  \ref{IandK}{Iwasaki,I. and Katase,K. {\it Proc.Japan Acad. Ser} {\bf A55}
  (1979) 141.}
  \ref{IandT}{Ikeda,A. and Taniguchi,Y. {\it Osaka J. Math.} {\bf 15} (1978)
  515.}
  \ref{GandM}{Gallot,S. and Meyer,D. \jmpa{54}{1975}{259}.}
  \ref{Flatto}{Flatto,L. {\it Enseign. Math.} {\bf 24} (1978) 237.}
  \ref{OandT}{Orlik,P and Terao,H. {\it Arrangements of Hyperplanes},
  Grundlehren der Math. Wiss. {\bf 300}, (Springer--Verlag, 1992).}
  \ref{Shepler}{Shepler,A.V. \joa{220}{1999}{314}.}
  \ref{SandT}{Solomon,L. and Terao,H. \cmh {73}{1998}{237}.}
  \ref{Vass}{Vassilevich, D.V. \plb {348}{1995}39.}
  \ref{Vass2}{Vassilevich, D.V. \jmp {36}{1995}3174.}
  \ref{CandH}{Camporesi,R. and Higuchi,A. {\it J.Geom. and Physics}
  {\bf 15} (1994) 57.}
  \ref{Solomon2}{Solomon,L. \tams{113}{1964}{274}.}
  \ref{Solomon}{Solomon,L. {\it Nagoya Math. J.} {\bf 22} (1963) 57.}
  \ref{Obukhov}{Obukhov,Yu.N. \pl{109B}{1982}{195}.}
  \ref{BGH}{Bernasconi,F., Graf,G.M. and Hasler,D. {\it The heat kernel
  expansion for the electromagnetic field in a cavity}; math-ph/0302035.}
  \ref{Baltes}{Baltes,H.P. \prA {6}{1972}{2252}.}
  \ref{BaandH}{Baltes.H.P and Hilf,E.R. {\it Spectra of Finite Systems}
  (Bibliographisches Institut, Mannheim, 1976).}
  \ref{Ray}{Ray,D.B. \aim{4}{1970}{109}.}
  \ref{Hirzebruch}{Hirzebruch,F. {\it Topological methods in algebraic
  geometry} (Springer-- Verlag,\break  Berlin, 1978). }
  \ref{BBG}{Bla\v{z}i\'c,N., Bokan,N. and Gilkey, P.B. {\it Ind.J.Pure and
  Appl.Math.} {\bf 23} (1992) 103.}
  \ref{WandWi}{Weck,N. and Witsch,K.J. {\it Math.Meth.Appl.Sci.} {\bf 17}
  (1994) 1017.}
  \ref{Norlund}{N\"orlund,N.E. \am{43}{1922}{121}.}
   \ref{Norlund1}{N\"orlund,N.E. {\it Differenzenrechnung} (Springer--Verlag, 1924, Berlin.)}
  \ref{Duff}{Duff,G.F.D. \aom{56}{1952}{115}.}
  \ref{DandS}{Duff,G.F.D. and Spencer,D.C. \aom{45}{1951}{128}.}
  \ref{BGM}{Berger, M., Gauduchon, P. and Mazet, E. {\it Lect.Notes.Math.}
  {\bf 194} (1971) 1. }
  \ref{Patodi}{Patodi,V.K. \jdg{5}{1971}{233}.}
  \ref{GandS}{G\"unther,P. and Schimming,R. \jdg{12}{1977}{599}.}
  \ref{MandS}{McKean,H.P. and Singer,I.M. \jdg{1}{1967}{43}.}
  \ref{Conner}{Conner,P.E. {\it Mem.Am.Math.Soc.} {\bf 20} (1956).}
  \ref{Gilkey2}{Gilkey,P.B. \aim {15}{1975}{334}.}
  \ref{MandP}{Moss,I.G. and Poletti,S.J. \plb{333}{1994}{326}.}
  \ref{BKD}{Bordag,M., Kirsten,K. and Dowker,J.S. \cmp{182}{1996}{371}.}
  \ref{RandO}{Rubin,M.A. and Ordonez,C. \jmp{25}{1984}{2888}.}
  \ref{BaandD}{Balian,R. and Duplantier,B. \aop {112}{1978}{165}.}
  \ref{Kennedy2}{Kennedy,G. \aop{138}{1982}{353}.}
  \ref{DandKi2}{Dowker,J.S. and Kirsten, K. {\it Analysis and Appl.}
 {\bf 3} (2005) 45.}
  \ref{Dow40}{Dowker,J.S. \cqg{23}{2006}{1}.}
  \ref{BandHe}{Br\"uning,J. and Heintze,E. {\it Duke Math.J.} {\bf 51} (1984)
   959.}
  \ref{Dowl}{Dowker,J.S. {\it Functional determinants on M\"obius corners};
    Proceedings, `Quantum field theory under
    the influence of external conditions', 111-121,Leipzig 1995.}
  \ref{Dowqg}{Dowker,J.S. in {\it Quantum Gravity}, edited by
  S. C. Christensen (Hilger, Bristol, 1984).}
  \ref{Dowit}{Dowker,J.S. \jpa{11}{1978}{347}.}
  \ref{Kane}{Kane,R. {\it Reflection Groups and Invariant Theory} (Springer,
  New York, 2001).}
  \ref{Sturmfels}{Sturmfels,B. {\it Algorithms in Invariant Theory}
  (Springer, Vienna, 1993).}
  \ref{Bourbaki}{Bourbaki,N. {\it Groupes et Alg\`ebres de Lie}  Chap.III, IV
  (Hermann, Paris, 1968).}
  \ref{SandTy}{Schwarz,A.S. and Tyupkin, Yu.S. \np{242}{1984}{436}.}
  \ref{Reuter}{Reuter,M. \prD{37}{1988}{1456}.}
  \ref{EGH}{Eguchi,T. Gilkey,P.B. and Hanson,A.J. \prp{66}{1980}{213}.}
  \ref{DandCh}{Dowker,J.S. and Chang,Peter, \prD{46}{1992}{3458}.}
  \ref{APS}{Atiyah M., Patodi and Singer,I.\mpcps{77}{1975}{43}.}
  \ref{Donnelly}{Donnelly.H. {\it Indiana U. Math.J.} {\bf 27} (1978) 889.}
  \ref{Katase}{Katase,K. {\it Proc.Jap.Acad.} {\bf 57} (1981) 233.}
  \ref{Gilkey3}{Gilkey,P.B.\invm{76}{1984}{309}.}
  \ref{Degeratu}{Degeratu.A. {\it Eta--Invariants and Molien Series for
  Unimodular Groups}, Thesis MIT, 2001.}
  \ref{Seeley}{Seeley,R. \ijmp {A\bf18}{2003}{2197}.}
  \ref{Seeley2}{Seeley,R. .}
  \ref{melrose}{Melrose}
  \ref{DandW}{Douglas,R.G. and Wojciekowski,K.P. \cmp{142}{1991}{139}.}
  \ref{Dai}{Dai,X. \tams{354}{2001}{107}.}
\end{putreferences}

\bye